\documentclass[notoc]{JHEP3}
\def\[{\left [} \def\]{\right ]}
\def\({\left (}
\def\){\right )}
\def\lbr{\left\{}
\def\rbr{\right\}}
\def\l{\left.}
\def\r{\right.}
\def\nn{\nonumber}

\title{Gaugino and Higgsino Coannihilations I:  Neutralino-Neutralino Interactions}
\author{Andreas Birkedal-Hansen\\ Department of Physics, University of California, Berkeley, California, 94720 and Theoretical Physics Group, Ernest Orlando Lawrence National Laboratory, University Of California, Berkeley, California, 94720\\ E-mail: \email{andreas1@uclink4.berkeley.edu}}
\author{Eunhwa Jeong\\ Department of Physics, University of California, Berkeley, California, 94720\\ E-mail: \email{ehjeong@socrates.berkeley.edu}}
\abstract{We present exact analytic cross sections for all neutralino-neutralino coannihilations into two-body tree level final states.  These expressions allow the calculation of important contributions to the neutralino relic abundance over large regions of mSUGRA parameter space and are particularly useful in theories with non-universal gaugino masses.}
\preprint{UCB-PTH-02/41 \\
LBNL-51547 \\}
\begin{document}
\section{Introduction}
The lightest neutralino is one of the strongest candidates to explain dark matter.  It naturally fits the definition of a weakly interacting massive particle (WIMP) and is the lightest supersymmetric particle in many models in the literature.  Furthermore, the lightest supersymmetric particle (LSP) is stable in most viable models involving supersymmetry.  These properties allow the neutralino to provide a significant fraction of critical density to the universe.  Throughout this paper, we assume that the lightest neutralino is the LSP and that it is the dark matter candidate whose relic density we will be calculating.

Measurements of the dark matter relic density are becoming a powerful constraint on all extensions of the standard model of particle physics.  Recent improvements in these measurements allow dark matter densities to be even more discriminating~\cite{dmmeasure}.  In the most studied extension, mSUGRA, only a few viable regions of parameter space are left~\cite{mSUGRA,B-H&Nelson01}.  Thus, it is becoming increasingly important to have accurate calculations of the dark matter density.

In this paper, we present analytic cross sections for all tree level neutralino-neutralino coannihilations~\cite{Griest:1990kh,Binetruy:1983jf} in the MSSM for general neutralinos.  Since our calculations allow the possibility that the two interacting neutralinos are not identical, our results include and generalize previously published expressions.  In an effort to standardize the notation of neutralino cross section calculations, and for ease of comparison, we closely follow the notation of~\cite{NRR}.  We also make no simplifying assumptions about a given MSSM model beyond neglecting most of the CP-violating phases in the SUSY sector.  That is, we make no additional assumptions about the neutralino, chargino, or sfermion sectors.  We also include all s-channel widths.  Thus, these results provide a step forward towards a complete analytic calculation of the neutralino relic density without using the normal velocity expansion.  Additionally, these results are valid regardless of proximity to resonances and final state thresholds.  However, coannihilations between the neutralino and a chargino, a neutralino and a sfermion, and two charginos can also be important in relic density calculations.  Results for squared matrix elements involving stau and stop coannihilation can be found elsewhere in the literature~\cite{NRR2,Ellis:2001nx,Ellis:1999mm}.  Complete cross section expressions involving charginos will be presented by the authors in the future.

The calculated expressions are voluminous.  The size of the expressions and the varying conventions in the literature forbid most direct analytical checks.  However, as we have taken steps to follow most conventions of~\cite{NRR}, we are able to check our answers in the limit of the two initial state neutralinos being identical.  Our results match exactly.  

In addition to including the standard neutralino relic density contributions, these expressions also include contributions which are important when the mass of the next-to-lightest neutralino is approximately degenerate with the lightest neutralino.  In mSUGRA, this occurs in the higgsino-dominated region of parameter space characterized by large $m_{0}$~\cite{Feng:2000gh}.  As $m_{0}$ is increased, the $\mu$ parameter eventually falls, leading to high higgsino content.  If $m_{0}$ is increased even more, it quickly becomes impossible to achieve correct electroweak symmetry breaking.  This difficulty in acquiring neutralino mass degeneracy in mSUGRA is directly related to having universal gaugino masses at the SUSY-breaking scale.  In contrast to mSUGRA, many models in the literature, including models derived from the heterotic string~\cite{Binetruy:1996nx,Binetruy:1997vr,Gaillard:1999et,Gaillard:1999yb} and the Type-I superstring~\cite{Ibanez:1998rf}, generically have non-universal gaugino masses.  Non-universal gaugino masses allow for larger sections of parameter space where the two lightest neutralinos are almost degenerate.  The dark matter prospects of non-universal gaugino masses have been studied in~\cite{B-H&Nelson01}, where it was pointed out that such models compare quite favorably with mSUGRA in terms of both motivation and neutralino relic density.

The outline of the paper is as follows.  In Section 2 we review the relic density calculation.  Here, again, we follow the format of~\cite{NRR}, which uses the simplifying work of~\cite{Gondolo:1990dk,Edsjo:1997bg} to motivate the format of the cross sections.  In Section 3 we present the cross section expressions for all tree level coannihilations of two neutralinos.  These expressions are exact and include integration over the final state angular distribution.  Since they are exact, they can be used even when the masses of the two neutralinos are far from degenerate.  This allows the expressions to be used for purposes in addition to relic density calculations.  We conclude in Section 4.  

\section{Review of the Relic Density Calculation}

The calculation of a relic density from thermal considerations is standard~\cite{Kolb&Turner}.  One starts with a theory that contains a particle expected to have a significant relic density.  One wishes to calculate the particle's contribution to the total matter density of the universe, $\Omega_{\chi}=\rho_{\chi}/\rho_{tot}=m_{\chi} n_{\chi}/\rho_{tot}$.  This number is then compared to the current experimental bounds on a particle's relic density.  The limit for the dark matter density is generally taken to be $0.1 \leq \Omega_{\chi} h^2 \leq 0.3$ though more aggressive limits have been promoted~\cite{dmmeasure}.

The initial number density of the neutralino is assumed to be determined because the particle is in thermal equilibrium.  When the particle begins in thermal equilibrium with its surroundings, interactions that create neutralinos usually happen as frequently as reverse interactions which destroy neutralinos.  Once the temperature drops below $T\simeq m_{\chi}$, most particles no longer have sufficient energy to create neutralinos.  Now neutralinos can only annihilate, and these annihilations occur until about the time when the Hubble expansion parameter becomes larger than the annihilation rate, $H\geq \Gamma_{ann}$.  When expansion dwarfs annihilation, neutralinos are being separated apart from each other too quickly for annihilation to be efficient.  This happens at the freeze-out temperature, usually at $T_{F}\simeq m_{\chi}/20$.

In most relic density calculations, the only interaction cross sections that need to be calculated are annihilations of the type $\chi \chi \rightarrow X$ where $X$ is any final state involving only standard model particles.  However, there are scenarios in which other particles in the thermal bath have important effects on the evolution of the neutralino relic density.   Such a particle, called a coannihilator, usually must have direct interactions with the neutralino and also must be very nearly degenerate in mass.  When this occurs, the neutralino and all relevant coannihilators form a coupled system.  In this section,  we will denote particles belonging to that coupled system by $\chi_{i}$.  Now all interactions involving particles in this coupled system come into play, including $\chi_{i} \chi_{j}\rightarrow X$, $\chi_{i} X \rightarrow \chi_{j} Y$, and $\chi_{i} \rightarrow \chi_{j} X$.  Here both $X$ and $Y$ denote states including standard model particles.  Decays once again enter the calculation because the coannihilators are generally not stable and eventually decay into the lightest neutralino.  

The effect of this coupled system on the relic density of neutralinos can best be understood through a consideration of the freeze-out temperature.  Because these coannihilators have essentially the same mass as the neutralino and share common interactions, they remain in thermal equilibrium for about as long as the neutralinos.  The neutralino can thus interact through more channels with more particles than in the usual single-species annihilation scenario.  This raises the effective annihilation rate, so the determining equation for freeze-out ($H\geq \Gamma_{ann}$) is satisfied at a later time and a lower temperature.  The neutralino is nonrelativistic during freeze-out, so its equilibrium number density is decreasing exponentially with decreasing temperature.  Since coannihilation keeps the neutralino in equilibrium down to a lower temperature, the number density at freeze-out is lower.  This lowering of the number density is the most relevant effect of coannihilation~\cite{Griest:1990kh} and can have drastic consequences for the viable parameter space of many supersymmetric models.  In the MSSM, many particles can serve as the coannihilator: the lightest stop, the lightest stau, any of the other neutralinos, or either of the charginos.  The cases of the stop and stau have been studied extensively in the literature~\cite{NRR2,Ellis:2001nx,Ellis:1999mm}.  Here we study the case of coannihilation between the lightest neutralino and another neutralino, reserving presentation of the results for coannihilation between the lightest neutralino and a chargino for future work.

For the case without coannihilations, evolution of the particle number density happens in accordance with the single species Boltzmann equation:

\begin{eqnarray}
\frac{dn}{dt} = - 3 H n -\langle \sigma v \rangle \left[n^2-\left(n^{eq}\right)^2\right]
\end{eqnarray}
The number density is reduced by Hubble expansion and annihilations  of the relic particle.  The number density of the relic is also affected by its own decays and inverse annihilation processes.  In most cases, including the versions of the MSSM studied here, the relic particle is assumed to be stable, so relic decay is neglected.  In the above expression, we have also assumed T invariance to relate annihilation and inverse annihilation processes.  As noted earlier, the thermally averaged cross section $\langle \sigma v \rangle$ involves only processes such as $\chi_{1}^{0} \chi_{1}^{0}\rightarrow X$.

In the presence of coannihilators, the Boltzmann equation becomes more complicated (Equation (27) from~\cite{Edsjo:1997bg}):

\begin{eqnarray}
\frac{d n_{i}}{dt}=&-&3 H n_{i} -\sum_{j=1}^{N} \langle\sigma_{ij} v_{ij}\rangle \left(n_{i} n_{j} - n_{i}^{eq} n_{j}^{eq}\right)\nn\\
&-&\sum_{j\neq i}\left[\langle\sigma_{X ij}' v_{ij}\rangle\left(n_{i} n_{X} - n_{i}^{eq} n_{X}^{eq}\right)-\langle\sigma_{X ji}' v_{ij}\rangle\left(n_{j} n_{X} - n_{j}^{eq} n_{X}^{eq}\right)\right]\nn\\
&-&\sum_{j\neq i} \left[\Gamma_{ij}\left(n_{i} -n_{i}^{eq}\right)-\Gamma_{ji}\left(n_{j} -n_{j}^{eq}\right)\right]
\end{eqnarray}
Here $i$ and $j$ can denote the relic particle or any possible coannihilators.  The velocity $v_{ij}$ is the relative velocity between particle $i$ and particle $j$.  The first cross section $\sigma_{ij}$ is a sum over final states of processes such as $\chi_{i} \chi_{j} \rightarrow X$.  $\sigma_{X ij}'$ is a sum over standard model final state particles of $\chi_{i} X \rightarrow \chi_{j} Y$.  The $\Gamma$ terms describe possible decays of the coannihilators $\chi_{i} \rightarrow \chi_{j} X$.

One uses the different stability properties of the relic particle and the coannihilators to simplify the expression.  As stated above, the relic particle is generally assumed to be stable (usually from R-parity in the MSSM).  The coannihilators' decay rates are also generally quite fast compared to the age of the universe.  In the MSSM, R-parity also guarantees that each coannihilator will eventually decay into standard model particles and the lightest neutralino (being the LSP).  Since the LSP is the relic particle that we are interested in, we can sum the Boltzmann equation over the index $i$ to get an equation for the final number density of the relic particle~\cite{Griest:1990kh}.  Now we use $n=\sum_{i=1}^{N} n_{i}$.  The terms in the second and third lines now cancel, so we have:

\begin{eqnarray}
\frac{d n}{dt}=&-&3 H n -\sum_{i,j=1}^{N} \langle\sigma_{ij} v_{ij}\rangle \left(n_{i} n_{j} - n_{i}^{eq} n_{j}^{eq}\right)
\end{eqnarray}

To a very good approximation, one can use the usual single species Boltzmann equation for the case of coannihilations if the following replacement is made for the thermally averaged cross section:

\begin{eqnarray}
\langle \sigma v\rangle = \sum_{i,j} \langle \sigma_{ij} v_{ij} \rangle \frac{n_{i}^{eq}}{n^{eq}}\frac{n_{j}^{eq}}{n^{eq}}
\end{eqnarray}

The thermally averaged cross section deserves careful treatment.  For scenarios involving coannihilations, the expression for the thermally averaged cross section begins as a six dimensional integral, seven dimensional including the integration over the final state angle.  This integral has been conveniently put into the form of a one-dimensional integral over the total squared center of mass energy~\cite{Edsjo:1997bg}:

\begin{equation}
\langle \sigma v\rangle = \frac{1}{8 m_{\chi}^4 T}\frac{\int_{4 m_{\chi}^2}^{\infty} ds\; s^{3/2} \sum_{i,j} \beta_{f}^{2} \left(s,m_{i},m_{j}\right)\frac{g_{i} g_{j}}{g_{\chi}^2}\sigma_{ij}\left(s\right) K_{1}\left(\frac{\sqrt{s}}{T}\right)}{\left(\sum_{i}\frac{g_{i}}{g_{\chi}}\frac{m_{i}^2}{m_{\chi}^2}K_{2}\left(\frac{m_{i}}{T}\right)\right)^2}
\end{equation}
Here $\sigma_{ij}\left(s\right)$ is the sum of all cross sections for any process $\chi_{i} \chi_{j}\rightarrow X$ and $\beta_{f} \left(s,m_{i},m_{j}\right)$ is a function related to the incoming and outgoing momenta:

\begin{equation}
\beta_{f} \left(s,m_{i},m_{j}\right) = \frac{1}{s}\sqrt{\left(s-\left(m_{i}+m_{j}\right)^2\right)\left(s-\left(m_{i}-m_{j}\right)^2\right)}
\end{equation}
Consideration of 2-body final states is sufficient for almost all relic density calculations, that is what we consider in this paper.  Part of the integrand can be written in the form:

\begin{equation}
\frac{s}{2} \beta_{f} \left(s,m_{i},m_{j}\right) \sigma_{ij}\left(s\right)=\frac{1}{32 \pi}\sum_{f_{1} f_{2}}\left[c\; \theta\!\left(s-\left(m_{f_{1}}+m_{f_{2}}\right)^2\right)\beta_{f}\left(s,m_{f_{1}},m_{f_{2}}\right)\tilde{\omega}_{ij\rightarrow f_{1} f_{2}} \left(s\right)\right]
\end{equation}
Here $c$ is a color factor, equal to three if the final state is a quark-antiquark pair, equal to one otherwise.  $\theta\!\left(s-\left(m_{f_{1}}+m_{f_{2}}\right)^2\right)$ is the usual step function to enforce the hard lower limit on the squared center of mass energy.  The $\tilde{\omega}$ functions are the squared amplitudes averaged over the final state angle:

\begin{equation}
\tilde{\omega}_{ij\rightarrow f_{1} f_{2}} \left(s\right)= \frac{1}{2} \int_{-1}^{+1} d \cos \theta_{CM} \left|A\left(ij\rightarrow f_{1} f_{2}\right)\right|^2
\end{equation}
Computer codes exist which numerically perform this final state angular integration.  Some codes include only a subset of all possible coannihilation channels~\cite{Gondolo:2000ee,Jungman:1995df} while the program {\it micrOMEGAs} includes all relevant coannihilation channels~\cite{Belanger:2001fz}.  We have performed the integration analytically, thus reducing the number of necessary numerical integrations and allowing for easier analysis of relic density properties.  The next section lists these $\tilde{\omega}_{ij\rightarrow f_{1} f_{2}} \left(s\right)$ functions for all possible 2-body final states from neutralino (co)annihilation.

Once the effective thermally averaged cross section has been determined as a function of temperature, it is used to iteratively calculate the freeze-out temperature.  In practice, the dimensionless inverse freeze-out temperature $x_{F} = m_{\chi}/T_{F}$ is calculated using:

\begin{equation}
x_{F} = \ln \left(\frac{0.038\, g\, m_{Pl} \,m_{\chi} \,\langle \sigma v \rangle }{\sqrt{g_{*} \,x_{F}}}\right)
\end{equation} 
Here $m_{Pl}$ is the Planck mass, $g$ is the total number of degrees of freedom of the $\chi$ particle (spin, color, etc.), $g_{*}$ is the total number of effective relativistic degrees of freedom at freeze-out, and the thermally averaged cross section is evaluated at the freeze-out temperature.  For most cases in the MSSM, $x_{F}\simeq 20$.

When the freeze-out temperature has been determined, the total (co)annihilation depletion of the neutralino number density can be calculated by integrating the thermally averaged cross section from freeze-out to the present temperature (essentially $T = 0$).  The thermally averaged cross section appears in the formula for the relic density in the form of the integral $J\left(x_{F}\right) = \int_{x_{F}}^{\infty} \langle\sigma v \rangle \, x^{-2} dx$ as

\begin{equation}
\Omega_{\chi} h^2 = 40 \sqrt{\frac{\pi}{5}}\frac{h^2}{H_{0}^2}\frac{s_{0}}{m_{Pl}^{3}}\frac{1}{\left(g_{*S}/g_{*}^{1/2}\right) \, J\left(x_{F}\right)},
\end{equation}
which is commonly given as

\begin{equation}
\Omega_{\chi} h^2 = \frac{1.07 \times 10^9 \, GeV^{-1}}{g_{*}^{1/2} \, m_{Pl}^{} \, J\left(x_{F}\right)}.
\end{equation}
This is the expression that one compares with the relic density limit of $0.1 \leq \Omega h^2 \leq 0.3$.  Once a freeze out temperature has been determined, one must still integrate over $x$ to determine the relic density.  In practice, these integrations are usually done numerically.  Previously, it was also necessary to integrate over the final state angular distribution.  The results listed in the next section, along with those of~\cite{NRR, NRR2}, have already performed this angular integration.   Our results also represent a significant step towards a full tabulation of all analytic cross sections relevant in the accurate determination of a relic neutralino density.

\section{Exact Cross Section Expressions}
In this section we list all possible $\tilde{\omega}$ functions for two neutralinos (co)annihilating.  All of these processes involve exchange via $s$, $t$ and $u$-channels.  We use the following definitions for the sum and difference of the neutralino masses:

\begin{eqnarray}
\sigma = m_{\chi^{0}_{i}}+ m_{\chi^{0}_{j}}\\
\delta = m_{\chi^{0}_{i}}- m_{\chi^{0}_{j}}\nonumber
\end{eqnarray}

We refer the reader to~\cite{NRR} for the definitions of the $\mathcal{T}_{i}$ and $\mathcal{Y}_{i}$ functions.  Since we are studying a generalization of the processes studied in~\cite{NRR}, their definitions for particle couplings are also used here.  Furthermore, we suppress most of the functional dependence of the $\mathcal{F}$ functions.  The suppressed dependencies are apparent from the $\mathcal{T}_{i}$ and $\mathcal{Y}_{i}$ functions.  We also refer the reader to~\cite{NRR} for an explicit listing.  However, there are some places where the difference between the two neutralino masses comes into play.  If the function $F$ (listed in~\cite{NRR}) is defined in terms of the incoming ($p$) and outgoing ($k$) CM momenta as $F=2\, k\, p$, then the only important difference is in the function $D$:

\begin{equation}
D\left(s,x_{1},x_{2},y_{1},y_{2}\right) = \frac{x_{1}+x_{2}}{2}+\frac{y_{1}+y_{2}}{2}-\frac{s}{2}-\frac{\left(x_{1}-x_{2}\right)\left(y_{1}-y_{2}\right)}{2s}
\end{equation}

Here $x_{1}$ and $x_{2}$ are the incoming squared particle masses and $y_{1}$ and $y_{2}$ are the outgoing squared particle masses.  

\subsection*{\mbox{{\Large$\underline{\bf{\chi^{0}_{i} \chi^{0}_{j}\rightarrow h H}}$}}}

Contributions to $\tilde{\omega}$ come from s-channel Higgs boson exchange, t- and u-channel neutralino exchange and cross terms

\begin{equation}
\tilde{\omega}_{\chi^{0}_{i} \chi^{0}_{j}\rightarrow H h}=\tilde{\omega}^{\(h,H\)}_{Hh}+\tilde{\omega}^{\(\chi^{0}\)}_{Hh}+\tilde{\omega}^{\(h,H-\chi^{0}\)}_{hH}:
\end{equation}
\begin{flushleft}
\textbf{S-channel CP-even Higgs bosons (h,H):}
\end{flushleft}

\begin{equation}
\tilde{\omega}^{\(h,H\)}_{Hh} = \frac{1}{2}\sum_{r=h,H}\left|\frac{C^{r h H} C^{\chi_{i} \chi_{j} r}_{S}}{s-m_{r}^{2} + i m_{r} \Gamma_{r}} \right|^{2}\(s-\sigma^{2}\);
\end{equation}
\begin{flushleft}
\textbf{T- and U-channel neutralino:}
\end{flushleft}

\begin{eqnarray}
\tilde{\omega}^{\(\chi^{0}\)}_{Hh} =\frac{1}{2}\sum_{k,l=1}^{4} \(m_{\chi_{k}^{0}} m_{\chi_{l}^{0}} I_{k l}^{h H}+m_{\chi_{k}^{0}} J_{k l}^{h H}+ K_{k l}^{h H}\)\, ,\nn\\
\end{eqnarray}

where

\begin{eqnarray}
I_{k l}^{h H} =&& \frac{s^2-s\(\sigma^2+\delta^2\)+\sigma^2 \delta^2}{s}\[ C_{Hh}^{\mathcal{T}} \mathcal{T}_{0}-C_{Hh}^{\mathcal{Y}} \mathcal{Y}_{0}\]\big(s, \sigma,\delta,m_{h}^2,m_{H}^2,m_{\chi_{k}^0}^2,m_{\chi_{l}^0}^2\big)\, ,\\
J_{k l}^{h H} =&& \frac{1}{4s} \(-8 C_{Hh}^{\mathcal{T}} s \sigma \mathcal{T}_{1}+ G_{Hh}^{J,T\(0\)} \mathcal{T}_{0} \r\nn\\
&-&\l  4 C_{Hh}^{\mathcal{Y}} s \sigma\(\mathcal{Y}_{1}+\(s-\sigma^2\)\mathcal{Y}_{0}\)\)\big(s, \sigma,\delta,m_{h}^2,m_{H}^2,m_{\chi_{k}^0}^2,m_{\chi_{l}^0}^2\big)\, ,\\
K_{k l}^{h H} =&-&\frac{1}{16 s}\[16 C_{Hh}^{\mathcal{T}} s \mathcal{T}_{2}+G_{Hh}^{K,T\(1\)}\mathcal{T}_{1}+ G_{Hh}^{K,T\(0\)}\mathcal{T}_{0}\r\nn\\
&+&\l 16 C_{Hh}^{\mathcal{Y}} s  \mathcal{Y}_{2}- 8\(C_{Hh}^{\mathcal{Y}}-2 C_{Hh}^{\mathcal{Y},1}\)\sigma \delta \(m_{H}^2-m_{h}^2\) \mathcal{Y}_{1}\r\nn\\
&+&\l C_{Hh}^{\mathcal{Y}} G_{Hh}^{K,Y\(0\)}\mathcal{Y}_{0}\]\big(s, \sigma,\delta,m_{h}^2,m_{H}^2,m_{\chi_{k}^{0}}^2,m_{\chi_{l}^{0}}^2\big),
\end{eqnarray}

and

\begin{eqnarray}
G_{Hh}^{J, T\(0\)} =&& 8 C_{Hh}^{\mathcal{T},1}\delta\(m_{H}^2-m_{h}^2\)\(s-\sigma^2\)\nn\\
&+& 2 C_{Hh}^{\mathcal{T}} s \(2\(m_{H}^2\(\sigma-\delta\)+m_{h}^2\(\sigma+\delta\)\)-\sigma\(\sigma^2-\delta^2\)\)\, ,\\
G_{Hh}^{K,T\(1\)} =&& 8 C_{Hh}^{\mathcal{T}}s\(2\(s-\(m_{H}^2+m_{h}^2\)\)+\sigma^2-\delta^2\) + 16 C_{Hh}^{\mathcal{T},1} \sigma \delta \(m_{H}^2-m_{h}^2\)\, ,\\
G_{Hh}^{K,T\(0\)} =&& \frac{1}{s}\(C_{Hh}^{\mathcal{T}} s^2\(\(\sigma-\delta\)^2-4m_{h}^2\)\(\(\sigma+\delta\)^2-4 m_{H}^2\)\r\nn\\
&-&\l 8 C_{Hh}^{\mathcal{T},1} \sigma \delta \(m_{H}^2-m_{h}^2\)\(s\(2\(m_{H}^2+m_{h}^2\)-\sigma^2+\delta^2\)\r\r\nn\\
&-&\l\l 2 \sigma \delta\(m_{H}^2-m_{h}^2\)\)\)\, ,\\
G_{Hh}^{K,Y\(0\)} =&& s\(\(\delta^2-4 m_{H}^2\)\(\delta^2-4 m_{h}^2\)+8 \sigma \delta \(m_{H}^2-m_{h}^2\)\r\nn\\
&+&\l 2 \sigma^2\(2\(m_{H}^2+m_{h}^2\)+\delta^2\)-3 \sigma^4\)\nn\\
&-&  4 \sigma \delta \(m_{H}^2-m_{h}^2\)\(2\(m_{H}^2+m_{h}^2\)+\sigma^2+\delta^2\),
\end{eqnarray}
and here we have defined the coupling functions as

\begin{eqnarray}
C_{Hh}^{\mathcal{T}}=&&C_{Hh}^{\mathcal{T},1}+C_{Hh}^{\mathcal{T},2}\, ,\\
C_{Hh}^{\mathcal{T},1}=&&C_{S}^{\chi_{i}^{0} \chi_{k}^{0} h} C_{S}^{\chi_{k}^{0} \chi_{j}^{0} H} \(C_{S}^{\chi_{i}^{0} \chi_{l}^{0} h}\)^{*} \(C_{S}^{\chi_{l}^{0} \chi_{j}^{0} H}\)^{*}\, ,\\
C_{Hh}^{\mathcal{T},2}=&&C_{S}^{\chi_{i}^{0} \chi_{k}^{0} H} C_{S}^{\chi_{k}^{0} \chi_{j}^{0} h} \(C_{S}^{\chi_{i}^{0} \chi_{l}^{0} H}\)^{*} \(C_{S}^{\chi_{l}^{0} \chi_{j}^{0} h}\)^{*},
\end{eqnarray}

\begin{eqnarray}
C_{Hh}^{\mathcal{Y}}=&&C_{Hh}^{\mathcal{Y},1}+C_{Hh}^{\mathcal{Y},2}\, ,\\
C_{Hh}^{\mathcal{Y},1}=&&C_{S}^{\chi_{i}^{0} \chi_{k}^{0} h} C_{S}^{\chi_{k}^{0} \chi_{j}^{0} H} \(C_{S}^{\chi_{i}^{0} \chi_{l}^{0} H}\)^{*} \(C_{S}^{\chi_{l}^{0} \chi_{j}^{0} h}\)^{*}\, ,\\
C_{Hh}^{\mathcal{Y},2}=&&C_{S}^{\chi_{i}^{0} \chi_{k}^{0} H} C_{S}^{\chi_{k}^{0} \chi_{j}^{0} h} \(C_{S}^{\chi_{i}^{0} \chi_{l}^{0} h}\)^{*} \(C_{S}^{\chi_{l}^{0} \chi_{j}^{0} H}\)^{*};
\end{eqnarray}
\\
\begin{flushleft}
\textbf{Higgs (h, H)-neutralino cross term:}
\end{flushleft}

\begin{eqnarray}
\tilde{\omega}^{\(h,H-\chi^{0}\)}_{hH}=\frac{1}{4s} \sum_{l=1}^{4} \sum_{r=h,H} Re\[\(\frac{C_{S}^{h H r} C_{S}^{\chi^{0}_{i} \chi^{0}_{j} r}}{s-m_{r}^{2} + i m_{r} \Gamma_{r}}\)^{*}\( C_{+}^{i j l} G_{h H}^{C_{+}}+ C_{-}^{i j l} G_{h H}^{C_{-}} \)\],
\end{eqnarray}

\begin{eqnarray}
G_{h H}^{C_{+}} =&-&4s \sigma+\(4 m_{\chi_{l}^{0}}\(s^2-s \sigma\(\sigma+m_{\chi_{l}^{0}}\)\)+ s\(\sigma\(2\(m_{H}^2+m_{h}^2\)+\delta^2\)-\sigma^2\)\r\nn\\
&-&\l 2\sigma \delta\(m_{H}^2-m_{h}^2\)\)\mathcal{F}\[s, \sigma,\delta,m_{h}^2,m_{H}^2,m_{\chi^{0}_{l}}^2\]\, ,\\
G_{h H}^{C_{-}} =&& \delta \(s-\sigma^2\)\(m_{h}^2-m_{H}^2\)\mathcal{F}\[s, \sigma,\delta,m_{h}^2,m_{H}^2,m_{\chi^{0}_{l}}^2\],
\end{eqnarray}
where the coupling functions are

\begin{eqnarray}
C_{+}^{i j l}=&& C_{S}^{\chi^{0}_{i} \chi^{0}_{l} H} C_{S}^{\chi^{0}_{l} \chi^{0}_{j} h}+C_{S}^{\chi^{0}_{i} \chi^{0}_{l} h} C_{S}^{\chi^{0}_{l} \chi^{0}_{j} H}\, ,\\
C_{-}^{i j l}=&& C_{S}^{\chi^{0}_{i} \chi^{0}_{l} H}C_{S}^{\chi^{0}_{l} \chi^{0}_{j} h}-C_{S}^{\chi^{0}_{i} \chi^{0}_{l} h}C_{S}^{\chi^{0}_{l} \chi^{0}_{j} H}.
\end{eqnarray}

\subsection*{\mbox{{\Large$\underline{\bf{\chi^{0}_{i} \chi^{0}_{j}\rightarrow A A}}$}}}

Contributions to $\tilde{\omega}$ come from s-channel Higgs boson exchange, t- and u-channel neutralino exchange and cross terms

\begin{equation}
\tilde{\omega}_{\chi^{0}_{i} \chi^{0}_{j}\rightarrow A A}=\tilde{\omega}^{\(h,H\)}_{AA}+\tilde{\omega}^{\(\chi^{0}\)}_{AA}+\tilde{\omega}^{\(h,H-\chi^{0}\)}_{AA}:
\end{equation}

\begin{flushleft}
\textbf{S-channel CP-even Higgs bosons (h,H):}
\end{flushleft}

\begin{equation}
\tilde{\omega}^{\(h,H\)}_{AA} = \frac{1}{4}\sum_{r=h,H}\left|\frac{C^{r A A} C^{\chi_{i} \chi_{j} r}_{S}}{s-m_{r}^{2} + i m_{r} \Gamma_{r}}\right|^{2} \(s-\sigma^{2}\);
\end{equation}
\begin{flushleft}
\textbf{T- and U-channel neutralino:}
\end{flushleft}

\begin{equation}
\tilde{\omega}^{\(\chi^{0}\)}_{AA} = \frac{1}{2}\sum_{k,l=1}^{4} C_{P}^{\chi^{0}_{i} \chi^{0}_{k} A} \(C_{P}^{\chi^{0}_{i} \chi^{0}_{l} A}\)^{*} C_{P}^{\chi^{0}_{k} \chi^{0}_{j} A} \(C_{P}^{\chi^{0}_{l} \chi^{0}_{j} A}\)^{*}\(m_{\chi^{0}_{k}} m_{\chi^{0}_{l}} I^{AA}_{k l} + m_{\chi^{0}_{k}}J^{AA}_{k l} + K^{AA}_{k l}\),
\end{equation}

where
\begin{eqnarray}
I^{AA}_{k l} =&& \(s-\sigma^2\)\(\mathcal{T}_{0}-\mathcal{Y}_{0}\)\big(s, \sigma,\delta,m_{A}^2,m_{A}^2,m_{\chi_{k}^{0}}^2,m_{\chi_{l}^{0}}^2\big)\, ,\\
J^{AA}_{k l} =&& \(2 \sigma \mathcal{T}_{1}+ \frac{\sigma}{2}\(\sigma^2-4m_{A}^2-\delta^2\)\mathcal{T}_{0}\r\nn\\
&+&\l \sigma \mathcal{Y}_{1}+\sigma\(s-\sigma^2\)\mathcal{Y}_{0}\)\big(s, \sigma,\delta,m_{A}^2,m_{A}^2,m_{\chi_{k}^{0}}^2,m_{\chi_{l}^{0}}^2\big)\, ,\\
K^{AA}_{k l} =&&\(-\mathcal{T}_{2}-\frac{1}{2} \(2s + \sigma^2-4m_{A}^2-\delta^2\) \mathcal{T}_{1}\r\nn\\
&-&\l\frac{1}{16}\(\(\sigma^2-\delta^2\)^2-8m_{A}^2\(\sigma^2+\delta^2\)+16 m_{A}^4\)\(\mathcal{T}_{0}+\mathcal{Y}_{0}\) -\mathcal{Y}_{2}\r\nn\\
&+&\l\frac{1}{4}\(\sigma^4-4 m_{A}^2 \sigma^2-\delta^2 \sigma^2\)\mathcal{Y}_{0}\)\big(s, \sigma,\delta,m_{A}^2,m_{A}^2,m_{\chi_{k}^{0}}^2,m_{\chi_{l}^{0}}^2\big);
\end{eqnarray}

\begin{flushleft}
\textbf{Higgs (h,H)-neutralino cross term:}
\end{flushleft}

\begin{eqnarray}
\tilde{\omega}^{\(h,H-\chi^{0}\)}_{AA}=&&\frac{1}{4} \sum_{l=1}^{4} Re\[\sum_{r=h,H} \(\frac{C^{A A r} C^{\chi^{0}_{i} \chi^{0}_{j} r}_{S}}{s-m_{r}^{2} + i m_{r} \Gamma_{r}}\)^{*} C_{P}^{\chi^{0}_{i} \chi^{0}_{l} A} C_{P}^{\chi^{0}_{l} \chi^{0}_{j} A}\]\nn\\
&\times &\[4 \sigma +\(4 m_{\chi^{0}_{l}}\(s-\sigma^2+\sigma m_{\chi^{0}_{l}} \)+\sigma^3-\sigma\(4 m_{A}^2+\delta^2\)\)\r\nn\\
&\times&\l\mathcal{F}\[s, \sigma,\delta,m_{A}^2,m_{A}^2,m_{\chi^{0}_{l}}^2\]\].\nn\\
\end{eqnarray}

\subsection*{\mbox{{\Large$\underline{\bf{\chi^{0}_{i} \chi^{0}_{j}\rightarrow h A}}$}}}

Contributions to $\tilde{\omega}$ come from s-channel Z and Higgs boson exchanges, t- and u-channel neutralino exchange and cross terms

\begin{equation}
\tilde{\omega}_{\chi^{0}_{i} \chi^{0}_{j}\rightarrow h A} = \tilde{\omega}^{\(A\)}_{hA}+\tilde{\omega}^{\(Z\)}_{hA}+\tilde{\omega}^{\(\chi^{0}\)}_{hA}+\tilde{\omega}^{\(A-Z\)}_{hA}+\tilde{\omega}^{\(A-\chi^{0}\)}_{hA}+\tilde{\omega}^{\(Z-\chi^{0}\)}_{hA}:
\end{equation}

\begin{flushleft}
\textbf{S-channel CP-odd Higgs boson A:}
\end{flushleft}

\begin{equation}
\tilde{\omega}^{\(A\)}_{hA} = \frac{1}{2}\left|\frac{C^{h A A} C^{\chi_{i} \chi_{j} A}_{P}}{s-m_{A}^{2} + i m_{A} \Gamma_{A}}\right|^{2} \(s+\delta^{2}\);
\end{equation}
\begin{flushleft}
\textbf{S-channel Z boson:}
\end{flushleft}

\begin{eqnarray}
\tilde{\omega}^{\(Z\)}_{hA} =&& \frac{1}{6}\left|\frac{C^{h A Z} C^{\chi_{i} \chi_{j} Z}_{A}}{s-m_{Z}^{2} + i m_{Z} \Gamma_{Z}}\right|^{2} \frac{1}{m_{Z}^{4} s^{2}} \nn\\
&\times&\lbr m_{Z}^4 s\(2s+\delta^2\)\(m_{A}^4 + \(s-m_{h}^2\)^2-2 m_{A}^2\(s+m_{h}^2\)\)\r\nn\\
&-&\l\(4 \delta^2\(m_{A}^2-m_{h}^2\)^2 m_{Z}^4-m_{Z}^2 \(\(m_{A}^2-m_{h}^2\)^2 m_{Z}^2\r\r\r\nn\\
&+&\l\l\l 2 \delta^2\(3\(m_{A}^2-m_{h}^2\)^2+\(m_{A}^2+m_{h}^2\)m_{Z}^2\)\)s\r\r\nn\\
&+&\l\l\(6\(m_{A}^2-m_{h}^2\)^2 m_{Z}^2-4\(m_{A}^2+m_{h}^2\)m_{Z}^4+\delta^2\(3\(m_{A}^2-m_{h}^2\)^2+m_{Z}^4\)\)s^2\r\r\nn\\
&-&\l\l\(3\(m_{A}^2-m_{h}^2\)^2-2m_{Z}^4\)s^3\)\sigma^2\rbr ;\nn\\
\end{eqnarray}

\begin{flushleft}
\textbf{T- and U-channel neutralino:}
\end{flushleft}

\begin{equation}
\tilde{\omega}^{\(\chi^{0}\)}_{hA} = \sum_{k,l=1}^{4} \(m_{\chi^{0}_{k}} m_{\chi^{0}_{l}} I^{hA}_{k l} + m_{\chi^{0}_{k}}J^{hA}_{k l} + K^{hA}_{k l}\),
\end{equation}

where

\begin{eqnarray}
I^{hA}_{k l} =&& \frac{1}{2}\(s-\delta^2\) \(C_{hA,T}^{+} \mathcal{T}_{0}-C_{hA,Y}^{+}\mathcal{Y}_{0}\)\big(s, \sigma,\delta,m_{h}^2,m_{A}^2,m_{\chi_{k}^0}^2,m_{\chi_{l}^0}^2\big)\, ,\\
J^{hA}_{k l} =&& \(C_{hA,T}^{-} \delta \mathcal{T}_{1} +\frac{1}{4s}\(2 C_{hA,T}^{+} \sigma\(m_{A}^2-m_{h}^2\)\(s-\delta^2\)\r\r\nn\\
&-&\l\l C_{hA,T}^{-}\delta\(s\(2\(m_{A}^2 + m_{h}^2\)+\sigma^2-\delta^2\)+2\sigma \delta\(m_{h}^2-m_{A}^2\)\)\)\mathcal{T}_{0}\r\nn\\
&+&\l\frac{1}{2s} C_{hA,Y}^{+}\sigma\(m_{h}^2-m_{A}^2\)\(s-\delta^2\)\mathcal{Y}_{0}\)\big(s, \sigma,\delta,m_{h}^2,m_{A}^2,m_{\chi_{k}^0}^2,m_{\chi_{l}^0}^2\big)\, ,\\
K^{hA}_{k l} =&&\(-\frac{1}{2} C_{hA,T}^{+} \mathcal{T}_{2} +G_{h A}^{K,T\(1\)}\mathcal{T}_{1}+G_{h A}^{K,T\(0\)}\mathcal{T}_{0} +\frac{1}{2} C_{hA,Y}^{+} \mathcal{Y}_{2}\r\nn\\
&+&\l \frac{1}{4s} C_{hA,Y}^{-} \sigma\delta \(m_{A}^2-m_{h}^2\)\mathcal{Y}_{1}+G_{h A}^{K,Y\(0\)}\mathcal{Y}_{0}\)\big(s, \sigma,\delta,m_{h}^2,m_{A}^2,m_{\chi_{k}^0}^2,m_{\chi_{l}^0}^2\big)\, ,\nn\\
\end{eqnarray}

and where

\begin{eqnarray}
G_{h A}^{K,T\(1\)} =&& \frac{1}{4s} \(2 C_{hA,T}^{-} \sigma \delta \(m_{A}^2 - m_{h}^2\)+ C_{hA,T}^{+} \(-2 s^2 +s\(2\(m_{A}^2+m_{h}^2\)+\sigma^2-\delta^2\)\r\r\nn\\
&+&\l\l 2\sigma \delta \(m_{h}^2-m_{A}^2\)\)\)\, ,\\
G_{h A}^{K,T\(0\)} =&& \frac{1}{32 s^2}\(-4 C_{hA,T}^{-} \sigma \delta \(m_{A}^2 - m_{h}^2\)\(s\(2\(m_{A}^2+m_{h}^2\)+\sigma^2-\delta^2\)\r\r\nn\\
&+&\l\l2\sigma\delta\(m_{h}^2-m_{A}^2\)\)+ C_{hA,T}^{+}\(8 \sigma \delta\(m_{A}^4\(s-\sigma \delta\)-m_{h}^4\(s+\sigma\delta\)\)-s^2\(\sigma^2-\delta^2\)^2 \r\r\nn\\
&+&\l\l 4 m_{h}^2 s \(\sigma+\delta\)\(s\(\sigma+\delta\)+\sigma\delta\(\delta-\sigma\)\)\r\r\nn\\
&-&\l\l 4 m_{A}^2\(4m_{h}^2\(s^2-\sigma^2\delta^2\)-s\(s\(\sigma-\delta\)^2+\delta \sigma^3-\delta^3 \sigma\)\)\)\)\, ,\\
G_{h A}^{K,Y\(0\)} =&&\frac{C_{hA,Y}^{+}}{32s^2}\(8 \sigma \delta \(s\(m_{h}^4-m_{A}^4\)+\sigma \delta \(m_{h}^4+m_{A}^4\)\)+s^2\(\sigma^4+2 \sigma^2 \delta^2 -3\delta^4\)\r\nn\\
&+&\l 4 m_{h}^2 s \(s\(\delta^2-2\sigma \delta-\sigma^2\)+\sigma \delta \(\sigma^2+\delta^2\)\)\r\nn\\
&+&\l 4 m_{A}^2\(4 m_{h}^2\(s^2 -\sigma^2 \delta^2\)+s\(s\(\delta^2+2\sigma \delta-\sigma^2\)-\sigma\delta\(\sigma^2+\delta^2\)\)\)\),
\end{eqnarray}

and the coupling functions are

\begin{eqnarray}
C_{hA,T}^{\pm} =&& C_{S}^{\chi_{j}^{0} \chi_{k}^{0} h} \(C_{S}^{\chi_{j}^{0} \chi_{l}^{0} h}\)^{*} C_{P}^{\chi_{k}^{0} \chi_{i}^{0} A} \(C_{P}^{\chi_{l}^{0} \chi_{i}^{0} A}\)^{*}\nn\\
&\pm& C_{S}^{\chi_{k}^{0} \chi_{i}^{0} h} \(C_{S}^{\chi_{l}^{0} \chi_{i}^{0} h}\)^{*} C_{P}^{\chi_{j}^{0} \chi_{k}^{0} A} \(C_{P}^{\chi_{j}^{0} \chi_{l}^{0} A}\)^{*},\,\,\,\,\,\,\,\,\,\,\,\,\,\,\,\,\,\,\,\,\,\\
C_{hA,Y}^{\pm} =&& C_{P}^{\chi_{j}^{0} \chi_{k}^{0} A} \(C_{S}^{\chi_{j}^{0} \chi_{l}^{0} h}\)^{*} C_{S}^{\chi_{k}^{0} \chi_{i}^{0} h} \(C_{P}^{\chi_{l}^{0} \chi_{i}^{0} A}\)^{*}\nn\\
&\pm& C_{P}^{\chi_{k}^{0} \chi_{i}^{0} A} \(C_{S}^{\chi_{l}^{0} \chi_{i}^{0} h}\)^{*} C_{S}^{\chi_{j}^{0} \chi_{k}^{0} h} \(C_{P}^{\chi_{j}^{0} \chi_{l}^{0} A}\)^{*}.
\end{eqnarray}

\begin{flushleft}
\textbf{Higgs (A)-Z cross term:}
\end{flushleft}

\begin{equation}
\tilde{\omega}^{\(A-Z\)}_{hA} = Re\[\(\frac{C^{h A A} C_{P}^{\chi^{0}_{i} \chi^{0}_{j} A}}{s-m_{A}^2+i m_{A} \Gamma_{A}}\)^{*}\frac{C^{h A Z} C_{A}^{\chi^{0}_{i} \chi^{0}_{j} Z}}{s-m_{Z}^2+i m_{Z} \Gamma_{Z}}\]\frac{\sigma\(s-\delta^2\)\(s-m_{Z}^2\)\(m_{A}^2-m_{h}^2\)}{m_{Z}^2 s}
\end{equation}

\begin{flushleft}
\textbf{Higgs (A)-neutralino cross term:}
\end{flushleft}

\begin{eqnarray}
\tilde{\omega}^{\(A-\chi^{0}\)}_{hA} =&& \sum_{l=1}^{4}Re\[\(\frac{C^{h A A} C^{\chi^{0}_{i} \chi^{0}_{j} A}}{s-m_{A}^2+i m_{A} \Gamma_{A}}\)^{*}\(C_{hA}^{+} G_{A,h A}^{C_{hA}^{+}}+C_{hA}^{-} G_{A,h A}^{C_{hA}^{-}}\)\]
\end{eqnarray}

\begin{eqnarray}
G_{A,h A}^{C_{hA}^{+}} =&& \frac{1}{2 s}\(s-\delta^2\)\(2 s m_{\chi_{l}^{0}}+\sigma\(m_{A}^2-m_{h}^2\)\)\mathcal{F}\[s, \sigma,\delta,m_{h}^2,m_{A}^2,m_{\chi_{l}^{0}}^2\]\\
G_{A,h A}^{C_{hA}^{-}} =&& \frac{1}{4 s}\delta\(4s+\(s\(4m_{\chi_{l}^{0}}^2-2\(m_{A}^2+m_{h}^2\)+\delta^2-\sigma^2\)\r\r\nn\\
&+&\l\l 2\sigma \delta \(m_{A}^2-m_{h}^2\)\)\mathcal{F}\[s, \sigma,\delta,m_{h}^2,m_{A}^2,m_{\chi_{l}^{0}}^2\]\)
\end{eqnarray}
\begin{flushleft}
\textbf{Z-neutralino cross term:}
\end{flushleft}

\begin{eqnarray}
\tilde{\omega}^{\(Z-\chi^{0}\)}_{hA} =&& \sum_{l=1}^{4}Re\[\(\frac{C^{h A Z} C^{\chi^{0}_{i} \chi^{0}_{j} Z}}{s-m_{Z}^2+i m_{Z} \Gamma_{Z}}\)^{*}\(C_{hA}^{+} G_{Z, h A}^{C_{hA}^{+}}+C_{hA}^{-}G_{Z, h A}^{C_{hA}^{-}}\)\]
\end{eqnarray}

\begin{eqnarray}
G_{Z, h A}^{C_{hA}^{+}} =&& \frac{1}{2 s}\(s\(2\(s-m_{A}^2-m_{h}^2\)+4m_{\chi_{l}^{0}}^2+\delta^2\)+ 2\sigma\delta\(m_{A}^2-m_{h}^2\)-\sigma^2 s\)\nn\\
&+&\frac{1}{8 m_{Z}^2 s^2}\(m_{Z}^2 s^2\(16\(m_{\chi_{l}^{0}}^2-m_{A}^2\)\(m_{\chi_{l}^{0}}^2-m_{h}^2\)+16 m_{\chi_{l}^{0}}^2 s -\delta^4\)\r\nn\\
&-&\l 8 \sigma s\(m_{A}^2-m_{h}^2\)\(m_{\chi_{l}^{0}}\(m_{Z}^2-s\)\(s-\delta^2\)\r\r\nn\\
&+&\l\l m_{Z}^2\delta\(m_{A}^2+m_{h}^2-s\)-2m_{\chi_{l}^{0}}^2 m_{Z}^2 \delta\)\r\nn\\
&+&\l 4\sigma^2\(s^2\(\(m_{A}^2-m_{h}^2\)^2-m_{Z}^2\(m_{A}^2+m_{h}^2+2m_{\chi_{l}^{0}}^2\)\)\r\r\nn\\
&-&\l\l \(m_{A}^2-m_{h}^2\)^2\(s-2m_{Z}^2\)\delta^2\)\r\nn\\
&-&\l 4m_{Z}^2 s \sigma^3 \delta\(m_{A}^2-m_{h}^2\)+m_{Z}^2 s^2 \sigma^4\)\mathcal{F}\[s, \sigma,\delta,m_{h}^2,m_{A}^2,m_{\chi_{l}^{0}}^2\]\\
G_{Z, h A}^{C_{hA}^{-}} =&& \frac{1}{m_{Z}^2 s}\delta\(\sigma\(m_{A}^2-m_{h}^2\)\(s-m_{Z}^2\)-2 m_{\chi_{l}^{0}} m_{Z}^2 s\)\nn\\
&-& \frac{1}{4 m_{Z}^2 s^2}\delta\(2 m_{\chi_{l}^{0}} m_{Z}^2 s^2\(4 m_{\chi_{l}^{0}}^2 - 2\(m_{A}^2+m_{h}^2-s\)-\delta^2\)\r\nn\\
&+&\l \sigma s \(m_{A}^2-m_{h}^2\)\(2\(-2 m_{Z}^2\(m_{A}^2+m_{h}^2-2 m_{\chi_{l}^{0}}^2\)\r\r\r\nn\\
&+&\l\l\l s \(m_{A}^2+m_{h}^2+m_{Z}^2-2 m_{\chi_{l}^{0}}^2\)\)+ 4 m_{\chi_{l}^{0}} m_{Z}^2 \delta-s\delta^2\)\r\nn\\
&-&\l 2\sigma^2\(s^2 m_{\chi_{l}^{0}} m_{Z}^2-\(m_{A}^2-m_{h}^2\)^2\(2m_{Z}^2-s\)\delta\)\r\nn\\
&+&\l s \sigma^3\(m_{A}^2-m_{h}^2\)\(s-2m_{Z}^2\)\)\mathcal{F}\[s, \sigma,\delta,m_{h}^2,m_{A}^2,m_{\chi_{l}^{0}}^2\]
\end{eqnarray}

where the coupling functions are:

\begin{eqnarray}
C_{hA}^{\pm} =&& C_{S}^{\chi_{j}^{0} \chi_{l}^{0} h} C_{P}^{\chi_{l}^{0} \chi_{i}^{0} A} \pm C_{P}^{\chi_{j}^{0} \chi_{l}^{0} A} C_{S}^{\chi_{l}^{0} \chi_{i}^{0} h}
\end{eqnarray}
Contributions to the $H A$ final state use the same expressions as above, but with $m_{h}$ replaced with $m_{H}$ and $h$ replaced with $H$ in the couplings.

\subsection*{\mbox{{\Large$\underline{\bf{\chi^{0}_{i} \chi^{0}_{j}\rightarrow H^{+} H^{-}}}$}}}

Contributions to $\tilde{\omega}$ come from s-channel Z and Higgs boson exchanges, t- and u-channel chargino exchange and cross terms

\begin{equation}
\tilde{\omega}_{\chi^{0}_{i} \chi^{0}_{j}\rightarrow H^{+} H^{-}}=\tilde{\omega}^{\(h,H\)}_{H^{+} H^{-}}+\tilde{\omega}^{\(Z\)}_{H^{+} H^{-}}+\tilde{\omega}^{\(\chi^{\pm}\)}_{H^{+} H^{-}}+\tilde{\omega}^{\(h,H-\chi^{\pm}\)}_{H^{+} H^{-}}+\tilde{\omega}^{\(Z-\chi^{\pm}\)}_{H^{+} H^{-}}:
\end{equation}
\begin{flushleft}
\textbf{S-channel CP-even Higgs boson (h,H):}
\end{flushleft}

\begin{equation}
\tilde{\omega}^{\(h,H\)}_{H^{+} H^{-}} = \frac{1}{2}\sum_{r=h,H}\left|\frac{C^{r H^{+} H^{-}} C^{\chi_{i} \chi_{j} r}}{s-m_{r}^{2} + i m_{r} \Gamma_{r}}\right|^{2} \(s-\sigma^{2}\);
\end{equation}

\begin{flushleft}
\textbf{S-channel Z boson:}
\end{flushleft}

\begin{equation}
\tilde{\omega}^{\(Z\)}_{H^{+} H^{-}} = \frac{1}{6}\left|\frac{C^{Z H^{+} H^{-}} C^{\chi_{i} \chi_{j} Z}_{A}}{s-m_{Z}^{2} + i m_{Z} \Gamma_{Z}}\right|^{2} \frac{\(s-4 m_{H^{\pm}}^{2}\)\(2 s + \delta^2\)\(s-\sigma^{2}\)}{s};
\end{equation}

\begin{flushleft}
\textbf{T- and U-channel chargino:}
\end{flushleft}

\begin{equation}
\tilde{\omega}^{\(\chi^{\pm}\)}_{H^{+} H^{-}} = \sum_{k,l=1}^{2}\[m_{\chi_{k}^{\pm}} m_{\chi_{l}^{\pm}} I_{k l}^{H^{+} H^{-}}+ m_{\chi_{k}^{\pm}} J_{k l}^{H^{+} H^{-}}+K_{k l}^{H^{+} H^{-}}\],
\end{equation}

where

\begin{eqnarray}
I_{k l}^{H^{+} H^{-}} =&&\frac{1}{2}\(\(-\bar{D}_{H^{+} H^{-}}^{\lbr ij\rbr kl -}\(s-\delta^2\)+\bar{C}_{H^{+} H^{-}}^{\lbr ij\rbr kl -}\(s-\sigma^2\)\)\mathcal{T}_{0}\r\nn\\
&+&\l\(D_{H^{+} H^{-}}^{ijkl -}\(s-\delta^2\)- C_{H^{+} H^{-}}^{ijkl -}\(s-\sigma^2\)\)\r\nn\\
&\times&\l \mathcal{Y}_{0}\)\big(s, \sigma,\delta,m_{H^{\pm}}^2,m_{H^{\pm}}^2,m_{\chi_{k}^\pm}^2,m_{\chi_{l}^\pm}^2\big)\, ,\\
J_{k l}^{H^{+} H^{-}} =&&\frac{1}{2}\(\(-\(\bar{C}_{H^{+} H^{-}}^{\lbr ij\rbr kl -+}+\bar{C}_{H^{+} H^{-}}^{\lbr ij\rbr kl +-}\)\sigma\r\r\nn\\
&+&\l\l \(\bar{D}_{H^{+} H^{-}}^{\[ ji \] kl +-}-\bar{D}_{H^{+} H^{-}}^{\[ ij \] kl +-}\)\delta\)\mathcal{T}_{1}\r\nn\\
&+&\l\frac{1}{4}\(\(\bar{C}_{H^{+} H^{-}}^{\lbr ij\rbr kl -+}+\bar{C}_{H^{+} H^{-}}^{\lbr ij\rbr kl +-}\)\sigma\(4m_{H^{\pm}}^2-\sigma^2+\delta^2\)\r\r\nn\\
&-&\l\l\(\bar{D}_{H^{+} H^{-}}^{\[ ji \] kl +-}-\bar{D}_{H^{+} H^{-}}^{\[ ij \] kl +-}\)\delta\(4m_{H^{\pm}}^2+\sigma^2-\delta^2\)\)\mathcal{T}_{0}\r\nn\\
&+&\l\frac{1}{2}\(\(C_{H^{+} H^{-}}^{ijkl +-}+C_{H^{+} H^{-}}^{ijkl -+}\)\sigma\r\r\nn\\
&+&\l\l\(D_{H^{+} H^{-}}^{ijkl +-}-D_{H^{+} H^{-}}^{jikl +-}\)\delta\)\mathcal{Y}_{1}\r\nn\\
&-&\l\frac{1}{2}\(\(C_{H^{+} H^{-}}^{ijkl +-}+C_{H^{+} H^{-}}^{ijkl -+}\)\sigma\(s-\sigma^2\)\r\r\nn\\
&+&\l\l\(D_{H^{+} H^{-}}^{ijkl +-}-D_{H^{+} H^{-}}^{jikl +-}\)\delta\(s-\delta^2\)\)\r\nn\\
&\times&\l\mathcal{Y}_{0}\)\big(s, \sigma,\delta,m_{H^{\pm}}^2,m_{H^{\pm}}^2,m_{\chi_{k}^\pm}^2,m_{\chi_{l}^\pm}^2\big)\, ,\\
K_{k l}^{H^{+} H^{-}} =&&\frac{1}{2}\(-\(\bar{D}_{H^{+} H^{-}}^{\lbr ij\rbr kl +}+\bar{C}_{H^{+} H^{-}}^{\lbr ij\rbr kl +}\) \mathcal{T}_{2}\r\nn\\
&+&\l\frac{1}{2}\(\(\bar{D}_{H^{+} H^{-}}^{\lbr ij\rbr kl +}-\bar{C}_{H^{+} H^{-}}^{\lbr ij\rbr kl +}\)\(\sigma^2-\delta^2\)\r\r\nn\\
&-&\l\l 2\(\bar{D}_{H^{+} H^{-}}^{\lbr ij\rbr kl +}+\bar{C}_{H^{+} H^{-}}^{\lbr ij\rbr kl +}\)\(s-2m_{H^{\pm}}^{2}\)\) \mathcal{T}_{1}\r\nn\\
&-&\l\frac{1}{16}\(\bar{D}_{H^{+} H^{-}}^{\lbr ij\rbr kl +}+\bar{C}_{H^{+} H^{-}}^{\lbr ij\rbr kl +}\)\r\nn\\
&\times&\l\(16 m_{H^{\pm}}^{4}-8 m_{H^{\pm}}^{2}\(\sigma^2+\delta^2\)+\(\sigma^2-\delta^2\)^{2}\)\mathcal{T}_{0}\r\nn\\
&+&\l\(D_{H^{+} H^{-}}^{ijkl +}-C_{H^{+} H^{-}}^{ijkl +}\) \mathcal{Y}_{2}\r\nn\\
&+&\l\frac{1}{16}\(\(D_{H^{+} H^{-}}^{ijkl +}+C_{H^{+} H^{-}}^{ijkl +}\)\(8 m_{H^{\pm}}^{2} \(\delta^2-\sigma^2\)+2\(\sigma^4-\delta^4\)\)\r\r\nn\\
&+&\l\l\(D_{H^{+} H^{-}}^{ijkl +}-C_{H^{+} H^{-}}^{ijkl +}\)\(16 m_{H^{\pm}}^{2}+2\delta^2 \sigma^2 - \sigma^4-\delta^4\)\)\r\nn\\
&\times&\l \mathcal{Y}_{0}\)\big(s, \sigma,\delta,m_{H^{\pm}}^2,m_{H^{\pm}}^2,m_{\chi_{k}^\pm}^2,m_{\chi_{l}^\pm}^2\big);\nn\\
\end{eqnarray}

\begin{flushleft}
\textbf{Higgs (h,H)-chargino cross term:}
\end{flushleft}

\begin{eqnarray}
\tilde{\omega}^{\(h,H-\chi^{\pm}\)}_{H^{+} H^{-}} =&& \frac{1}{2}\sum_{k=1}^{2} Re\[\(\frac{C^{r H^{+} H^{-}} C^{\chi_{i} \chi_{j} r}}{s-m_{r}^{2} + i m_{r} \Gamma_{r}}\)^{*} \r\nn\\
&\times&\l \(C_{H^{+} H^{-}}^{ijk,+}\(-4\sigma +\sigma\(4 m_{H^{\pm}}^2-4 m_{\chi_{k}^{\pm}}^2-\sigma^2+\delta^2\)\r\r\r\nn\\
&\times&\l\l\l\mathcal{F}\[s, \sigma,\delta,m_{H^{\pm}}^2,m_{H^{\pm}}^2,m_{\chi_{k}^{\pm}}^2\]\)\r\r\nn\\
&+&\l\l C_{H^{+} H^{-}}^{ijk,-} 4 m_{\chi_{k}^{\pm}}\(s-\sigma^2\)\mathcal{F}\[s, \sigma,\delta,m_{H^{\pm}}^2,m_{H^{\pm}}^2,m_{\chi_{k}^{\pm}}^2\]\)\];
\end{eqnarray}

\begin{flushleft}
\textbf{Z-chargino cross term:}
\end{flushleft}

\begin{eqnarray}
\tilde{\omega}^{\(Z-\chi^{\pm}\)}_{H^{+} H^{-}} =&&\sum_{k=1}^{2} Re\[\(\frac{C^{Z H^{+} H^{-}} C_{A}^{\chi_{i} \chi_{j} Z}}{s-m_{Z}^{2} + i m_{Z} \Gamma_{Z}}\)^{*}\]D_{i j k}^{+}\nn\\
&\times&\(\frac{s}{s-\delta^2}\(2 s -4 m_{H^{\pm}}^2+4m_{\chi_{k}^{\pm}}^2-\sigma^2-\delta^2\)\r\nn\\
&+&\l \frac{1}{4}\(-4\(s-\sigma^2\)\(s-4m_{H^{\pm}}^{2}\) \r\r\nn\\
&+&\l\l \frac{s}{s-\delta^2}\(2 s -4 m_{H^{\pm}}^2+4m_{\chi_{k}^{\pm}}^2-\sigma^2-\delta^2\)^2\)\mathcal{F}\[s, \sigma,\delta,m_{H^{\pm}}^2,m_{H^{\pm}}^2,m_{\chi_{k}^{\pm}}^2\]\)\, ,\nn\\
\end{eqnarray}

where the coupling functions are

\begin{eqnarray}
C_{H^{+} H^{-}}^{ijkl \pm}= C_{H^{+} H^{-}}^{ijk,\pm} C_{H^{+} H^{-}}^{ijl,\pm}+ C_{H^{+} H^{-}}^{jik,\pm}C_{H^{+} H^{-}}^{jil,\pm}\, ,\\
\bar{C}_{H^{+} H^{-}}^{\lbr ij\rbr k l \pm}= C_{H^{+} H^{-}}^{ijk,\pm} C_{H^{+} H^{-}}^{jil,\pm}+ C_{H^{+} H^{-}}^{jik,\pm}C_{H^{+} H^{-}}^{ijl,\pm}\, ,\\
C_{H^{+} H^{-}}^{ ijk l +-}= C_{H^{+} H^{-}}^{ijk,+} C_{H^{+} H^{-}}^{ijl,-}+ C_{H^{+} H^{-}}^{jik,+}C_{H^{+} H^{-}}^{jil,-}\, ,\\
C_{H^{+} H^{-}}^{ ijk l -+}= C_{H^{+} H^{-}}^{ijk,-} C_{H^{+} H^{-}}^{ijl,+}+ C_{H^{+} H^{-}}^{jik,-}C_{H^{+} H^{-}}^{jil,+}\, ,\\
\bar{C}_{H^{+} H^{-}}^{\lbr ij\rbr k l +-}= C_{H^{+} H^{-}}^{ijk,+} C_{H^{+} H^{-}}^{jil,-}+ C_{H^{+} H^{-}}^{jik,+}C_{H^{+} H^{-}}^{ijl,-}\, ,\\
\bar{C}_{H^{+} H^{-}}^{\lbr ij\rbr k l -+}= C_{H^{+} H^{-}}^{ijk,-} C_{H^{+} H^{-}}^{jil,+}+ C_{H^{+} H^{-}}^{jik,-}C_{H^{+} H^{-}}^{ijl,+}\, .
\end{eqnarray}

The same expressions are used to define the $D$ couplings, but with the replacement $C\rightarrow D$.  We also need
\begin{eqnarray}
\bar{D}_{H^{+} H^{-}}^{\[ ij\] k l +-}= D_{H^{+} H^{-}}^{ijk,+} D_{H^{+} H^{-}}^{jil,-}- D_{H^{+} H^{-}}^{ijk,-}D_{H^{+} H^{-}}^{jil,+},\\
\bar{D}_{H^{+} H^{-}}^{\[ ji \] k l +-}= D_{H^{+} H^{-}}^{jik,+} D_{H^{+} H^{-}}^{ijl,-}- D_{H^{+} H^{-}}^{jik,-}D_{H^{+} H^{-}}^{ijl,+}.
\end{eqnarray}

All of these souplings are based on the combinations
\begin{eqnarray}
C_{H^{+} H^{-}}^{ijk,\pm} =&& \(C_{S}^{\chi^{0}_{i} \chi^{+}_{k} H^{-}}\)^{*} C_{S}^{\chi^{0}_{j} \chi^{+}_{k} H^{-}}\pm\(C_{P}^{\chi^{0}_{i} \chi^{+}_{k} H^{-}}\)^{*} C_{P}^{\chi^{0}_{j} \chi^{+}_{k} H^{-}}\, ,\\
D_{H^{+} H^{-}}^{ijk,\pm} =&& \(C_{P}^{\chi^{0}_{i} \chi^{+}_{k} H^{-}}\)^{*} C_{S}^{\chi^{0}_{j} \chi^{+}_{k} H^{-}}\pm\(C_{S}^{\chi^{0}_{i} \chi^{+}_{k} H^{-}}\)^{*} C_{P}^{\chi^{0}_{j} \chi^{+}_{k} H^{-}}\, .
\end{eqnarray}

\subsection*{\mbox{{\Large$\underline{\bf{\chi^{0}_{i} \chi^{0}_{j}\rightarrow W^{+} H^{-}}}$}}}

Contributions to $\tilde{\omega}$ come from s-channel Higgs boson exchanges (both CP-odd and -even), t- and u-channel chargino exchange and cross terms

\begin{equation}
\tilde{\omega}_{\chi^{0}_{i} \chi^{0}_{j}\rightarrow W^{+} H^{-}}=\tilde{\omega}^{\(A\)}_{W^{+} H^{-}}+\tilde{\omega}^{\(h,H\)}_{W^{+} H^{-}}+\tilde{\omega}^{\(\chi^{\pm}\)}_{W^{+} H^{-}}+\tilde{\omega}^{\(A-\chi^{\pm}\)}_{W^{+} H^{-}}+\tilde{\omega}^{\(h,H-\chi^{\pm}\)}_{W^{+} H^{-}}:
\end{equation}

\begin{flushleft}
\textbf{S-channel CP-odd Higgs boson A:}
\end{flushleft}
\begin{equation}
\tilde{\omega}^{\(A\)}_{W^{+} H^{-}} = \left|\frac{C^{A H^{-} W^{+}} C^{\chi_{i} \chi_{j} A}_{P}}{s-m_{A}^{2} + i m_{A} \Gamma_{A}}\right|^{2} \frac{\(s-\delta^{2}\)\(s^2-2\(m_{H^{\pm}}^2+m_{W}^2\)s+\(m_{H^{\pm}}^2-m_{W}^2\)^2\)}{2m_{W}^2};
\end{equation}

\begin{flushleft}
\textbf{S-channel CP-even Higgs boson (h,H):}
\end{flushleft}
 \begin{equation}
\tilde{\omega}^{\(h,H\)}_{W^{+} H^{-}} = \sum_{r=h,H}\left|\frac{C^{r H^{-} W^{+}} C^{\chi_{i} \chi_{j} r}_{S}}{s-m_{r}^{2} + i m_{r} \Gamma_{r}}\right|^{2} \frac{\(s-\sigma^{2}\)\(s^2-2\(m_{H^{\pm}}^2+m_{W}^2\)s+\(m_{H^{\pm}}^2-m_{W}^2\)^2\)}{2m_{W}^2};
\end{equation}

\begin{flushleft}
\textbf{T- and U-channel chargino:}
\end{flushleft}
\begin{equation}
\tilde{\omega}^{\(\chi^{\pm}\)}_{W^{+} H^{-}} = \frac{1}{m_{W}^2}\sum_{k,l=1}^{2}\[m_{\chi_{k}^{+}} m_{\chi_{l}^{+}} I_{k l}^{WH}+m_{\chi_{k}^{+}} J_{k l}^{WH}+ K_{k l}^{WH}\],
\end{equation}

where

\begin{eqnarray}
I_{kl}^{WH} =&& \(G_{WH}^{I,T\(2\)} \mathcal{T}_{2} + G_{WH}^{I,T\(1\)} \mathcal{T}_{1}+ G_{WH}^{I,T\(0\)} \mathcal{T}_{0} \r\nn\\
&+&\l G_{WH}^{I,Y\(2\)} \mathcal{Y}_{2} + G_{WH}^{I,Y\(1\)} \mathcal{Y}_{1}+ G_{WH}^{I,Y\(0\)} \mathcal{Y}_{0}\)\big(s, \sigma,\delta,m_{H^{\pm}}^2,m_{W}^2,m_{\chi_{k}^\pm}^2,m_{\chi_{l}^\pm}^2\big)\, ,\\
J_{kl}^{WH} =&& \(G_{WH}^{J,T\(2\)} \mathcal{T}_{2} + G_{WH}^{J,T\(1\)} \mathcal{T}_{1}+ G_{WH}^{J,T\(0\)} \mathcal{T}_{0} \r\nn\\
&+&\l G_{WH}^{J,Y\(2\)} \mathcal{Y}_{2} + G_{WH}^{J,Y\(1\)} \mathcal{Y}_{1}+ G_{WH}^{J,Y\(0\)} \mathcal{Y}_{0}\)\big(s, \sigma,\delta,m_{H^{\pm}}^2,m_{W}^2,m_{\chi_{k}^\pm}^2,m_{\chi_{l}^\pm}^2\big)\, ,\\
K_{kl}^{WH} =&& \(G_{WH}^{K,T\(2\)} \mathcal{T}_{2} + G_{WH}^{K,T\(1\)} \mathcal{T}_{1}+ G_{WH}^{K,T\(0\)} \mathcal{T}_{0} + G_{WH}^{K,Y\(2\)} \mathcal{Y}_{2} \r\nn\\
&+&\l G_{WH}^{K,Y\(1\)} \mathcal{Y}_{1}+ G_{WH}^{K,Y\(0\)} \mathcal{Y}_{0}\)\big(s, \sigma,\delta,m_{H^{\pm}}^2,m_{W}^2,m_{\chi_{k}^\pm}^2,m_{\chi_{l}^\pm}^2\big),
\end{eqnarray}

where

\begin{eqnarray}
G_{WH}^{I,T\(2\)} =&-&\frac{1}{2} \(D_{WH,ij}^{C,+}+D_{WH, ji}^{C,+}\)\, ,\\
G_{WH}^{I,T\(1\)} =&& \frac{1}{4 s} \(s\, D_{WH,ij}^{C,+} \(-2\(s-m_{W}^2-m_{H^{\pm}}^2\)+\(\sigma+\delta\)^2\)\r\nn\\
&+&\l D_{WH, ji}^{C,+}\(s\(-2s+\(\delta-\sigma\)^2\)\r\r\nn\\
&+&\l\l 2\(m_{W}^2\(s-2\sigma\delta\)+m_{H^{\pm}}^2\(s+2\sigma\delta\)\)\)\)\, ,\\
G_{WH}^{I,T\(0\)} =&& \frac{1}{32 s^2}\(-s^2 D_{WH,ij}^{C,+}\(-4m_{W}^2\(8s+3\sigma^2-2\sigma\delta-9\delta^2\)\r\r\nn\\
&+&\l\l 4 m_{H^{\pm}}^2\(4m_{W}^2+\(\sigma+\delta\)^2\)+ \(\sigma+\delta\)^2\(-4s+\(\sigma+\delta\)^2\)\)\r\nn\\
&-&\l D_{WH, ji}^{C,+}\(s^2\(-4s+\(\delta-\sigma\)^2\)\(\delta-\sigma\)^2\r\r\nn\\
&+&\l\l 16 m_{W}^4 \sigma\delta\(-s+\sigma\delta\)+ 16 m_{H^{\pm}}^2\sigma\delta\(s+\sigma\delta\)\r\r\nn\\
&-&\l\l 4m_{W}^2 s\(8s^2+2\sigma\delta\(\delta-\sigma\)^2+s\(3\sigma^2-2\sigma\delta-9\delta^2\)\)\r\r\nn\\
&+&\l\l 4m_{H^{\pm}}^2\(4m_{W}^2\(s^2-2\sigma^2\delta^2\)+s\(s\(\sigma^2-6\sigma\delta+\delta^2\)+2\sigma\delta\(\sigma-\delta\)^2\)\)\)\r\nn\\
&-&\l 48 m_{W}^2 s^2\(\sigma^2-\delta^2\)\(\(D_{WH,i j k}^{+}\)^{*} D_{WH,i j l}^{+}+\(D_{WH,j i k}^{+}\)^{*} D_{WH,j i l}^{+}\)\)\, ,\\
G_{WH}^{I,Y\(2\)} =&-&\frac{1}{2}\(C_{WH,\{ij\}}^{D,+}-2 D_{WH,\{ij\}}^{2,+}\)\, ,\\
G_{WH}^{I,Y\(1\)} =&-&\frac{\sigma\delta}{4 s}\(C_{WH,\[ij\]}^{D,+}-2 D_{WH,\[ij\]}^{2,+}\)\(s+m_{W}^2-m_{H^{\pm}}^2\)\, ,\\
G_{WH}^{I,Y\(0\)} =&& \frac{1}{32 s^2}\(\(2 D_{WH,\{ij\}}^{2,+}-C_{WH,\{ij\}}^{D,+}\)\(-8m_{W}^4 \sigma\delta\(s-\sigma\delta\)+8m_{H^{\pm}}^4\sigma\delta\(s+\sigma\delta\)\r\r\nn\\
&+&\l\l 4m_{H^{\pm}}^2\(s+\sigma\delta\)\(s\(4m_{W}^2+\(\sigma-\delta\)^2\)-4m_{W}^2\sigma\delta\)\r\r\nn\\
&+&\l\l s^2\(\sigma^4+6\sigma^2\delta^2+\delta^4-4s\(\sigma^2+\delta^2\)\)\r\r\nn\\
&-&\l\l 4m_{W}^2 s\(8s^2-s\(3\sigma^2+2\sigma\delta+3\delta^2\)+\sigma\delta\(\sigma-\delta\)^2\)\)\r\nn\\
&+&\l 24 C_{WH,\{ij\}}^{D,+} m_{W}^2 s^2 \(\sigma^2-\delta^2\)\),
\end{eqnarray}

\begin{eqnarray}
G_{WH}^{J,T\(2\)} =&& \frac{1}{2}\(\sigma Re\(C_{WH,\{ij\}}^{-+}-D_{WH,\{ij\}}^{-+}\)-\delta Re\(C_{WH,\[ij\]}^{-+}-D_{WH,\[ij\]}^{-+}\)\)\, ,\\
G_{WH}^{J,T\(1\)} =&& \frac{1}{4s}\(Re\(C_{WH,\[ij\]}^{-+}-D_{WH,\[ij\]}^{-+}\)\delta\(\sigma+\delta\)\(s\(\delta-\sigma\)+2m_{H^{\pm}}^2 \sigma\)\r\nn\\
&-&\l Re\(C_{WH,\{ij\}}^{-+}-D_{WH,\{ij\}}^{-+}\)\sigma\(\sigma+\delta\)\(s\(\sigma-\delta\)+2m_{H^{\pm}}^2 \delta\)\r\nn\\
&-&\l 8m_{W}^2 s Re\(\delta\,C_{WH,\[ij\]}^{-+}+\sigma\,D_{WH,\{ij\}}^{-+}\)- 4 m_{W}^2 s  Re\(\sigma\,C_{WH,\{ij\}}^{-+}+\delta\,D_{WH,\[ij\]}^{-+}\)\r\nn\\
&-&\l 2m_{W}^2 Re\(\sigma \(C_{WH,\[ij\]}^{-+}-D_{WH,\[ij\]}^{-+}\)\r\r\nn\\
&+&\l\l \delta \(D_{WH,\{ij\}}^{-+}-C_{WH,\{ij\}}^{-+}\)\)\sigma\,\delta\,\(\sigma+\delta\)\)\, ,\\
G_{WH}^{J,T\(0\)} =&& \frac{1}{32 s^2}\( Re\(C_{WH,\{ij\}}^{-+}-D_{WH,\{ij\}}^{-+}\) \sigma\(8 m_{W}^4 \(-4 s^2+s\delta\(\sigma+\delta\)+\sigma\delta^2\(\sigma+\delta\)\)\r\r\nn\\
&-&\l\l 4m_{W}^2\(\sigma+\delta\)\(s\(\delta-\sigma\)\(s+\delta\(\delta-\sigma\)\)+2m_{H^{\pm}}^2\delta\(s+2\sigma\delta\) \)\r\r\nn\\
&+&\l\l \(\sigma+\delta\)\(4 m_{H^{\pm}}^2 s \delta\(\sigma-\delta\)^2+8 m_{H^{\pm}}^4 \delta^2\sigma+s^2 \(\sigma-\delta\)\(\sigma^2+3\delta^2\)\)\)\r\nn\\
&+&\l 12 Re\(C_{WH,\{ij\}}^{-+}+D_{WH,\{ij\}}^{-+}\) m_{W}^2 s \sigma\(2 m_{H^{\pm}}^2 \(2 s + \delta\(\sigma-\delta\)\)\r\r\nn\\
&-&\l\l \(\sigma-\delta\)\(s\(\sigma+\delta\)+2m_{W}^2 \delta\)\)\r\nn\\
&-&\l Re\(C_{WH,\[ij\]}^{-+}-D_{WH,\[ij\]}^{-+}\)\delta\(-4m_{W}^2\(\sigma+\delta\)\(s\sigma\(2 m_{H^{\pm}}^2+\(\sigma-\delta\)^2\)\r\r\r\nn\\
&+&\l\l\l 4m_{H^{\pm}}^2\sigma^2\delta+s^2\(\sigma-\delta\)\)\r\r\nn\\
&+&\l\l 8m_{W}^4\(-4s^2+s\sigma\(\sigma+\delta\)+\sigma^2\delta\(\sigma+\delta\)\)\r\r\nn\\
&+&\l\l \(\sigma+\delta\)\(4 m_{H^{\pm}}^2 s \sigma \(\sigma-\delta\)^2+8m_{H^{\pm}}^4\sigma^2\delta+ s^2\(\delta-\sigma\)\(3\sigma^2+\delta^2\)\)\)\r\nn\\
&+&\l 12 Re\(C_{WH,\[ij\]}^{-+}+D_{WH,\[ij\]}^{-+}\)m_{W}^2 s \delta\(2 m_{H^{\pm}}^2\(2 s+\sigma\(\delta-\sigma\)\)\r\r\nn\\
&+&\l\l \(\sigma-\delta\)\(s\(\sigma+\delta\)+2m_{W}^2\sigma\)\)\)\, ,\\
G_{WH}^{J,Y\(2\)} =&& \frac{1}{4}\(\sigma\(C_{WH,\{ij\}}^{1,\{+-\}}+D_{WH,\{ij\}}^{1,\{+-\}}\)+\delta\((C_{WH,\[ij\]}^{1,\[+-\]}+D_{WH,\[ij\]}^{1,\[+-\]}\)\)\, ,\\
G_{WH}^{J,Y\(1\)} =&-&\frac{1}{8 s}\(\(\delta C_{WH,\[ij\]}^{1,\[+-\]}+\sigma D_{WH,\{ij\}}^{1,\{+-\}}\)\(-s^2+s\(m_{H^{\pm}}^2+5m_{W}^2\)\)\r\nn\\
&+&\l \sigma\,\delta\,\(s+m_{W}^2-m_{H^{\pm}}^2\)\(\sigma \,C_{WH,\[ij\]}^{1,\[+-\]}+\delta\, D_{WH,\{ij\}}^{1,\{+-\}}\)\r\nn\\
&-&\l \(\sigma \(s-\delta^2\)C_{WH,\{ij\}}^{1,\{+-\}}+\delta \(s-\sigma^2\)D_{WH,\[ij\]}^{1,\[+-\]}\)\(s+m_{W}^2-m_{H^{\pm}}^2\)\)\, ,\\
G_{WH}^{J,Y\(0\)} =&&\frac{1}{64 s^2}\(\sigma\(-8m_{W}^4\(5 s^2+s\delta\(\sigma-5\delta\)-\delta^2\sigma^2\)\r\r\nn\\
&+&\l\l 8 m_{H^{\pm}}^4\(s^2+s\delta\(\sigma-\delta\)+\delta^2 \sigma^2\)\r\r\nn\\
&+&\l\l 4m_{W}^2 s \(\delta-\sigma\)\(\delta\sigma\(\sigma-\delta\)+s\(3\sigma+\delta\)\)\r\r\nn\\
&+&\l\l s^2\(8s^2-4s\(\sigma^2+3\delta^2\)+\sigma^4+6\delta^2\sigma^2+\delta^4\)\r\r\nn\\
&+&\l\l 4m_{H^{\pm}}^2\(4m_{W}^2\(3s^2-2s\delta^2-\sigma^2\delta^2\)+s\(-4s^2\r\r\r\r\nn\\
&+&\l\l\l\l s\(\sigma^2-2\delta\sigma+5\delta^2\)+\sigma\delta\(\delta-\sigma\)^2\)\)\)C_{WH,\{ij\}}^{1,\{+-\}}\r\nn\\
&+&\l \delta\(s^2\(8\(m_{H^{\pm}}^4+m_{W}^4-2s\(m_{H^{\pm}}^2+3m_{W}^2\)+s^2\)\r\r\r\nn\\
&+&\l\l\l 4\delta^2\(m_{H^{\pm}}^2+9m_{W}^2-s\)+\delta^4\)\r\r\nn\\
&+&\l\l 4s\sigma\delta\(m_{H^{\pm}}^2-m_{W}^2\)\(2\(m_{H^{\pm}}^2+m_{W}^2-s\)+\delta^2\)\r\r\nn\\
&+&\l\l 2\sigma^2\(2s\(-3s^2+s\(5m_{H^{\pm}}^2+m_{W}^2\)-2\(m_{H^{\pm}}^2-m_{W}^2\)^2\)\r\r\r\nn\\
&+&\l\l\l \delta^2\(3s^2 -4s \(m_{H^{\pm}}^2-m_{W}^2\)+4\(m_{H^{\pm}}^2-m_{W}^2\)^2\)\)+s^2\sigma^4\r\r\nn\\
&+&\l\l 4s\sigma^3\delta\(m_{H^{\pm}}^2-m_{W}^2\)\)C_{WH,\[ij\]}^{1,\[+-\]}\r\nn\\
&+&\l \sigma\(s\(8s\(m_{H^{\pm}}^4+m_{W}^4-2s\(m_{H^{\pm}}^2+3m_{W}^2\)+s^2\)\r\r\r\nn\\
&-&\l\l\l 4\delta^2\(2\(m_{H^{\pm}}^2-m_{W}^2\)^2-s\(5m_{H^{\pm}}^2+m_{W}^2\)+3s^2\)+s\delta^4\)\r\r\nn\\
&+&\l\l 4s\sigma\delta\(m_{H^{\pm}}^2-m_{W}^2\)\(2\(m_{H^{\pm}}^2+m_{W}^2-s\)+\delta^2\)\r\r\nn\\
&+&\l\l 2\sigma^2\(2s^2\(m_{H^{\pm}}^2+9m_{W}^2-s\)\r\r\r\nn\\
&+&\l\l\l \delta^2\(3s^2-4s\(m_{H^{\pm}}^2-m_{W}^2\)+4\(m_{H^{\pm}}^2-m_{W}^2\)^2\)\)\r\r\nn\\
&+&\l\l 4s\sigma^3\delta \(m_{H^{\pm}}^2-m_{W}^2\)+s^2\sigma^4\)D_{WH,\{ij\}}^{1,\{+-\}}\r\nn\\
&+&\l \delta\(8m_{H^{\pm}}^4\(s^2+s\sigma\(\delta-\sigma\)+\sigma^2\delta^2\)-8m_{W}^4\(5s^2+s\sigma\(\delta-5\sigma\)-\sigma^2\delta^2\)\r\r\nn\\
&-&\l\l 4m_{W}^2 s \(\delta-\sigma\)\(s\(\sigma+3\delta\)+\sigma\delta\(\delta-\sigma\)\)\r\r\nn\\
&+&\l\l s^2\(8s^2-4s\(3\sigma^2+\delta^2\)+\sigma^4+6\sigma^2\delta^2+\delta^4\)\r\r\nn\\
&+&\l\l 4m_{H^{\pm}}^2\(4m_{W}^2\(3s^2-\sigma^2\(2s+\delta^2\)\)\r\r\r\nn\\
&+&\l\l\l s\(-4s^2+s\(5\sigma^2-2\sigma\delta+\delta^2\)+\sigma\delta\(\delta-\sigma\)^2\)\)\)D_{WH,\[ij\]}^{1,\[+-\]}\),
\end{eqnarray}

\begin{eqnarray}
G_{WH}^{K,T\(2\)} =&-&\frac{1}{8}\(\(D_{WH,ij}^{C,-}+D_{WH, ji}^{C,-}\)\(-4s+8 m_{W}^2 +\sigma^2+\delta^2\)\r\nn\\
&-&\l 2\(D_{WH,ij}^{C,-}-D_{WH, ji}^{C,-}\)\sigma \delta\)\, ,\\
G_{WH}^{K,T\(1\)} =&& \frac{1}{16 s} \(D_{WH,ij}^{C,-} s \(16 m_{W}^4 + \(-2s + \(\delta-\sigma\)^2\)\(\delta+\sigma\)^2\r\r\nn\\
&-&\l\l 2m_{W}^2\(8s + \(5\delta-7\sigma\)\(\delta+\sigma\)\)+ 2m_{H^{\pm}}^2\(8 m_{W}^2-\(\delta+\sigma\)^2\)\)\r\nn\\
&+&\l D_{WH, ji}^{C,-}\(16 m_{W}^4\(s-2\delta\sigma\)+s\(\delta-\sigma\)^2\(-2s+\(\delta+\sigma\)^2\)\r\r\nn\\
&-&\l\l 2m_{W}^2\(8s^2+2\delta\sigma\(\delta+\sigma\)^2+s\(5\delta^2-6\delta\sigma-7\sigma^2\)\)\r\r\nn\\
&+&\l\l 2m_{H^{\pm}}^2\(2\delta\sigma\(\delta+\sigma\)^2+8m_{W}^2\(s+2\delta\sigma\)-s\(\delta^2+6\delta\sigma+\sigma^2\)\)\)\r\nn\\
&+&\l 24 m_{W}^2 s \(\delta^2-\sigma^2\)\(\(D_{WH,i j k}^{-}\)^{*} D_{WH,i j l}^{-}+\(D_{WH,j i k}^{-}\)^{*} D_{WH,j i l}^{-}\)\)\, ,\\
G_{WH}^{K,T\(0\)} =&& \frac{1}{128 s^2} \(-D_{WH, ji}^{C,-}\(s^2\(\delta-\sigma\)^4\(\delta+\sigma\)^2+128 m_{W}^6 \delta \sigma\(-s+\delta\sigma\)\r\r\nn\\
&-&\l\l 16 m_{W}^4 \(2s^2\(\delta-\sigma\)^2-s\delta\sigma\(5\delta-7\sigma\)\(\delta+\sigma\)-\delta^2\sigma^2\(\delta+\sigma\)^2\)\r\r\nn\\
&+&\l\l 16 m_{H^{\pm}}^4 \delta\sigma\(\(\delta+\sigma\)^2\(-s+\delta\sigma\)+8m_{W}^2\(s+\delta\sigma\)\)\r\r\nn\\
&+&\l\l 4m_{W}^2 s \(\delta-\sigma\)^2\(-2\delta\sigma\(\delta+\sigma\)^2+s\(\delta^2+6\delta\sigma+\sigma^2\)\)\r\r\nn\\
&+&\l\l 4m_{H^{\pm}}^2\(-s\(s-2\delta\sigma\)\(\delta^2-\sigma^2\)^2+32 m_{W}^4\(s^2-2\delta^2\sigma^2\)\r\r\r\nn\\
&-&\l\l\l 4m_{W}^2\(4 s \delta\sigma\(\delta-2\sigma\)\(\delta+\sigma\)+2\delta^2\sigma^2\(\delta+\sigma\)^2+s^2\(\delta^2+6\delta\sigma+\sigma^2\)\)\)\)\r\nn\\
&+&\l D_{WH,ij}^{C,-} s^2\(4 m_{H^{\pm}}^2 - \(\delta-\sigma\)^2\)\(-32m_{W}^4+4m_{W}^2\(\delta+\sigma\)^2+\(\delta+\sigma\)^4\)\r\nn\\
&-&\l 192\(D_{WH,j i k}^{-}\)^{*}D_{WH,j i l}^{-} s \delta\sigma\(m_{H^{\pm}}^2-m_{W}^2\)\(\delta^2-\sigma^2\)\)\, ,\\
G_{WH}^{K,Y\(2\)} =&& \frac{1}{8}\(-8 m_{W}^2 C_{WH,\{ij\}}^{D-} + C_{WH,\{ij\}}^{D,-}\(4s-\sigma^2+\delta^2\)+2 D_{WH,\{ij\}}^{2,-} \(\sigma^2-\delta^2\)\)\, ,\nn\\
\\
G_{WH}^{K,Y\(1\)} =&& \frac{\sigma\delta}{s}\(-8m_{W}^4 C_{WH,\[ij\]}^{D,-}+\(2 D_{WH,\[ij\]}^{2,-}-C_{WH,\[ij\]}^{D,-}\)s\(\sigma^2-\delta^2\)\r\nn\\
&+&\l m_{W}^2\(-4 s C_{WH,\[ij\]}^{D,-}+\(\sigma^2 -\delta^2\)\(2 D_{WH,\[ij\]}^{2,-}-C_{WH,\[ij\]}^{D,-}\)\)\r\nn\\
&+&\l m_{H^{\pm}}^2\(8 m_{W}^2 C_{WH,\[ij\]}^{D,-}-\(\sigma^2 -\delta^2\)\(2 D_{WH,\[ij\]}^{2,-}-C_{WH,\[ij\]}^{D,-}\)\)\)\, ,\\
G_{WH}^{K,Y\(0\)} =&& \frac{1}{128 s^2}\(s^2\(128 m_{H^{\pm}}^2 m_{W}^2 C_{WH,\{ij\}}^{D,-} \(s-m_{W}^2\)\r\r\nn\\
&-&\l\l 16\(-m_{H^{\pm}}^4 C_{WH,\{ij\}}^{D,-}+m_{H^{\pm}}^2 \(s C_{WH,\{ij\}}^{D,-}+ m_{W}^2\(7 C_{WH,\{ij\}}^{D,-}-4 D_{WH,\{ij\}}^{2,-}\)\r\r\r\r\nn\\
&+&\l\l\l\l m_{W}^2\(m_{W}^2\(C_{WH,\{ij\}}^{D,-}-6 D_{WH,\{ij\}}^{2,-}\)+s\(C_{WH,\{ij\}}^{D,-}\r\r\r\r\r\r\nn\\
&+&\l\l\l\l\l\l 2 D_{WH,\{ij\}}^{2,-}\)\)\)\)\delta^2 +4 \delta^4\(m_{W}^2\(7 C_{WH,\{ij\}}^{D,-}-6 D_{WH,\{ij\}}^{2,-}\)\r\r\r\nn\\
&+&\l\l\l m_{H^{\pm}}^2\(C_{WH,\{ij\}}^{D,-}-2 D_{WH,\{ij\}}^{2,-}\)+2s D_{WH,\{ij\}}^{2,-}\)\r\r\nn\\
&+&\l\l \delta^6\(C_{WH,\{ij\}}^{D,-}-2 D_{WH,\{ij\}}^{2,-}\)\)\r\nn\\
&+&\l 4s \sigma \delta\(m_{H^{\pm}}^2-m_{W}^2\)\(2\(s-m_{W}^2-m_{H^{\pm}}^2\)-\delta^2\)\(8 m_{W}^2 C_{WH,\{ij\}}^{D,-}\r\r\nn\\
&+&\l\l 2 \delta^2 D_{WH,\{ij\}}^{2,-}- C_{WH,\{ij\}}^{D,-}\(4s + \delta^2\)\) + \sigma^2\(16 s^2 \( C_{WH,\{ij\}}^{D,-}\(m_{H^{\pm}}^4+5m_{W}^4\r\r\r\r\nn\\
&-&\l\l\l\l 3m_{W}^2 s -m_{H^{\pm}}^2 \(s+3m_{W}^2\)\)+ 2m_{W}^2 D_{WH,\{ij\}}^{2,-} \(s-3m_{W}^2-2 m_{H^{\pm}}^2\)\)\r\r\nn\\
&-&\l\l 8 \delta^2 C_{WH,\{ij\}}^{D,-}\(8 m_{W}^2\(m_{W}^2-m_{H^{\pm}}^2\)^2\r\r\r\nn\\
&+&\l\l\l 4m_{W}^2 s\(m_{W}^2-m_{H^{\pm}}^2\)-4 s^2 \(m_{W}^2+m_{H^{\pm}}^2\)+s^3\)\) \r\nn\\
&+&\l \delta^4\(C_{WH,\{ij\}}^{D,-}-2 D_{WH,\{ij\}}^{2,-}\)\(5 s^2 +8s \(m_{W}^2-m_{H^{\pm}}^2\)+8\(m_{W}^2-m_{H^{\pm}}^2\)^2\) \r\nn\\
&+&\l 8 s \sigma^3 \delta\(m_{W}^2-m_{H^{\pm}}^2\)\(-C_{WH,\{ij\}}^{D,-}\(3s-5 m_{W}^2-m_{H^{\pm}}^2\)\r\r\nn\\
&+&\l\l 2 D_{WH,\{ij\}}^{2,-}\(s-m_{W}^2-m_{H^{\pm}}^2\)\) \r\nn\\
&+&\l \sigma^4 \(4s^2\(-m_{H^{\pm}}^2\(C_{WH,\{ij\}}^{D,-}-2 D_{WH,\{ij\}}^{2,-}\)+2s \(C_{WH,\{ij\}}^{D,-}- D_{WH,\{ij\}}^{2,-}\)\r\r\r\nn\\
&+&\l\l\l m_{W}^2\(C_{WH,\{ij\}}^{D,-}+6 D_{WH,\{ij\}}^{2,-}\)\)\r\r\nn\\
&-&\l\l \delta^2 \(C_{WH,\{ij\}}^{D,-}-2 D_{WH,\{ij\}}^{2,-}\)\(5s^2+8s \(m_{W}^2-m_{H^{\pm}}^2\)+8\(m_{W}^2-m_{H^{\pm}}^2\)^2\)\)\r\nn\\
&+&\l \(-s^2 \sigma^6+4s \sigma^5 \delta\(m_{W}^2-m_{H^{\pm}}^2\)\)\(C_{WH,\{ij\}}^{D,-}-2 D_{WH,\{ij\}}^{2,-}\)\),
\end{eqnarray}

where we have used the following combinations of gauge functions:
\begin{eqnarray}
C_{WH,\{ij\}}^{1,\{+-\}} = C_{WH,\{ij\}}^{1,+-}+C_{WH,\{ij\}}^{1,-+}\, ,\\
C_{WH,\[ij\]}^{1,\[+-\]} = C_{WH,\{ij\}}^{1,+-}-C_{WH,\{ij\}}^{1,-+}\, ,\\
D_{WH,\{ij\}}^{1,\{+-\}} = D_{WH,\{ij\}}^{1,+-}+D_{WH,\{ij\}}^{1,-+}\, ,\\
D_{WH,\[ij\]}^{1,\[+-\]} = D_{WH,\{ij\}}^{1,+-}-D_{WH,\{ij\}}^{1,-+},
\end{eqnarray}

\begin{eqnarray}
C_{WH,\{ij\}}^{1,-+} = \(C_{WH,j i k}^{+}\)^{*} C_{WH,i j l}^{-}+ \(C_{WH,i j k}^{+}\)^{*} C_{WH,j i l}^{-}\, ,\\
C_{WH,\[ij\]}^{1,-+} = \(C_{WH,j i k}^{+}\)^{*} C_{WH,i j l}^{-}- \(C_{WH,i j k}^{+}\)^{*} C_{WH,j i l}^{-}\, ,\\
C_{WH,\{ij\}}^{1,+-} = \(C_{WH,j i k}^{-}\)^{*} C_{WH,i j l}^{+}+ \(C_{WH,i j k}^{-}\)^{*} C_{WH,j i l}^{+}\, ,\\
C_{WH,\[ij\]}^{1,+-} = \(C_{WH,j i k}^{-}\)^{*} C_{WH,i j l}^{+}- \(C_{WH,i j k}^{-}\)^{*} C_{WH,j i l}^{+}\, ,\\
D_{WH,\{ij\}}^{1,-+} = \(D_{WH,j i k}^{+}\)^{*} D_{WH,i j l}^{-}+ \(D_{WH,i j k}^{+}\)^{*} D_{WH,j i l}^{-}\, ,\\
D_{WH,\[ij\]}^{1,-+} = \(D_{WH,j i k}^{+}\)^{*} D_{WH,i j l}^{-}- \(D_{WH,i j k}^{+}\)^{*} D_{WH,j i l}^{-}\, ,\\
D_{WH,\{ij\}}^{1,+-} = \(D_{WH,j i k}^{-}\)^{*} D_{WH,i j l}^{+}+ \(D_{WH,i j k}^{-}\)^{*} D_{WH,j i l}^{+}\, ,\\
D_{WH,\[ij\]}^{1,+-} = \(D_{WH,j i k}^{-}\)^{*} D_{WH,i j l}^{+}- \(D_{WH,i j k}^{-}\)^{*} D_{WH,j i l}^{+},
\end{eqnarray}

\begin{eqnarray}
C_{WH,\{ij\}}^{+-} = \(C_{WH,j i k}^{+}\)^{*} C_{WH,j i l}^{-}+ \(C_{WH,i j k}^{-}\)^{*} C_{WH,i j l}^{+}\, ,\\
C_{WH,\[ij\]}^{+-} = \(C_{WH,j i k}^{+}\)^{*} C_{WH,j i l}^{-}- \(C_{WH,i j k}^{-}\)^{*} C_{WH,i j l}^{+}\, ,\\
C_{WH,\{ij\}}^{-+} = \(C_{WH,j i k}^{-}\)^{*} C_{WH,j i l}^{+}+ \(C_{WH,i j k}^{+}\)^{*} C_{WH,i j l}^{-}\, ,\\
C_{WH,\[ij\]}^{-+} = \(C_{WH,j i k}^{-}\)^{*} C_{WH,j i l}^{+}- \(C_{WH,i j k}^{+}\)^{*} C_{WH,i j l}^{-}\, ,\\
D_{WH,\{ij\}}^{+-} = \(D_{WH,j i k}^{+}\)^{*} D_{WH,j i l}^{-}+ \(D_{WH,i j k}^{-}\)^{*} D_{WH,i j l}^{+}\, ,\\
D_{WH,\[ij\]}^{+-} = \(D_{WH,j i k}^{+}\)^{*} D_{WH,j i l}^{-}- \(D_{WH,i j k}^{-}\)^{*} D_{WH,i j l}^{+}\, ,\\
D_{WH,\{ij\}}^{-+} = \(D_{WH,j i k}^{-}\)^{*} D_{WH,j i l}^{+}+ \(D_{WH,i j k}^{+}\)^{*} D_{WH,i j l}^{-}\, ,\\
D_{WH,\[ij\]}^{-+} = \(D_{WH,j i k}^{-}\)^{*} D_{WH,j i l}^{+}- \(D_{WH,i j k}^{+}\)^{*} D_{WH,i j l}^{-},
\end{eqnarray}

\begin{eqnarray}
D_{WH,\{ij\}}^{2,\pm} =&& \(D_{WH,j i k}^{\pm}\)^{*} D_{WH,i j l}^{\pm}+\(D_{WH,i j k}^{\pm}\)^{*} D_{WH,j i l}^{\pm}\, ,\\
D_{WH,\[ij\]}^{2,\pm} =&& \(D_{WH,j i k}^{\pm}\)^{*} D_{WH,i j l}^{\pm}-\(D_{WH,i j k}^{\pm}\)^{*} D_{WH,j i l}^{\pm}\, ,\\
C_{WH,\{ij\}}^{D,\pm} =&& P_{i j k l}^{\pm}+P_{j i k l}^{\pm}\, ,\\
C_{WH,\[ij\]}^{D,\pm} =&& P_{i j k l}^{\pm}-P_{j i k l}^{\pm},
\end{eqnarray}

\begin{eqnarray}
D_{WH, ji}^{C,\pm} =&& \(C_{WH,j i k}^{\pm}\)^{*} C_{WH,j i l}^{\pm}+\(D_{WH,j i k}^{\pm}\)^{*} D_{WH,j i l}^{\pm}\, ,\\
P_{i j k l}^{\pm} =&& \(C_{WH,j i k}^{\pm}\)^{*} C_{WH,i j l}^{\pm}+\(D_{WH,j i k}^{\pm}\)^{*} D_{WH,i j l}^{\pm},
\end{eqnarray}

\begin{eqnarray}
C_{WH,jik}^{\pm} =&& C_{P}^{\chi_{k}^{\pm} \chi_{i}^{0} H^{\pm}} \(C_{A}^{\chi_{j}^{\pm} \chi_{k}^{0} W^{\pm}}\)^{*} \pm C_{S}^{\chi_{k}^{\pm} \chi_{i}^{0} H^{\pm}} \(C_{V}^{\chi_{j}^{\pm} \chi_{k}^{0} W^{\pm}}\)^{*}\, ,\\
D_{WH,jik}^{\pm} =&& C_{S}^{\chi_{k}^{\pm} \chi_{i}^{0} H^{\pm}} \(C_{A}^{\chi_{j}^{\pm} \chi_{k}^{0} W^{\pm}}\)^{*} \pm C_{P}^{\chi_{k}^{\pm} \chi_{i}^{0} H^{\pm}} \(C_{V}^{\chi_{j}^{\pm} \chi_{k}^{0} W^{\pm}}\)^{*};
\end{eqnarray}

\begin{flushleft}
\textbf{Higgs (A)-chargino cross term:}
\end{flushleft}
\begin{eqnarray}
\tilde{\omega}^{\(A-\chi^{\pm}\)}_{W^{+} H^{-}} =&-&\sum_{l=1}^{2}Re\(\(\frac{C^{A H^{-} W^{+}} C^{\chi_{i} \chi_{j} A}_{P}}{s-m_{A}^{2} + i m_{A} \Gamma_{A}}\)^{*} \frac{1}{2m_{W}^2}\(D_{WH,\{ij\}}^{H,-} G_{WH}^{D_{WH,\{ij\}}^{H,-}} \r\r\nn\\
&+&\l\l D_{WH,\[ij\]}^{H,-} G_{WH}^{D_{WH,\[ij\]}^{H,-}}+ D_{WH,\{ij\}}^{H,+}G_{WH}^{D_{WH,\{ij\}}^{H,+}}+D_{WH,\[ij\]}^{H,+}G_{WH}^{D_{WH,\[ij\]}^{H,+}}\)\),\,\,\,\,\,\,\,\,\,\,\,\,\,\,
\end{eqnarray}

where

\begin{eqnarray}
G_{WH}^{D_{WH,\{ij\}}^{H,+}} =&& \frac{1}{s} m_{\chi_{l}^{+}}\sigma\(s-\delta^2\)\(\(s-m_{W}^2\)^2+m_{H^{\pm}}^4\r\nn\\
&-&\l 2m_{H^{\pm}}^2\(s+m_{W}^2\)\)\mathcal{F}\[s, \sigma,\delta,m_{H^{\pm}}^2,m_{W}^2,m_{\chi_{l}^{+}}^2\]\, ,\\
G_{WH}^{D_{WH,\[ij\]}^{H,+}} =&& \frac{1}{2s} m_{\chi_{l}^{+}}\delta\(s+m_{W}^2-m_{H^{\pm}}^2\)\nn\\
&\times&\(4 s +\(4  m_{\chi_{l}^{+}}^2 s +s\(2\(s-m_{W}^2-m_{H^{\pm}}^2\)-\delta^2\)\r\r\nn\\
&-&\l\l 2\sigma\delta\(m_{H^{\pm}}^2-m_{W}^2\)-s \sigma^2\)\mathcal{F}\[s, \sigma,\delta,m_{H^{\pm}}^2,m_{W}^2,m_{\chi_{l}^{+}}^2\]\)\, ,\\
G_{WH}^{D_{WH,\{ij\}}^{H,-}} =&& \frac{1}{4s}\(4s\(2s\(s-m_{W}^2-m_{H^{\pm}}^2\)-\delta^2\(s+m_{W}^2-m_{H^{\pm}}^2\)\)\r\nn\\
&-&\l \(s\(-8s\(2m_{W}^2 m_{H^{\pm}}^2+m_{\chi_{l}^{+}}^2\(s-m_{W}^2-m_{H^{\pm}}^2\)\)\r\r\r\nn\\
&+&\l\l\l 2\delta^2\(\(m_{W}^2-m_{H^{\pm}}^2\)\(2 m_{\chi_{l}^{+}}^2+m_{H^{\pm}}^2-3m_{W}^2\)\r\r\r\r\nn\\
&+&\l\l\l\l s\(2 m_{\chi_{l}^{+}}^2+m_{H^{\pm}}^2+3m_{W}^2\)\)-\delta^4\(s+m_{W}^2-m_{H^{\pm}}^2\)\)\r\r\nn\\
&+&\l\l 2\sigma\delta\(m_{W}^2-m_{H^{\pm}}^2\)\(2s\(m_{H^{\pm}}^2+m_{W}^2-s\)\r\r\r\nn\\
&+&\l\l\l \delta^2\(s+m_{W}^2-m_{H^{\pm}}^2+m_{W}^2\)\)\r\r\nn\\
&-&\l\l \sigma^2\(2s\(\(m_{H^{\pm}}^2-m_{W}^2\)^2-s\(m_{H^{\pm}}^2+m_{W}^2\)\)\r\r\r\nn\\
&-&\l\l\l \delta^2\(2\(m_{H^{\pm}}^2-m_{W}^2\)^2-\(3 m_{H^{\pm}}^2+5m_{W}^2\)+s^2\)\)\) \r\nn\\
&\times&\l\mathcal{F}\[s, \sigma,\delta,m_{H^{\pm}}^2,m_{W}^2,m_{\chi_{l}^{+}}^2\]\)\, ,\\
G_{WH}^{D_{WH,\[ij\]}^{H,-}} =&& \frac{1}{4s} \sigma \delta\(-4m_{W}^4 s+4m_{W}^2 s^2 + 4m_{\chi_{l}^{+}}^2 s \(s+m_{W}^2-m_{H^{\pm}}^2\)+2m_{W}^4 \delta^2\r\nn\\
&-&\l 5 m_{W}^2 \delta^2 + s^2 \delta^2 + 2 m_{W}^2 \sigma \delta \(s+m_{W}^2\)-s \sigma^2\(s+m_{W}^2\)+2 m_{H^{\pm}}^4 \delta\(\sigma+\delta\)\r\nn\\
&+&\l m_{H^{\pm}}^2\(s\(\sigma-3\delta\)\(\sigma+\delta\)+4m_{W}^2\(s-\delta\(\sigma+\delta\)\)\)\);
\end{eqnarray}

\begin{flushleft}
\textbf{Higgs (h,H)-chargino cross term:}
\end{flushleft}

\begin{eqnarray}
\tilde{\omega}^{\(h,H-\chi^{\pm}\)}_{W^{+} H^{-}} =&-&\sum_{l=1}^{2}Re\(\(\frac{C^{r H^{-} W^{+}} C^{\chi_{i} \chi_{j} r}_{S}}{s-m_{r}^{2} + i m_{r} \Gamma_{r}}\)^{*} \frac{1}{m_{W}^2},\r\nn\\
&\times&\l\(C_{WH,\{ij\}}^{H,-} G_{WH}^{C_{WH,\{ij\}}^{H,-}} + C_{WH,\[ij\]}^{H,-} G_{WH}^{C_{WH,\[ij\]}^{H,-}}\r\r\nn\\
&+&\l\l C_{WH,\{ij\}}^{H,+} G_{WH}^{C_{WH,\{ij\}}^{H,+}} +C_{WH,\[ij\]}^{H,+} G_{WH}^{C_{WH,\[ij\]}^{H,+}}\)\),
\end{eqnarray}

where

\begin{eqnarray}
G_{WH}^{C_{WH,\{ij\}}^{H,+}} =&& \frac{1}{4s} m_{\chi_{l}^{\pm}}\sigma\(s+m_{W}^2-m_{H^{\pm}}^2\)\(4s+\(4 m_{\chi_{l}^{\pm}}^2 s - 2m_{W}^2 s + 2s^2-s\delta^2\r\r\nn\\
&+&\l\l 2m_{W} \sigma\delta-s\sigma^2 -2 m_{H^{\pm}}^2\(s+\sigma\delta\)\)\mathcal{F}\[s, \sigma,\delta,m_{H^{\pm}}^2,m_{W}^2, m_{\chi_{l}^{\pm}}^2\]\)\, ,\\
G_{WH}^{C_{WH,\[ij\]}^{H,+}} =&& \frac{1}{4s} m_{\chi_{l}^{\pm}}\delta\(2s-\sigma\(\sigma+\delta\)\)\(m_{H^{\pm}}^4+\(s-m_{W}^2\)^2\r\nn\\
&-&\l 2 m_{H^{\pm}}^2\(s+m_{W}^2\)\)\mathcal{F}\[s, \sigma,\delta,m_{H^{\pm}}^2,m_{W}^2, m_{\chi_{l}^{\pm}}^2\]\, ,\\
G_{WH}^{C_{WH,\{ij\}}^{H,-}} =&-&\frac{1}{32s^2}\(4s\(8s^2\(m_{H^{\pm}}^2+m_{W}^2-s\)-2s \sigma \delta\(3m_{H^{\pm}}^2+m_{W}^2-3s\)\r\r\nn\\
&+&\l\l \sigma^2\(s+m_{W}^2-m_{H^{\pm}}^2\)\(4s-3\delta^2\)\)\r\nn\\
&+&\l \(8s^2\(-8m_{H^{\pm}}^2  m_{W}^2 s+4m_{\chi_{l}^{\pm}}^2 s\(m_{H^{\pm}}^2+m_{W}^2-s\)\r\r\r\nn\\
&+&\l\l\l \delta^2\(s\(m_{H^{\pm}}^2+m_{W}^2\)-\(m_{H^{\pm}}^2-m_{W}^2\)^2\)\)\r\r\nn\\
&-&\l\l 2s \sigma\delta\(4m_{\chi_{l}^{\pm}}^2 s \(3m_{H^{\pm}}^2+m_{W}^2-3s\)\r\r\r\nn\\
&-&\l\l\l 4s \(-2 m_{H^{\pm}}^4+m_{W}^4-m_{W}^2 s +m_{H^{\pm}}^2\(5m_{W}^2+2s\)\)\r\r\r\nn\\
&+&\l\l\l \delta^2\(s^2+s\(3m_{W}^2+m_{H^{\pm}}^2\)- 2\(m_{H^{\pm}}^2-m_{W}^2\)^2\)\) \r\r\nn\\
&+&\l\l s\sigma^2\(8s\(\(m_{W}^2-m_{H^{\pm}}^2\)\(2m_{\chi_{l}^{\pm}}^2+m_{H^{\pm}}^2-3m_{W}^2\)\r\r\r\r\nn\\
&+&\l\l\l\l s\(2m_{\chi_{l}^{\pm}}^2+m_{H^{\pm}}^2+3m_{W}^2\)\)\r\r\r\nn\\
&+&\l\l\l4\delta^2\(\(m_{H^{\pm}}^2-m_{W}^2\)\(3 m_{\chi_{l}^{\pm}}^2+5 m_{H^{\pm}}^2-4m_{W}^2\)\r\r\r\r\nn\\
&-&\l\l\l\l s\(3 m_{\chi_{l}^{\pm}}^2+6 m_{H^{\pm}}^2+5m_{W}^2\)+s^2\)+3\delta^4\(s+m_{W}^2-m_{H^{\pm}}^2\)\) \r\r\nn\\
&+&\l\l 2\sigma^3 \delta\(s\(8\(m_{H^{\pm}}^2-m_{W}^2\)^2-3s\(3m_{H^{\pm}}^2+m_{W}^2\)+s^2\)\r\r\r\nn\\
&-&\l\l\l 3\delta^2\(2\(m_{H^{\pm}}^2-m_{W}^2\)^2-s\(3m_{H^{\pm}}^2+m_{W}^2\)+s^2\)\) \r\r\nn\\
&+&\l\l s\sigma^4\(m_{H^{\pm}}^2-m_{W}^2-s\)\(4s-3\delta^2\)\)\mathcal{F}\[s, \sigma,\delta,m_{H^{\pm}}^2,m_{W}^2, m_{\chi_{l}^{\pm}}^2\]\)\, ,\\
G_{WH}^{C_{WH,\[ij\]}^{H,-}} =&& \frac{1}{32 s^2}\sigma \delta\(4 s \(2s - \sigma \delta\)\(s+m_{W}^2-m_{H^{\pm}}^2\)\r\nn\\
&+&\l \(2s^2\(4 m_{W}^2\(s-m_{W}^2+m_{H^{\pm}}^2\)-\delta^2\(s+m_{W}^2-m_{H^{\pm}}^2\)\)\r\r\nn\\
&+&\l\l s \sigma \delta \(4 m_{H^{\pm}}^4+8 m_{W}^4+\delta^2\(s+m_{W}^2\)- m_{H^{\pm}}^2\(4s+12 m_{W}^2+\delta^2\)\) \r\r\nn\\
&+&\l\l 2\sigma^2\(s\(s^2-s\(3 m_{H^{\pm}}^2+5m_{W}^2\)+2\(m_{H^{\pm}}^2-m_{W}^2\)^2\)\r\r\r\nn\\
&-&\l\l\l \delta^2\(s^2-s\(3 m_{H^{\pm}}^2+m_{W}^2\)+2\(m_{H^{\pm}}^2-m_{W}^2\)^2\)\)+s \sigma^3 \delta \(s+m_{W}^2-m_{H^{\pm}}^2\)\r\r\nn\\
&+&\l\l 4 m_{\chi_{l}^{\pm}}^2 s\(2s - \sigma \delta\)\(s+m_{W}^2-m_{H^{\pm}}^2\)\)\mathcal{F}\[s, \sigma,\delta,m_{H^{\pm}}^2,m_{W}^2, m_{\chi_{l}^{\pm}}^2\]\),
\end{eqnarray}

where the coupling functions are:

\begin{eqnarray}
D_{WH,\{ij\}}^{H,-} =&&D_{WH,i j l}^{-}+D_{WH,j i l}^{-}\, ,\\
D_{WH,\[ij\]}^{H,-} =&&D_{WH,i j l}^{-}-D_{WH,j i l}^{-}\, ,\\
D_{WH,\{ij\}}^{H,+} =&&D_{WH,i j l}^{+}+D_{WH,j i l}^{+}\, ,\\
D_{WH,\[ij\]}^{H,+} =&&D_{WH,i j l}^{+}-D_{WH,j i l}^{+}\, ,\\
C_{WH,\{ij\}}^{H,-} =&&C_{WH,i j l}^{-}+C_{WH,j i l}^{-}\, ,\\
C_{WH,\[ij\]}^{H,-} =&&C_{WH,i j l}^{-}-C_{WH,j i l}^{-}\, ,\\
C_{WH,\{ij\}}^{H,+} =&&C_{WH,i j l}^{+}+C_{WH,j i l}^{+}\, ,\\
C_{WH,\[ij\]}^{H,+} =&&C_{WH,i j l}^{+}-C_{WH,j i l}^{+},
\end{eqnarray}

\begin{eqnarray}
D_{WH,i j l}^{\pm} =&& C_{S}^{\chi_{l}^{+} \chi_{j}^{0} H^{-}} \(C_{A}^{\chi_{l}^{+} \chi_{i}^{0} W^{-}}\)^{*}\pm C_{P}^{\chi_{l}^{+} \chi_{j}^{0} H^{-}} \(C_{V}^{\chi_{l}^{+} \chi_{i}^{0} W^{-}}\)^{*}\, ,\\
C_{WH,i j l}^{\pm} =&&C_{P}^{\chi_{l}^{+} \chi_{j}^{0} H^{-}} \(C_{A}^{\chi_{l}^{+} \chi_{i}^{0} W^{-}}\)^{*}\pm C_{S}^{\chi_{l}^{+} \chi_{j}^{0} H^{-}} \(C_{V}^{\chi_{l}^{+} \chi_{i}^{0} W^{-}}\)^{*}.
\end{eqnarray}

\subsection*{\mbox{{\Large$\underline{\bf{\chi^{0}_{i} \chi^{0}_{j}\rightarrow Z h}}$}}}
Contributions to $\tilde{\omega}$ come from s-channel Z and Higgs boson exchanges, t- and u-channel neutralino exchange and cross terms

\begin{equation}
\tilde{\omega}_{\chi^{0}_{i} \chi^{0}_{j}\rightarrow Z h}=\tilde{\omega}^{\(A\)}_{Z h}+ \tilde{\omega}^{\(Z\)}_{Z h}+ \tilde{\omega}^{\(\chi^{0}\)}_{Z h}+\tilde{\omega}^{\(A-Z\)}_{Z h}+\tilde{\omega}^{\(A-\chi^{0}\)}_{Z h}+\tilde{\omega}^{\(Z-\chi^{0}\)}_{Z h}:
\end{equation}

\begin{flushleft}
\textbf{S-channel CP-odd Higgs boson A:}
\end{flushleft}

\begin{equation}
\tilde{\omega}^{\(A\)}_{Z h} = \left|\frac{C^{A Z h} C^{\chi_{i} \chi_{j} A}_{P}}{s-m_{A}^{2} + i m_{A} \Gamma_{A}}\right|^{2} \frac{\(s-\delta^{2}\)\(s^2-2\(m_{Z}^2+m_{h}^2\)s+\(m_{Z}^2-m_{h}^2\)^2\)}{2m_{Z}^2};
\end{equation}
\\
\\
\begin{flushleft}
\textbf{S-channel Z boson:}
\end{flushleft}

\begin{eqnarray}
\tilde{\omega}^{\(Z\)}_{Z h} =&& \left|\frac{C^{Z Z h} C^{\chi_{i} \chi_{j} Z}_{A}}{s-m_{Z}^{2} + i m_{Z} \Gamma_{Z}}\right|^{2} \frac{1}{24 m_{Z}^6 s}\(2 s \, m_{Z}^4 \(m_{h}^4+m_{Z}^4+10 s\, m_{Z}^2 + s^2-2 m_{h} \(s+m_{Z}^2\)\)\r\nn\\
&+&\l\sigma^2 \(3 s^4 -12 s^3 m_{Z}^2+16 s^2 m_{Z}^4-32 s\, m_{Z}^6+m_{Z}^8 \r\r\nn\\
&+&\l\l \(m_{h}^4 -2 m_{h}^2 \(s+m_{Z}^2\)\)\(m_{Z}^4 -6 s\, m_{Z}^2 +3 s^2\)\)\r\nn\\
&+&\l\frac{\delta^2}{s}\(12 s^3 \sigma^2 m_{Z}^2-3 s^4 \sigma^2+s^2 m_{Z}^4\(s-19 \sigma^2\)+m_{Z}^8 \(s-4\sigma^2\)+2 s\, m_{Z}^6\(5s+\sigma^2\)\r\r\nn\\
&+&\l\l\(m_{h}^4 -2 m_{h}^2 \(s+m_{Z}^2\)\)\(m_{Z}^4\(s-4\sigma^2\) +6 s\, \sigma^2 m_{Z}^2 -3 s^2 \sigma^2\)\)\);
\end{eqnarray}
\newpage
\begin{flushleft}
\textbf{T- and U-channel neutralino:}
\end{flushleft}

\begin{eqnarray}
\tilde{\omega}^{\(\chi^{0}\)}_{Z h} =&& \frac{1}{2 m_{Z}^2}\sum_{k,l=1}^{4} \[m_{\chi_{k}^{0}}m_{\chi_{l}^{0}}I_{k l}^{Z h} + m_{\chi_{k}^{0}} J_{k l}^{Z h} + K_{k l}^{Z h}\],
\end{eqnarray}

where

\begin{eqnarray}
I_{k l}^{Z h} =&&\(-\(D_{Zh,A}^{\{i j\} k l}+D_{Zh,V}^{\{i j\} k l}\) \mathcal{T}_{2} + G_{Z h}^{I,T\(1\)} \mathcal{T}_{1}+ G_{Z h}^{I,T\(0\)} \mathcal{T}_{0}+ \(C_{Zh,A}^{\{i j\} k l}+C_{Zh,V}^{\{i j\} k l}\) \mathcal{Y}_{2} \r\nn\\
&-&\l\frac{1}{2s} \(C_{Zh,A}^{\[i j\] k l}+C_{Zh,V}^{\[i j\] k l}\)\sigma\delta\(s+m_{Z}^2-m_{h}^2\) \mathcal{Y}_{1}\r\nn\\
&+&\l G_{Z h}^{I,Y\(0\)} \mathcal{Y}_{0}\)\big(s, \sigma,\delta,m_{h}^2,m_{Z}^2,m_{\chi_{k}^0}^2,m_{\chi_{l}^0}^2\big)\, ,\\
J_{k l}^{Z h} =&&\(-\(\(D_{Zh,A}^{\{i j\} k l}+D_{Zh,V}^{\{i j\} k l}\) \sigma+\(D_{Zh,A}^{\[i j\] k l}+D_{Zh,V}^{\[i j\] k l}\)\delta\) \mathcal{T}_{2} + G_{Z h}^{J,T\(1\)} \mathcal{T}_{1}\r\nn\\
&+& \l G_{Z h}^{J,T\(0\)} \mathcal{T}_{0}+ \(C_{Zh,A}^{\{i j\} k l}+C_{Zh,V}^{\{i j\} k l}\)\sigma \mathcal{Y}_{2} +G_{Z h}^{J,Y\(1\)}\mathcal{Y}_{1}\r\nn\\
&+&\l G_{Z h}^{J,Y\(0\)} \mathcal{Y}_{0}\)\big(s, \sigma,\delta,m_{h}^2,m_{Z}^2,m_{\chi_{k}^0}^2,m_{\chi_{l}^0}^2\big)\, ,\\
K_{k l}^{Z h} =&&  \(G_{Z h}^{K,T\(2\)} \mathcal{T}_{2} + G_{Z h}^{K,T\(1\)} \mathcal{T}_{1}+ G_{Z h}^{K,T\(0\)} \mathcal{T}_{0}\r\nn\\
&+&\l G_{Z h}^{K,Y\(2\)}  \mathcal{Y}_{2} +G_{Z h}^{K,Y\(1\)} \mathcal{Y}_{1}+G_{Z h}^{K,Y\(0\)} \mathcal{Y}_{0}\)\big(s, \sigma,\delta,m_{h}^2,m_{Z}^2,m_{\chi_{k}^0}^2,m_{\chi_{l}^0}^2\big),
\end{eqnarray}

\begin{eqnarray}
G_{Z h}^{I,T\(1\)} =&& \frac{1}{2s}\(\left( D_{Zh,A}^{\{i j\} k l} + D_{Zh,V}^{\{i j\} k l} \right) \,s\,
   \left( 2\,\left( {m_{h}}^2 + {m_{Z}}^2 - s \right)  + {\delta }^2 \right)  \r\nn\\
&+& \l 2\,\left( \left( D_{Zh,A}^{\[i j\] k l} + D_{Zh,V}^{[i j] k l} + D_{Zh,A}^{\{i j\} k l} + D_{Zh,V}^{\{i j\} k l}
        \right) \,\left( m_{h} - m_{Z} \right) \,
      \left( m_{h} + m_{Z} \right)  \r\r\nn\\
&-&\l\l 
     \left( D_{Zh,A}^{\[i j\] k l} + D_{Zh,V}^{[i j] k l} \right) \,s \right) \,\delta \,\sigma+ \left( D_{Zh,A}^{\{i j\} k l} + D_{Zh,V}^{\{i j\} k l} \right) \,s\,{\sigma }^2\)\, ,\\
G_{Z h}^{I,T\(0\)} =&-&\frac{1}{16 s^2 \sigma}\(4\,\left( 6\,D_{Zh,8}^{\sigma}\,{m_{Z}}^2\,s^2\,\left( {\delta }^2 - {\sigma }^2 \right)  \r\r\nn\\
&+&\l\l 
     D_{Zh,5}^{\delta \sigma}\,\left( {m_{h}}^2 - {m_{Z}}^2 - s \right) \,{\sigma }^2\,
      \left( 2\,{m_{Z}}^2\,\left( s - \delta \,\sigma  \right)  \r\r\r\nn\\
&+&\l\l\l
        2\,{m_{h}}^2\,\left( s + \delta \,\sigma  \right)  + s\,\left( -2\,s + {\delta }^2 + {\sigma }^2 \right)  \right) 
     \right)  \r\nn\\
&+&\l D_{Zh,7}^{\sigma}\,\left( 8\,{m_{Z}}^4\,\delta \,\sigma \,\left( -s + \delta \,\sigma  \right)  + 
     8\,{m_{h}}^4\,\delta \,\sigma \,\left( s + \delta \,\sigma  \right)  \r\r\nn\\
&+&\l\l 
     4\,{m_{h}}^2\,\left( s + \delta \,\sigma  \right) \,
      \left( s\,{\left( \delta  - \sigma  \right) }^2 + 4\,{m_{Z}}^2\,\left( s - \delta \,\sigma  \right)  \right)  \r\r\nn\\
&+&\l\l  s^2\,\left( {\delta }^4 + 6\,{\delta }^2\,{\sigma }^2 + {\sigma }^4 - 4\,s\,\left( {\delta }^2 + {\sigma }^2 \right)  \right)  \r\r\nn\\
&-&\l\l 
     4\,{m_{Z}}^2\,s\,\left( 8\,s^2 + \delta \,{\left( \delta  - \sigma  \right) }^2\,\sigma  - 
        s\,\left( 3\,{\delta }^2 + 2\,\delta \,\sigma  + 3\,{\sigma }^2 \right)  \right)  \right)\)\, ,\\
G_{Z h}^{I,Y\(0\)} =&& \frac{1}{16 s^2} \(24\,C_{Zh,4}\,{m_{Z}}^2\,s^2\,\left( -{\delta }^2 + {\sigma }^2 \right)  \r\nn\\
&+&\l 
  C_{Zh,3}\,\left( 8\,{m_{Z}}^4\,\delta \,\sigma \,\left( -s + \delta \,\sigma  \right)  + 
     8\,{m_{h}}^4\,\delta \,\sigma \,\left( s + \delta \,\sigma  \right)  \r\r\nn\\
&+&\l\l 
     4\,{m_{h}}^2\,\left( s + \delta \,\sigma  \right) \,
      \left( s\,{\left( \delta  - \sigma  \right) }^2 + 4\,{m_{Z}}^2\,\left( s - \delta \,\sigma  \right)  \right)  \r\r\nn\\
&+&\l\l 
     s^2\,\left( {\delta }^4 + 6\,{\delta }^2\,{\sigma }^2 + {\sigma }^4 - 4\,s\,\left( {\delta }^2 + {\sigma }^2 \right)  \right)  \r\r\nn\\
&-&\l\l 
     4\,{m_{Z}}^2\,s\,\left( 8\,s^2 + \delta \,{\left( \delta  - \sigma  \right) }^2\,\sigma  - 
        s\,\left( 3\,{\delta }^2 + 2\,\delta \,\sigma  + 3\,{\sigma }^2 \right)  \right)  \right)\),
\end{eqnarray}

\begin{eqnarray}
G_{Z h}^{J,T\(1\)} =&& \frac{1}{2s}\(D_{Zh,1}^{\delta \sigma}\,\left( 2\,{m_{h}}^2\,\delta \,\sigma  + 
     {m_{Z}}^2\,\left( 4\,s - 2\,\delta \,\sigma  \right)  + s\,\left( {\delta }^2 + {\sigma }^2 \right)  \right)  \r\nn\\
&+&\l 
  D_{Zh,3}^{\delta \sigma}\,\left( 2\,{m_{h}}^2\,\delta \,\sigma  - 
     2\,{m_{Z}}^2\,\left( 4\,s + \delta \,\sigma  \right)  + s\,\left( {\delta }^2 + {\sigma }^2 \right)  \right)\)\, ,\\
G_{Z h}^{J,T\(0\)} =&-&\frac{1}{16 s}\(-24\,{m_{Z}}^2\,\left( D_{Zh,2}^{\delta \sigma}\,\left( -{m_{h}}^2 + {m_{Z}}^2 \right)+
     D_{Zh,4}^{\delta \sigma}\,s \right) \,\left( \delta  - \sigma  \right) \,\left( \delta  + \sigma  \right)  \r\nn\\
&+&\l 
  D_{Zh,1}^{\delta \sigma}\,\left( -8\,{m_{Z}}^4\,
      \left( 4\,s - 3\,{\delta }^2+ 2\,\delta \,\sigma  - 3\,{\sigma }^2 \right)  + 
     s\,{\left( {\delta }^2 - {\sigma }^2 \right) }^2 \r\r\nn\\
&+&\l\l 
     4\,{m_{h}}^2\,\left( \delta \,\sigma \,{\left( \delta  + \sigma  \right) }^2 + 
        2\,{m_{Z}}^2\,\left( 6\,s - 3\,{\delta }^2 + 2\,\delta \,\sigma  - 3\,{\sigma }^2 \right)  \right)  \r\r\nn\\
&-&\l\l
     4\,{m_{Z}}^2\,\left( 2\,s\,\left( {\delta }^2 + {\sigma }^2 \right)  + 
        \delta \,\sigma \,\left( {\delta }^2 - 6\,\delta \,\sigma  + {\sigma }^2 \right)  \right)  \right)  \r\nn\\
&+&\l 
  D_{Zh,3}^{\delta \sigma}\,\left( -32\,{m_{Z}}^4\,\left( s - \delta \,\sigma  \right)  + 
     s\,{\left( {\delta }^2 - {\sigma }^2 \right) }^2 \r\r\nn\\
&+&\l\l 
     4\,{m_{h}}^2\,\left( \delta \,\sigma \,{\left( \delta  + \sigma  \right) }^2 - 
        4\,{m_{Z}}^2\,\left( 3\,s + 2\,\delta \,\sigma  \right)  \right)  \r\r\nn\\
&+&\l\l 
     {m_{Z}}^2\,\left( 40\,s\,\left( {\delta }^2 + {\sigma }^2 \right)  - 
        4\,\delta \,\sigma \,\left( {\delta }^2 + 18\,\delta \,\sigma  + {\sigma }^2 \right)  \right)  \right)\)\, ,\\
G_{Z h}^{J,Y\(1\)} =&& \frac{1}{2}\(\left( C_{Zh,A}^{\{i j\} k l}\,\left( -{m_{h}}^2 - 5\,{m_{Z}}^2 + s \right)  \r\r\nn\\
&+&\l\l 
    C_{Zh,V}^{\{i j\} k l}\,\left( -{m_{h}}^2 + {m_{Z}}^2 + s \right)  \right) \,\sigma\)\, ,\\
G_{Z h}^{J,Y\(0\)} =&& \frac{1}{16 s}\(\sigma \,\left( -24\,C_{Zh,4}\,{m_{Z}}^2\,
     \left( {m_{h}}^2\,\left( s - {\delta }^2 \right)  + {m_{Z}}^2\,\left( -s + {\delta }^2 \right)  \r\r\r\nn\\
&+&\l\l\l  \left( s - 2\,{\delta }^2 \right) \,\left( s - {\sigma }^2 \right)  \right) +  C_{Zh,3}\,\left( 8\,{m_{h}}^4\,\left( s + \delta \,\sigma  \right)  \r\r\r\nn\\
&-&\l\l\l 
       8\,{m_{Z}}^4\,\left( 2\,s + \delta \,\left( -3\,\delta  + \sigma  \right)  \right) + s\,\left( 8\,s^2 + {\left( {\delta }^2 - {\sigma }^2 \right) }^2 - 4\,s\,\left( {\delta }^2 + {\sigma }^2 \right)  \right)  \r\r\r\nn\\
&-&\l\l\l 
       4\,{m_{Z}}^2\,\left( 6\,s^2 + \delta \,\sigma \,\left( {\delta }^2 + 6\,\delta \,\sigma  + {\sigma }^2 \right)  - 
          s\,\left( 9\,{\delta }^2 + 2\,\delta \,\sigma  + 3\,{\sigma }^2 \right)  \right)  \r\r\r\nn\\
&+&\l\l\l 
       4\,{m_{h}}^2\,\left( 6\,{m_{Z}}^2\,\left( s - {\delta }^2 \right)  - 
          \left( s + \delta \,\sigma  \right) \,\left( 4\,s - {\left( \delta  + \sigma  \right) }^2 \right)  \right)  \right)  \right)\),
\end{eqnarray}

\begin{eqnarray}
G_{Z h}^{K,T\(2\)} =&& \frac{1}{4s}\(-\left( \left( D_{Zh,A}^{\{i j\} k l} + D_{Zh,V}^{\{i j\} k l} \right) \,s\,
     \left( 8\,{m_{Z}}^2 - 4\,s + {\delta }^2 \right)  \right)  \r\nn\\
&+&\l 
  6\,\left( D_{Zh,A}^{\[i j\] k l} + D_{Zh,V}^{[i j] k l} \right) \,s\,\delta \,\sigma  \r\nn\\
&-&\l 
  \left( D_{Zh,A}^{\{i j\} k l} + D_{Zh,V}^{\{i j\} k l} \right) \,\left( s - 4\,{\delta }^2 \right) \,{\sigma }^2\)\, ,\\
G_{Z h}^{K,T\(1\)} =&& \frac{1}{8s^2 \sigma}\(2\,s\,\left( -6\,D_{Zh,8}^{\sigma}\,{m_{Z}}^2\,s\,\left( {\delta }^2 - {\sigma }^2 \right)  \r\r\nn\\
&-&\l\l 
     D_{Zh,5}^{\delta \sigma}\,{\sigma }^2\,\left( 8\,{m_{Z}}^4 + 
        {m_{Z}}^2\,\left( 6\,s - {\delta }^2 - 6\,\delta \,\sigma  - {\sigma }^2 \right)  \r\r\r\nn\\
&+&\l\l\l 
        2\,s\,\left( -s + {\delta }^2 + {\sigma }^2 \right)  + 
        {m_{h}}^2\,\left( -8\,{m_{Z}}^2 + 2\,s + {\delta }^2 + 6\,\delta \,\sigma  + {\sigma }^2 \right)  \right) 
     \right)  \r\nn\\
&+&\l D_{Zh,7}^{\sigma}\,\left( 16\,{m_{Z}}^4\,s\,\left( s - \delta \,\sigma  \right)  \r\r\nn\\
&+&\l\l 
     2\,{m_{h}}^2\,\left( s\,\delta \,{\left( \delta  - \sigma  \right) }^2\,\sigma  - 4\,{\delta }^3\,{\sigma }^3 + 
        8\,{m_{Z}}^2\,s\,\left( s + \delta \,\sigma  \right)  - 
        s^2\,\left( {\delta }^2 + 4\,\delta \,\sigma  + {\sigma }^2 \right)  \right)  \r\r\nn\\
&+&\l\l 
     2\,{m_{Z}}^2\,\left( -8\,s^3 + 4\,{\delta }^3\,{\sigma }^3 - 
        s\,\delta \,\sigma \,\left( {\delta }^2 - 10\,\delta \,\sigma  + {\sigma }^2 \right)  + 
        s^2\,\left( {\delta }^2 + 4\,\delta \,\sigma  + {\sigma }^2 \right)  \right)  \r\r\nn\\
&+&\l\l 
     s\,\left( -2\,s^2\,\left( {\delta }^2 + {\sigma }^2 \right)  - 
        4\,{\delta }^2\,{\sigma }^2\,\left( {\delta }^2 + {\sigma }^2 \right)  + 
        s\,\left( {\delta }^4 + 6\,{\delta }^2\,{\sigma }^2 + {\sigma }^4 \right)  \right)  \right)\)\, ,\\
G_{Z h}^{K,T\(0\)} =&& \frac{1}{64s^2 \sigma}\(D_{Zh,7}^{\sigma}\,\left( 64\,{m_{Z}}^6\,\delta \,\sigma \,\left( s - \delta \,\sigma  \right)  \r\r\nn\\
&+&\l\l 
     8\,{m_{h}}^4\,\delta \,\sigma \,\left( s + \delta \,\sigma  \right) \,
      \left( -8\,{m_{Z}}^2 + {\left( \delta  + \sigma  \right) }^2 \right)  \r\r\nn\\
&-&\l\l 
     s\,{\left( {\delta }^2 - {\sigma }^2 \right) }^2\,
      \left( -4\,{\delta }^2\,{\sigma }^2 + s\,\left( {\delta }^2 + {\sigma }^2 \right)  \right)  \r\r\nn\\
&+&\l\l 
     8\,{m_{Z}}^4\,\left( 4\,s^2\,{\left( \delta  - \sigma  \right) }^2 + 
        s\,\delta \,\sigma \,\left( {\delta }^2 - 14\,\delta \,\sigma  + {\sigma }^2 \right) + {\delta }^2\,{\sigma }^2\,\left( {\delta }^2 + 10\,\delta \,\sigma  + {\sigma }^2 \right)  \right)\r\r\nn\\
&-&\l\l
     4\,{m_{Z}}^2\,\left( s^2\,{\left( \delta  - \sigma  \right) }^2\,
         \left( {\delta }^2 + 4\,\delta \,\sigma  + {\sigma }^2 \right) +  4\,{\delta }^3\,{\sigma }^3\,\left( {\delta }^2 + 6\,\delta \,\sigma  + {\sigma }^2 \right)\r\r\r\nn\\
&-&\l\l\l s\,\delta \,\sigma \,\left( {\delta }^4 + 12\,{\delta }^3\,\sigma  + 6\,{\delta }^2\,{\sigma }^2 + 12\,\delta \,{\sigma }^3 + 
           {\sigma }^4 \right)  \right)  - 4\,{m_{h}}^2\,
      \left( 32\,{m_{Z}}^4\,\left( s^2 - {\delta }^2\,{\sigma }^2 \right)  \r\r\r\nn\\
&-&\l\l\l 
        {\left( \delta  + \sigma  \right) }^2\,\left( 4\,{\delta }^3\,{\sigma }^3 - 
           s\,\delta \,\sigma \,{\left( \delta  + \sigma  \right) }^2 + s^2\,\left( {\delta }^2 + {\sigma }^2 \right)  \right)  \r\r\r\nn\\
&-&\l\l\l 
        4\,{m_{Z}}^2\,\left( s^2\,\left( {\delta }^2 + 4\,\delta \,\sigma  + {\sigma }^2 \right)  - s\delta \sigma \left( {\delta }^2 + 6\delta \sigma  + {\sigma }^2 \right)  \r\r\r\r\nn\\
&-&\l\l\l\l 
           {\delta }^2{\sigma }^2\left( {\delta }^2 + 6\delta \sigma  + {\sigma }^2 \right)  \right)  \right)  \right)+ 2\sigma \left( 24D_{Zh,8}^{\sigma}{m_{Z}}^2\left( {m_{h}}^2 - {m_{Z}}^2 \right) s
      \delta \left( {\delta }^2 - {\sigma }^2 \right)  \r\r\nn\\
&+&\l\l 
     \sigma \,\left( 24\,D_{Zh,6}^{\delta \sigma}\,{m_{Z}}^2\,\left( {m_{h}}^2 - {m_{Z}}^2 \right) \,s\,
         \left( {\delta }^2 - {\sigma }^2 \right)  \r\r\r\nn\\
&+&\l\l\l D_{Zh,5}^{\delta \sigma}\,
         \left( 32\,{m_{Z}}^6\,\left( s - \delta \,\sigma  \right)  + s^2\,{\left( {\delta }^2 - {\sigma }^2 \right) }^2 \r\r\r\r\nn\\
&-&\l\l\l\l 
           2\,{m_{Z}}^2\,s\,{\left( \delta  - \sigma  \right) }^2\,
            \left( 2\,s + {\delta }^2 + 6\,\delta \,\sigma  + {\sigma }^2 \right)  \r\r\r\r\nn\\
&+&\l\l\l\l 
           4\,{m_{h}}^4\,\left( s + \delta \,\sigma  \right) \,
            \left( -8\,{m_{Z}}^2 + {\left( \delta  + \sigma  \right) }^2 \right)  \r\r\r\r\nn\\
&-&\l\l\l\l 
           4\,{m_{Z}}^4\,\left( -\left( \delta \,\sigma \,
                 \left( {\delta }^2 + 18\,\delta \,\sigma  + {\sigma }^2 \right)  \right)  + 
              s\,\left( 7\,{\delta }^2 + 6\,\delta \,\sigma  + 7\,{\sigma }^2 \right)  \right)  \r\r\r\r\nn\\
&+&\l\l\l\l  2\,{m_{h}}^2\,\left( 32\,{m_{Z}}^4\,\delta \,\sigma+ s\,{\left( \delta  + \sigma  \right) }^2\,\left( -2\,s + {\left( \delta  + \sigma  \right) }^2 \right)  \r\r\r\r\r\nn\\
&+&\l\l\l\l\l 4\,{m_{Z}}^2\,\left( 2\,s^2 - \delta \,\sigma \,
                  \left( {\delta }^2 + 10\,\delta \,\sigma  + {\sigma }^2 \right) +  s\,\left( 3\,{\delta }^2 + 2\,\delta \,\sigma  + 3\,{\sigma }^2 \right)  \right)  \right)  \right)  \right)  \right) \)\, ,\\
G_{Z h}^{K,Y\(2\)} =&& \frac{1}{4s}\(C_{Zh,A}^{\{i j\} k l}\,\left( -8\,{m_{Z}}^2\,s + 4\,s^2 - s\,{\delta }^2 + s\,{\sigma }^2 - 
     4\,{\delta }^2\,{\sigma }^2 \right)  \r\nn\\
&+&\l C_{Zh,V}^{\{i j\} k l}\,
   \left( 8\,{m_{Z}}^2\,s - 4\,s^2 - s\,{\delta }^2 + s\,{\sigma }^2 + 
     4\,{\delta }^2\,{\sigma }^2 \right)\)\, ,\\
G_{Z h}^{K,Y\(1\)} =&& \frac{1}{8s}\(\delta \,\sigma \,\left( -4\,C_{Zh,2}\,{m_{Z}}^2\,\left( 2\,{m_{h}}^2 - 2\,{m_{Z}}^2 + s \right)  \r\r\nn\\
&+&\l\l 
    C_{Zh,1}\,\left( {m_{h}}^2 - {m_{Z}}^2 - s \right) \,\left( {\delta }^2 - {\sigma }^2 \right)  \right)\)\, ,\\
G_{Z h}^{K,Y\(0\)} =&-&\frac{1}{64 s^2}\(C_{Zh,3}\,\left( {\delta }^2 - {\sigma }^2 \right) \,
   \left( 8\,{m_{h}}^4\,\delta \,\sigma \,\left( s + \delta \,\sigma  \right)  - 
     8\,{m_{Z}}^4\,\left( 6\,s^2 + s\,\delta \,\sigma  - {\delta }^2\,{\sigma }^2 \right)  \r\r\nn\\
&+&\l\l 
     s^2\,\left( {\delta }^4 + 6\,{\delta }^2\,{\sigma }^2 + {\sigma }^4 - 4\,s\,\left( {\delta }^2 + {\sigma }^2 \right)  \right)  \r\r\nn\\
&+&\l\l 
     4\,{m_{Z}}^2\,s\,\left( 4\,s^2 - \delta \,{\left( \delta  - \sigma  \right) }^2\,\sigma  + 
        s\,\left( 3\,{\delta }^2 + 2\,\delta \,\sigma  + 3\,{\sigma }^2 \right)  \right)  \r\r\nn\\
&-&\l\l 
     4\,{m_{h}}^2\,\left( -\left( s\,{\left( \delta  - \sigma  \right) }^2\,\left( s + \delta \,\sigma  \right)  \right)  + 
        4\,{m_{Z}}^2\,\left( 2\,s^2 + {\delta }^2\,{\sigma }^2 \right)  \right)  \right)  \r\nn\\
&+&\l 
  4\,C_{Zh,4}\,\left( 16\,{m_{Z}}^6\,\delta \,\sigma \,\left( -s + \delta \,\sigma  \right) +  s\,{\left( {\delta }^2 - {\sigma }^2 \right) }^2\,\left( -s^2 + {\delta }^2\,{\sigma }^2 \right)  \r\r\nn\\
&-&\l\l 
     4\,{m_{Z}}^4\,\left( -\left( {\delta }^2\,{\left( \delta  - \sigma  \right) }^2\,{\sigma }^2 \right)  + 
        2\,s^2\,\left( {\delta }^2 - 3\,\delta \,\sigma  + {\sigma }^2 \right)\r\r\r\nn\\
&+&\l\l\l 2\,s\,\delta \,\sigma \,\left( {\delta }^2 - \delta \,\sigma  + {\sigma }^2 \right)  \right)  \r\r\nn\\
&+&\l\l 
     4\,{m_{Z}}^2\,\left( 2\,s^3\,\left( {\delta }^2 - \delta \,\sigma  + {\sigma }^2 \right)  - 
        {\delta }^3\,{\sigma }^3\,\left( {\delta }^2 + 6\,\delta \,\sigma  + {\sigma }^2 \right)  \r\r\r\nn\\
&+&\l\l\l 
        s\,{\delta }^2\,{\sigma }^2\,\left( 7\,{\delta }^2 + 2\,\delta \,\sigma  + 7\,{\sigma }^2 \right)  - 
        s^2\,\left( {\delta }^4 - {\delta }^3\,\sigma  + 10\,{\delta }^2\,{\sigma }^2 - \delta \,{\sigma }^3 + {\sigma }^4 \right) 
        \right)  \r\r\nn\\
&-&\l\l 4\,{m_{h}}^4\,\left( -4\,{m_{Z}}^2\,s\,\delta \,\sigma +  s^2\,{\left( \delta  + \sigma  \right) }^2 - {\delta }^2\,{\sigma }^2\,
         \left( 4\,{m_{Z}}^2 + {\left( \delta  + \sigma  \right) }^2 \right)  \right)  \r\r\nn\\
&+&\l\l 
     4\,{m_{h}}^2\,\left( {\left( \delta  + \sigma  \right) }^2\,{\left( s - \delta \,\sigma  \right) }^2\,
         \left( s + \delta \,\sigma  \right)+ 8\,{m_{Z}}^4\,\left( s^2 - {\delta }^2\,{\sigma }^2 \right)  \r\r\r\nn\\
&+&\l\l\l
        {m_{Z}}^2\,\left( -8\,s^3 - 2\,{\delta }^2\,{\sigma }^2\,\left( {\delta }^2 + {\sigma }^2 \right)  + 
           2\,s\,\delta \,\sigma \,\left( {\delta }^2 + \delta \,\sigma  + {\sigma }^2 \right)  \r\r\r\r\nn\\
&+&\l\l\l\l 
           s^2\,\left( 5\,{\delta }^2 - 4\,\delta \,\sigma  + 5\,{\sigma }^2 \right)  \right)  \right)  \right)\),
\end{eqnarray}

where we have used the following coupling functions

\begin{eqnarray}
D_{Zh,1}^{\delta\sigma} =&& \delta D_{Zh,A}^{\[i j\] k l}+\sigma D_{Zh,V}^{\{i j\} k l}\, ,\\
D_{Zh,2}^{\delta\sigma} =&& \delta D_{Zh,A}^{\[i j\] k l}-\sigma D_{Zh,V}^{\{i j\} k l}\, ,\\
D_{Zh,3}^{\delta\sigma} =&& \delta D_{Zh,V}^{\[i j\] k l}+\sigma D_{Zh,A}^{\{i j\} k l}\, ,\\
D_{Zh,4}^{\delta\sigma} =&& \delta D_{Zh,V}^{\[i j\] k l}-\sigma D_{Zh,A}^{\{i j\} k l}\, ,\\
D_{Zh,5}^{\delta\sigma} =&& \delta D_{Zh,A}^{\[i j\] k l}+\delta D_{Zh,V}^{\[i j\] k l}\, ,\\
D_{Zh,6}^{\delta\sigma} =&& \delta D_{Zh,A}^{\[i j\] k l}-\delta D_{Zh,V}^{\[i j\] k l}\, ,\\
D_{Zh,7}^{\sigma} =&& \sigma D_{Zh,V}^{\{i j\} k l}+\sigma D_{Zh,A}^{\{i j\} k l}\, ,\\
D_{Zh,8}^{\sigma} =&& \sigma D_{Zh,V}^{\{i j\} k l}-\sigma D_{Zh,A}^{\{i j\} k l},
\end{eqnarray}

\begin{eqnarray}
C_{Zh,1} =&& C_{Zh,A}^{\[i j\] k l}+C_{Zh,V}^{\[i j\] k l}\, ,\\
C_{Zh,2} =&& C_{Zh,A}^{\[i j\] k l}-C_{Zh,V}^{\[i j\] k l}\, ,\\
C_{Zh,3} =&& C_{Zh,A}^{\{i j\} k l}+C_{Zh,V}^{\{i j\} k l}\, ,\\
C_{Zh,4} =&& C_{Zh,A}^{\{i j\} k l}-C_{Zh,V}^{\{i j\} k l},
\end{eqnarray}

\begin{eqnarray}
D_{Zh,A}^{\{i j\} k l} =&& C_{A}^{\chi_{j}^{0} \chi_{k}^{0} Z} \(C_{A}^{\chi_{j}^{0} \chi_{l}^{0} Z}\)^{*} C_{S}^{\chi_{k}^{0} \chi_{i}^{0} h} \(C_{S}^{\chi_{l}^{0} \chi_{i}^{0} h}\)^{*}\nn\\
&+&C_{A}^{\chi_{k}^{0} \chi_{i}^{0} Z} \(C_{A}^{\chi_{l}^{0} \chi_{i}^{0} Z}\)^{*} C_{S}^{\chi_{j}^{0} \chi_{k}^{0} h} \(C_{S}^{\chi_{j}^{0} \chi_{l}^{0} h}\)^{*}\, ,\\
D_{Zh,A}^{\[i j\] k l} =&& C_{A}^{\chi_{j}^{0} \chi_{k}^{0} Z} \(C_{A}^{\chi_{j}^{0} \chi_{l}^{0} Z}\)^{*} C_{S}^{\chi_{k}^{0} \chi_{i}^{0} h} \(C_{S}^{\chi_{l}^{0} \chi_{i}^{0} h}\)^{*}\nn\\
&-&C_{A}^{\chi_{k}^{0} \chi_{i}^{0} Z} \(C_{A}^{\chi_{l}^{0} \chi_{i}^{0} Z}\)^{*} C_{S}^{\chi_{j}^{0} \chi_{k}^{0} h} \(C_{S}^{\chi_{j}^{0} \chi_{l}^{0} h}\)^{*}\, ,\\
D_{Zh,V}^{\{i j\} k l} =&& C_{V}^{\chi_{j}^{0} \chi_{k}^{0} Z} \(C_{V}^{\chi_{j}^{0} \chi_{l}^{0} Z}\)^{*} C_{S}^{\chi_{k}^{0} \chi_{i}^{0} h} \(C_{S}^{\chi_{l}^{0} \chi_{i}^{0} h}\)^{*}\nn\\
&+&C_{V}^{\chi_{k}^{0} \chi_{i}^{0} Z} \(C_{V}^{\chi_{l}^{0} \chi_{i}^{0} Z}\)^{*} C_{S}^{\chi_{j}^{0} \chi_{k}^{0} h} \(C_{S}^{\chi_{j}^{0} \chi_{l}^{0} h}\)^{*}\, ,\\
D_{Zh,V}^{\[i j\] k l} =&& C_{V}^{\chi_{j}^{0} \chi_{k}^{0} Z} \(C_{V}^{\chi_{j}^{0} \chi_{l}^{0} Z}\)^{*} C_{S}^{\chi_{k}^{0} \chi_{i}^{0} h} \(C_{S}^{\chi_{l}^{0} \chi_{i}^{0} h}\)^{*}\nn\\
&-&C_{V}^{\chi_{k}^{0} \chi_{i}^{0} Z} \(C_{V}^{\chi_{l}^{0} \chi_{i}^{0} Z}\)^{*} C_{S}^{\chi_{j}^{0} \chi_{k}^{0} h} \(C_{S}^{\chi_{j}^{0} \chi_{l}^{0} h}\)^{*},
\end{eqnarray}

\begin{eqnarray}
C_{Zh,A}^{\{i j\} k l} =&& C_{A}^{\chi_{k}^{0} \chi_{i}^{0} Z} \(C_{A}^{\chi_{j}^{0} \chi_{l}^{0} Z}\)^{*} C_{S}^{\chi_{j}^{0} \chi_{k}^{0} h} \(C_{S}^{\chi_{l}^{0} \chi_{i}^{0} h}\)^{*}\nn\\
&+&C_{A}^{\chi_{j}^{0} \chi_{k}^{0} Z} \(C_{A}^{\chi_{l}^{0} \chi_{i}^{0} Z}\)^{*} C_{S}^{\chi_{k}^{0} \chi_{i}^{0} h} \(C_{S}^{\chi_{j}^{0} \chi_{l}^{0} h}\)^{*}\, ,\\
C_{Zh,A}^{\[i j\] k l} =&& C_{A}^{\chi_{k}^{0} \chi_{i}^{0} Z} \(C_{A}^{\chi_{j}^{0} \chi_{l}^{0} Z}\)^{*} C_{S}^{\chi_{j}^{0} \chi_{k}^{0} h} \(C_{S}^{\chi_{l}^{0} \chi_{i}^{0} h}\)^{*}\nn\\
&-&C_{A}^{\chi_{j}^{0} \chi_{k}^{0} Z} \(C_{A}^{\chi_{l}^{0} \chi_{i}^{0} Z}\)^{*} C_{S}^{\chi_{k}^{0} \chi_{i}^{0} h} \(C_{S}^{\chi_{j}^{0} \chi_{l}^{0} h}\)^{*}\, ,\\
C_{Zh,V}^{\{i j\} k l} =&& C_{V}^{\chi_{k}^{0} \chi_{i}^{0} Z} \(C_{V}^{\chi_{j}^{0} \chi_{l}^{0} Z}\)^{*} C_{S}^{\chi_{j}^{0} \chi_{k}^{0} h} \(C_{S}^{\chi_{l}^{0} \chi_{i}^{0} h}\)^{*}\nn\\
&+&C_{V}^{\chi_{j}^{0} \chi_{k}^{0} Z} \(C_{V}^{\chi_{l}^{0} \chi_{i}^{0} Z}\)^{*} C_{S}^{\chi_{k}^{0} \chi_{i}^{0} h} \(C_{S}^{\chi_{j}^{0} \chi_{l}^{0} h}\)^{*}\, ,\\
C_{Zh,V}^{\[i j\]k l} =&& C_{V}^{\chi_{k}^{0} \chi_{i}^{0} Z} \(C_{V}^{\chi_{j}^{0} \chi_{l}^{0} Z}\)^{*} C_{S}^{\chi_{j}^{0} \chi_{k}^{0} h} \(C_{S}^{\chi_{l}^{0} \chi_{i}^{0} h}\)^{*}\nn\\
&-&C_{V}^{\chi_{j}^{0} \chi_{k}^{0} Z} \(C_{V}^{\chi_{l}^{0} \chi_{i}^{0} Z}\)^{*} C_{S}^{\chi_{k}^{0} \chi_{i}^{0} h} \(C_{S}^{\chi_{j}^{0} \chi_{l}^{0} h}\)^{*};
\end{eqnarray}

\begin{flushleft}
\textbf{Higgs (A)-Z cross term:}
\end{flushleft}

\begin{eqnarray}
\tilde{\omega}^{\(A-Z\)}_{Z h} =&& Re\[\(\frac{C^{Z h A} C_{P}^{\chi_{i}^{0} \chi_{j}^{0} A}}{s-m_{A}^{2} + i m_{A} \Gamma_{A}}\)^{*} \frac{C^{Z Z h} C_{A}^{\chi_{i}^{0} \chi_{j}^{0} Z}}{s-m_{Z}^{2} + i m_{Z} \Gamma_{Z}}\]\nn\\
&\times& \frac{\(m_{Z}^2-s\)}{2 m_{Z}^4}\left( {m_{h}}^4 + {\left( {m_{Z}}^2 - s \right) }^2 - 
    2\,{m_{h}}^2\,\left( {m_{Z}}^2 + s \right)  \right) \,\sigma ;
\end{eqnarray}

\begin{flushleft}
\textbf{Higgs (A)-neutralino cross term:}
\end{flushleft}

\begin{eqnarray}
\tilde{\omega}^{\(A-\chi^{0}\)}_{Z h} =&& \frac{1}{8 m_{Z}^2 s}\sum_{k=1}^{4}Re\[\(\frac{C^{Z h A} C_{P}^{\chi_{i}^{0} \chi_{j}^{0} A}}{s-m_{A}^{2} + i m_{A} \Gamma_{A}}\)^{*}\r\nn\\
&\times&\l\(C_{Zh}^{\{i j\} k}G_{A,Z h}^{C_{Zh}^{\{i j\} k}}+C_{Zh}^{\[i j\] k}G_{A,Z h}^{C_{Zh}^{\[i j\] k}}\)\],
\end{eqnarray}

\begin{eqnarray}
G_{A,Z h}^{C_{Zh}^{\{i j\} k}} =&-&4\,s\,\left( -2\,\left( {m_{h}}^2 + {m_{Z}}^2 - s \right) \,s + 
     \left( {m_{h}}^2 - {m_{Z}}^2 - s \right) \,{\delta }^2 \right)  \nn\\
&+& 
  \left( s\,\left( -8\,\left( 2\,{m_{h}}^2\,{m_{Z}}^2-  {m_{\chi_{k}^{0}}}^2\,\left( {m_{h}}^2 + {m_{Z}}^2 - s \right)  \right) \,s \r\r\nn\\
&+&\l\l 
        2\,\left( -\left( \left( m_{h}^2 - m_{Z}^2 \right) 
              \left( 2\,{m_{\chi_{k}^{0}}}^2 + {m_{h}}^2 - 3\,{m_{Z}}^2 \right)  \right)  \r\r\r\nn\\
&+&\l\l\l 
           \left( 2\,{m_{\chi_{k}^{0}}}^2 + {m_{h}}^2 + 3\,{m_{Z}}^2 \right) \,s \right) \,{\delta }^2 - 
        \left( -{m_{h}}^2 + {m_{Z}}^2 + s \right) \,{\delta }^4 \right)  \r\nn\\
&-&\l 2\,\left( 2\,m_{\chi_{k}^{0}}\,\left( {m_{h}}^4 + {\left( {m_{Z}}^2 - s \right) }^2- 2\,{m_{h}}^2\,\left( {m_{Z}}^2 + s \right)  \right) \,\left( s - {\delta }^2 \right)  \r\r\nn\\
&+&\l\l 
        \left( m_{h}^2 - m_{Z}^2 \right)\delta \,
         \left( 2\,\left( {m_{h}}^2 + {m_{Z}}^2 - s \right) \,s+ \left( -{m_{h}}^2 + {m_{Z}}^2 + s \right) \,{\delta }^2 \right)  \right) \,\sigma \r\nn\\
&+&\l \left( 2\,s\,\left( -{\left( {m_{h}}^2 - {m_{Z}}^2 \right) }^2 + 
           \left( {m_{h}}^2 + {m_{Z}}^2 \right) \,s \right)  \r\r\nn\\
&+&\l\l 
        \left( 2\,{\left( {m_{h}}^2 - {m_{Z}}^2 \right) }^2 - 
           \left( 3\,{m_{h}}^2 + 5\,{m_{Z}}^2 \right) \,s + s^2 \right) \,{\delta }^2 \right) \,{\sigma }^2 \right) \nn\\
&\times&
    \mathcal{F}\[s, \sigma,\delta,m_{h}^2,m_{Z}^2,{m_{\chi_{k}^{0}}}^2\]\, ,\nn\\
\\
G_{A,Z h}^{C_{Zh}^{\[i j\] k}} =&-&\delta \,\left( 4\,\left( {m_{h}}^2 - {m_{Z}}^2 - s \right) \,s\,\left( 2\,m_{\chi_{k}^{0}} - \sigma  \right)  + 
    \left( 8\,{m_{\chi_{k}^{0}}}^3\,\left( {m_{h}}^2 - {m_{Z}}^2 - s \right) \,s \r\r\nn\\
&+&\l\l 
       4\,{m_{\chi_{k}^{0}}}^2\,s\,\left( -{m_{h}}^2 + {m_{Z}}^2 + s \right) \,\sigma  \r\r\nn\\
&-&\l\l 
       2\,m_{\chi_{k}^{0}}\,\left( {m_{h}}^2 - {m_{Z}}^2 - s \right) \,
        \left( 2\,{m_{Z}}^2\,\left( s - \delta \,\sigma  \right)  + 
          2\,{m_{h}}^2\,\left( s + \delta \,\sigma  \right)  \r\r\r\nn\\
&+&\l\l\l s\,\left( -2\,s + {\delta }^2 + {\sigma }^2 \right)  \right)  +
        \sigma \,\left( 2\,{m_{h}}^4\,\delta \,\left( \delta  + \sigma  \right)  + 
          s^2\,\left( \delta  - \sigma  \right) \,\left( \delta  + \sigma  \right)  \r\r\r\nn\\
&+&\l\l\l 
          {m_{Z}}^2\,s\,\left( 4\,s - 5\,{\delta }^2 + 2\,\delta \,\sigma  - {\sigma }^2 \right)  + 
          2\,{m_{Z}}^4\,\left( -2\,s + \delta \,\left( \delta  + \sigma  \right)  \right)  \r\r\r\nn\\
&+&\l\l\l 
          {m_{h}}^2\,\left( -\left( s\,\left( 3\,\delta  - \sigma  \right) \,\left( \delta  + \sigma  \right)  \right)  + 
             4\,{m_{Z}}^2\,\left( s - \delta \,\left( \delta  + \sigma  \right)  \right)  \right)  \right)  \right) \r\nn\\
&\times&\l
     \mathcal{F}\[s, \sigma,\delta,m_{h}^2,m_{Z}^2,{m_{\chi_{k}^{0}}}^2\] \right)\, ,\nn\\
\end{eqnarray}

where we have used the following coupling functions

\begin{eqnarray}
C_{Zh}^{\{i j\} k} =&& C_{A}^{\chi_{j}^{0} \chi_{k}^{0} Z} C_{S}^{\chi_{k}^{0} \chi_{i}^{0} H}+C_{A}^{\chi_{k}^{0} \chi_{i}^{0} Z} C_{S}^{\chi_{j}^{0} \chi_{k}^{0} H}\, ,\\
C_{Zh}^{\[i j\] k} =&& C_{A}^{\chi_{j}^{0} \chi_{k}^{0} Z} C_{S}^{\chi_{k}^{0} \chi_{i}^{0} H}-C_{A}^{\chi_{k}^{0} \chi_{i}^{0} Z} C_{S}^{\chi_{j}^{0} \chi_{k}^{0} H};
\end{eqnarray}

\begin{flushleft}
\textbf{Z-neutralino cross term:}
\end{flushleft}

\begin{eqnarray}
\tilde{\omega}^{\(Z-\chi^{0}\)}_{Z h} =&& \frac{1}{2}\sum_{k=1}^{4} Re\[\(\frac{C^{Z Z h}}{s-m_{Z}^{2} + i m_{Z} \Gamma_{Z}}\)\frac{1}{16 m_{Z}^4 s^2}\( C_{Zh,A}^{+} G_{Z, Zh}^{C_{Zh,A}^{+}}+C_{Zh,A}^{-} G_{Z, Zh}^{C_{Zh,A}^{-}}\r\r\nn\\
&+&\l\l C_{Zh,S}^{+} G_{Z, Zh}^{C_{Zh,S}^{+}}+C_{Zh,S}^{-} G_{Z, Zh}^{C_{Zh,S}^{-}}\)\]\, ,\nn\\
\end{eqnarray}

\begin{eqnarray}
G_{Z, Zh}^{C_{Zh,A}^{+}} =&-&\left( \delta \,\left( -4\,s\,\left( 4\,{m_{\chi_{k}^{0}}}^2\,{m_{Z}}^2\,s + 6\,{m_{Z}}^4\,s - 
         4\,{m_{Z}}^2\,s^2 + 2\,s^3 - {m_{Z}}^2\,s\,{\delta }^2 \r\r\r\nn\\
&-&\l\l\l 
         4\,m_{\chi_{k}^{0}}\,\left( {m_{h}}^2 - {m_{Z}}^2 - s \right) \,\left( 2\,{m_{Z}}^2 - s \right) \,
          \sigma  + 2\,{m_{Z}}^4\,\delta \,\sigma  - {m_{Z}}^2\,s\,{\sigma }^2 \r\r\r\nn\\
&-&\l\l\l 
         2\,{m_{h}}^2\,\left( s^2 + {m_{Z}}^2\,\left( -s + \delta \,\sigma  \right)  \right)  \right)  +
      \left( -16\,{m_{\chi_{k}^{0}}}^4\,{m_{Z}}^2\,s^2 \r\r\r\nn\\
&+&\l\l\l 
         16\,{m_{\chi_{k}^{0}}}^3\,\left( {m_{h}}^2 - {m_{Z}}^2 - s \right) \,
          \left( 2\,{m_{Z}}^2 - s \right) \,s\,\sigma \r\r\r\nn\\
&-&\l\l\l
         4\,m_{\chi_{k}^{0}}\,\left( {m_{h}}^2 - {m_{Z}}^2 - s \right) \,\left( 2\,{m_{Z}}^2 - s \right) \,
          \sigma \,\left( 2\,{m_{Z}}^2\,\left( s - \delta \,\sigma  \right)  + 
            2\,{m_{h}}^2\,\left( s + \delta \,\sigma  \right)  \r\r\r\r\nn\\
&+&\l\l\l\l s\,\left( -2\,s + {\delta }^2 + {\sigma }^2 \right)  \right) 
         - 8\,{m_{\chi_{k}^{0}}}^2\,s\,\left( s^2\,\left( -{m_{h}}^2 + s \right)  + 
            2\,{m_{Z}}^4\,\left( s + \delta \,\sigma  \right)  \r\r\r\r\nn\\
&-&\l\l\l\l
            {m_{Z}}^2\,\left( s^2 + 2\,{m_{h}}^2\,\delta \,\sigma  + 
               s\,\left( {\delta }^2 + {\sigma }^2 \right)  \right)  \right)  \r\r\r\nn\\
&+&\l\l\l 
         s\,\left( -2\,{m_{h}}^2\,\left( {m_{h}}^2 - s \right) \,s\,{\left( \delta  + \sigma  \right) }^2 + 
            8\,{m_{Z}}^6\,\left( 4\,s - \sigma \,\left( \delta  + 3\,\sigma  \right)  \right)  \r\r\r\r\nn\\
&-&\l\l\l\l 
            {m_{Z}}^2\,\left( s\,{\left( \delta  - \sigma  \right) }^2\,
                \left( -2\,s + {\left( \delta  + \sigma  \right) }^2 \right)  \r\r\r\r\r\nn\\
&+&\l\l\l\l\l 
               4\,{m_{h}}^2\,\left( 4\,s^2 + \delta \,\sigma \,{\left( \delta  + \sigma  \right) }^2 - 
                  s\,\left( {\delta }^2 + {\sigma }^2 \right)  \right)  \right)  \r\r\r\r\nn\\
&+&\l\l\l\l 
            {m_{Z}}^4\,\left( s\,\left( -6\,{\delta }^2 + 4\,\delta \,\sigma  - 30\,{\sigma }^2 \right)  + 
               4\,\sigma \,\left( 2\,{m_{h}}^2\,\left( \delta  + 3\,\sigma  \right)  \r\r\r\r\r\r\nn\\
&+&\l\l\l\l\l\l 
                  \delta \,\left( {\delta }^2 + 6\,\delta \,\sigma  + {\sigma }^2 \right)  \right)  \right)  \right)  \right) \,
       \mathcal{F}\[s, \sigma,\delta,m_{h}^2,m_{Z}^2,{m_{\chi_{k}^{0}}}^2\] \right)  \right)\, ,\\
G_{Z, Zh}^{C_{Zh,A}^{-}} =&-&2\,s\,\delta \,\left( -4\,s\,\left( -6\,{m_{Z}}^4 + 
       \left( {m_{h}}^2 - s \right) \,\left( s - {\sigma }^2 \right)  + {m_{Z}}^2\,\left( s + {\sigma }^2 \right) 
       \right)  \r\nn\\
&-&\l \left( -7\,{m_{Z}}^4\,s\,{\delta }^2 + {m_{Z}}^2\,s^2\,{\delta }^2 - 12\,{m_{Z}}^6\,\delta \,\sigma  + 2\,{m_{Z}}^4\,s\,\delta \,\sigma  - 
       2\,{m_{Z}}^2\,s^2\,\delta \,\sigma  \r\r\nn\\
&+&\l\l 12\,{m_{Z}}^6\,{\sigma }^2 - 23\,{m_{Z}}^4\,s\,{\sigma }^2 + 
       5\,{m_{Z}}^2\,s^2\,{\sigma }^2 + 26\,{m_{Z}}^4\,{\delta }^2\,{\sigma }^2 -
       5\,{m_{Z}}^2\,s\,{\delta }^2\,{\sigma }^2 \r\r\nn\\
&+&\l\l s^2\,{\delta }^2\,{\sigma }^2 + 
       2\,{m_{Z}}^4\,\delta \,{\sigma }^3 + 2\,{m_{Z}}^2\,s\,\delta \,{\sigma }^3 - 
       {m_{Z}}^2\,s\,{\sigma }^4 - s^2\,{\sigma }^4 \r\r\nn\\
&-&\l\l {m_{h}}^4\,\left( \delta  + \sigma  \right) \,
        \left( -2\,\delta \,{\sigma }^2 + s\,\left( \delta  + \sigma  \right)  \right)  \r\r\nn\\
&-&\l\l 
       4\,{m_{\chi_{k}^{0}}}^2\,s\,\left( 6\,{m_{Z}}^4 - 
          \left( {m_{h}}^2 - s \right) \,\left( s - {\sigma }^2 \right)  - {m_{Z}}^2\,\left( s + {\sigma }^2 \right) 
          \right)  \r\r\nn\\
&+&\l\l {m_{h}}^2\left( 12{m_{Z}}^4\,
           \left( 2\,s + \left( \delta  - \sigma  \right) \sigma  \right)  + 
          s\left( \delta  + \sigma  \right) \left( {\sigma }^2\left( -3\,\delta  + \sigma  \right)  + 
             s\left( \delta  + \sigma  \right)  \right)  \r\r\r\nn\\
&+&\l\l\l 
          {m_{Z}}^2\,\left( -8\,s^2 - 4\,\delta \,{\sigma }^2\,\left( \delta  + \sigma  \right)  + 
             2\,s\,\left( {\delta }^2 + 3\,{\sigma }^2 \right)  \right)  \right)  \right) \r\nn\\
&\times&\l \mathcal{F}\[s, \sigma,\delta,m_{h}^2,m_{Z}^2,{m_{\chi_{k}^{0}}}^2\]
    \right)\, ,\\
G_{Z, Zh}^{C_{Zh,S}^{+}} =&& 4\,s\,\left( 8\,{m_{\chi_{k}^{0}}}^3\,{m_{Z}}^2\,s + 4\,{m_{\chi_{k}^{0}}}^2\,{m_{Z}}^2\,s\,\sigma  \r\nn\\
&-&\l 
     2\,m_{\chi_{k}^{0}}\,{m_{Z}}^2\,\left( 2\,{m_{Z}}^2\,\left( s - \delta \,\sigma  \right)  + 
        2\,{m_{h}}^2\,\left( s + \delta \,\sigma  \right)  \r\r\nn\\
&+&\l\l s\,\left( -2\,s + {\delta }^2 + {\sigma }^2 \right)  \right)  - 
     \sigma \,\left( -2\,s^3 - 2\,{m_{Z}}^4\,\left( 3\,s + \delta \,\sigma  \right)  + 
        {m_{Z}}^2\,s\,\left( 4\,s + {\delta }^2 + {\sigma }^2 \right)  \r\r\nn\\
&+&\l\l 
        2\,{m_{h}}^2\,\left( s^2 + {m_{Z}}^2\,\left( -s + \delta \,\sigma  \right)  \right)  \right)  \right)  + 
  \left( 32\,{m_{\chi_{k}^{0}}}^5\,{m_{Z}}^2\,s^2 + 16\,{m_{\chi_{k}^{0}}}^4\,{m_{Z}}^2\,s^2\,\sigma  \r\nn\\
&-&\l 
     16\,{m_{\chi_{k}^{0}}}^3\,{m_{Z}}^2\,s\,\left( 2\,{m_{Z}}^2\,\left( s - \delta \,\sigma  \right)  + 
        2\,{m_{h}}^2\,\left( s + \delta \,\sigma  \right)  + s\,\left( -2\,s + {\delta }^2 + {\sigma }^2 \right)  \right)  \r\nn\\
&+&\l 
     8\,{m_{\chi_{k}^{0}}}^2\,s\,\sigma \,\left( s^2\,\left( -{m_{h}}^2 + s \right)  + 
        2\,{m_{Z}}^4\,\left( s + \delta \,\sigma  \right)  \r\r\nn\\
&-&\l\l 
        {m_{Z}}^2\,\left( s^2 + 2\,{m_{h}}^2\,\delta \,\sigma  + s\,\left( {\delta }^2 + {\sigma }^2 \right)  \right)
            \right)  \r\nn\\
&+&\l s\,\sigma \,\left( 2\,{m_{h}}^2\,\left( {m_{h}}^2 - s \right) \,s\,
         {\left( \delta  + \sigma  \right) }^2 + 8\,{m_{Z}}^6\,
         \left( -4\,s + \delta \,\left( 3\,\delta  + \sigma  \right)  \right)  \r\r\nn\\
&-&\l\l 
        2\,{m_{Z}}^4\,\left( 4\,{m_{h}}^2\,\delta \,\left( 3\,\delta  + \sigma  \right) + s\,\left( -15\,{\delta }^2 + 2\,\delta \,\sigma  - 3\,{\sigma }^2 \right)  \r\r\r\nn\\
&+&\l\l\l 
           2\,\delta \,\sigma \,\left( {\delta }^2 + 6\,\delta \,\sigma  + {\sigma }^2 \right)  \right) +  {m_{Z}}^2\,\left( s\,{\left( \delta  - \sigma  \right) }^2\,
            \left( -2\,s + {\left( \delta  + \sigma  \right) }^2 \right) \r\r\r\nn\\
&+&\l\l\l 4\,{m_{h}}^2\,\left( 4\,s^2 + \delta \,\sigma \,{\left( \delta  + \sigma  \right) }^2 - 
              s\,\left( {\delta }^2 + {\sigma }^2 \right)  \right)  \right)  \right) \r\nn\\
&+&\l 2\,m_{\chi_{k}^{0}}\,\left( 8\,{m_{Z}}^6\,\delta \,\sigma \,\left( -s + \delta \,\sigma  \right) + {m_{Z}}^2\,s^2\,\left( {\delta }^4 + 14\,{\delta }^2\,{\sigma }^2 + {\sigma }^4 - 
           8\,s\,\left( {\delta }^2 + {\sigma }^2 \right)  \right)  \r\r\nn\\
&+&\l\l 
        2\,s^3\,\left( -2\,{\delta }^2\,{\sigma }^2 + s\,\left( {\delta }^2 + {\sigma }^2 \right)  \right)  \r\r\nn\\
&+&\l\l 
        2\,{m_{h}}^4\,\left( 4\,{m_{Z}}^2\,{\delta }^2\,{\sigma }^2 - 
           2\,s\,\delta \,\sigma \,\left( -2\,{m_{Z}}^2 + \delta \,\sigma  \right)  + 
           s^2\,\left( {\delta }^2 + {\sigma }^2 \right)  \right)  \r\r\nn\\
&+&\l\l 
        2\,{m_{Z}}^4\,s\,\left( -16\,s^2 - 2\,\delta \,\sigma \,
            \left( {\delta }^2 - \delta \,\sigma  + {\sigma }^2 \right)  + 
           s\,\left( 7\,{\delta }^2 + 4\,\delta \,\sigma  + 7\,{\sigma }^2 \right)  \right) \r\r\nn\\
&+&\l\l 
        4\,{m_{h}}^2\,\left( {m_{Z}}^2\,s\,\delta \,\sigma \,\left( -2\,s + {\delta }^2 + {\sigma }^2 \right)  \r\r\r\nn\\
&+&\l\l\l 
           4\,{m_{Z}}^4\,\left( s^2 - {\delta }^2\,{\sigma }^2 \right)  - 
           s^2\,\left( -2\,{\delta }^2\,{\sigma }^2 + s\,\left( {\delta }^2 + {\sigma }^2 \right)  \right)  \right)  \right)  \right) \nn\\
&\times&
    \mathcal{F}\[s, \sigma,\delta,m_{h}^2,m_{Z}^2,{m_{\chi_{k}^{0}}}^2\]\, ,\\
G_{Z, Zh}^{C_{Zh,S}^{-}} =&-&2\,s\,\left( 4\,s\,\left( -6\,{m_{Z}}^4 + \left( {m_{h}}^2 - s \right) \,\left( s - {\delta }^2 \right)  + 
       {m_{Z}}^2\,\left( s + {\delta }^2 \right)  \right) \,\sigma  \r\nn\\
&+&\l 
    \left( -4\,{m_{\chi_{k}^{0}}}^2\,s\,\left( 6\,{m_{Z}}^4 - 
          \left( {m_{h}}^2 - s \right) \,\left( s - {\delta }^2 \right)  - {m_{Z}}^2\,\left( s + {\delta }^2 \right) 
          \right) \,\sigma  \r\r\nn\\
&+&\l\l 2\,m_{\chi_{k}^{0}}\,s\,\left( {m_{h}}^4 + 13\,{m_{Z}}^4 - 2\,{m_{Z}}^2\,s + 
          s^2 - 2\,{m_{h}}^2\,\left( {m_{Z}}^2 + s \right)  \right) \,\left( {\delta }^2 - {\sigma }^2 \right)  \r\r\nn\\
&+&\l\l 
       \sigma \,\left( 12\,{m_{Z}}^6\,\delta \,\left( \delta  - \sigma  \right)  + 
          s^2\,{\delta }^2\,\left( -{\delta }^2 + {\sigma }^2 \right) \r\r\r\nn\\
&-&\l\l\l
          {m_{h}}^4\,\left( \delta  + \sigma  \right) \,
           \left( -2\,{\delta }^2\,\sigma  + s\,\left( \delta  + \sigma  \right)  \right)  \r\r\r\nn\\
&+&\l\l\l {m_{Z}}^4\,\left( 2\,{\delta }^2\,\sigma \,\left( \delta  + 13\,\sigma  \right)  + 
             s\,\left( -23\,{\delta }^2 + 2\,\delta \,\sigma  - 7\,{\sigma }^2 \right)  \right)  \r\r\r\nn\\
&+&\l\l\l 
          {m_{Z}}^2\,s\,\left( s\,\left( 5\,{\delta }^2 - 2\,\delta \,\sigma  + {\sigma }^2 \right)  - 
             {\delta }^2\,\left( {\delta }^2 - 2\,\delta \,\sigma  + 5\,{\sigma }^2 \right)  \right)  \r\r\r\nn\\
&+&\l\l\l 
          {m_{h}}^2\,\left( 12\,{m_{Z}}^4\,\left( 2\,s + \delta \,\left( -\delta  + \sigma  \right)  \right) + s\,\left( \delta  + \sigma  \right) \,\left( {\delta }^2\,\left( \delta  - 3\,\sigma  \right)  + 
                s\,\left( \delta  + \sigma  \right)  \right)  \r\r\r\r\nn\\
&+&\l\l\l\l 
             {m_{Z}}^2\,\left( -8\,s^2 - 4\,{\delta }^2\,\sigma \,\left( \delta  + \sigma  \right)  + 
                2\,s\,\left( 3\,{\delta }^2 + {\sigma }^2 \right)  \right)  \right)  \right)  \right) \r\nn\\
&\times&\l
     \mathcal{F}\[s, \sigma,\delta,m_{h}^2,m_{Z}^2,{m_{\chi_{k}^{0}}}^2\] \right),
\end{eqnarray}

where we have used the following coupling functions

\begin{eqnarray}
C_{Zh,S}^{\pm} =&& C_{A}^{\chi_{i}^{0} \chi_{j}^{0} Z} C_{Zh,A}^{\{i j\} k}\pm C_{V}^{\chi_{i}^{0} \chi_{j}^{0} Z} C_{Zh,V}^{\{i j\} k}\, ,\\
C_{Zh,A}^{\pm} =&& C_{A}^{\chi_{i}^{0} \chi_{j}^{0} Z} C_{Zh,A}^{\[i j\] k}\pm C_{V}^{\chi_{i}^{0} \chi_{j}^{0} Z} C_{Zh,V}^{\[i j\] k},
\end{eqnarray}

\begin{eqnarray}
C_{Zh,A}^{\{i j\} k} =&& C_{A}^{\chi_{j}^{0} \chi_{k}^{0} Z} C_{S}^{\chi_{k}^{0} \chi_{i}^{0} r}+C_{A}^{\chi_{k}^{0} \chi_{i}^{0} Z} C_{S}^{\chi_{j}^{0} \chi_{k}^{0} r}\, ,\\
C_{Zh,A}^{\[i j\] k} =&& C_{A}^{\chi_{j}^{0} \chi_{k}^{0} Z} C_{S}^{\chi_{k}^{0} \chi_{i}^{0} r}-C_{A}^{\chi_{k}^{0} \chi_{i}^{0} Z} C_{S}^{\chi_{j}^{0} \chi_{k}^{0} r}\, ,\\
C_{Zh,V}^{\{i j\} k} =&& C_{V}^{\chi_{j}^{0} \chi_{k}^{0} Z} C_{S}^{\chi_{k}^{0} \chi_{i}^{0} r}+C_{V}^{\chi_{k}^{0} \chi_{i}^{0} Z} C_{S}^{\chi_{j}^{0} \chi_{k}^{0} r}\, ,\\
C_{Zh,V}^{\[i j\] k} =&& C_{V}^{\chi_{j}^{0} \chi_{k}^{0} Z} C_{S}^{\chi_{k}^{0} \chi_{i}^{0} r}-C_{V}^{\chi_{k}^{0} \chi_{i}^{0} Z} C_{S}^{\chi_{j}^{0} \chi_{k}^{0} r}.
\end{eqnarray}
\subsection*{\mbox{{\Large$\underline{\bf{\chi^{0}_{i} \chi^{0}_{j}\rightarrow Z A}}$}}}

Contributions to $\tilde{\omega}$ come from s-channel Higgs boson exchange, t- and u-channel neutralino exchange and cross terms

\begin{equation}
\tilde{\omega}_{\chi^{0}_{i} \chi^{0}_{j}\rightarrow Z A}= \tilde{\omega}^{\(h,H\)}_{Z A}+\tilde{\omega}^{\(\chi^{0}\)}_{Z A}+\tilde{\omega}^{\(h, H-\chi^{0}\)}_{Z A}\, :
\end{equation}
\begin{flushleft}
\textbf{S-channel CP-even Higgs boson (h,H):}
\end{flushleft}

\begin{equation}
\tilde{\omega}^{\(h,H\)}_{Z A} = \sum_{r=h,H}\left|\frac{C^{A Z r} C^{\chi_{i} \chi_{j} A}_{P}}{s-m_{r}^{2} + i m_{r} \Gamma_{r}}\right|^{2} \frac{\(s-\sigma^{2}\)\(s^2-2\(m_{Z}^2+m_{A}^2\)s+\(m_{Z}^2-m_{A}^2\)^2\)}{2m_{Z}^2}\, ;
\end{equation}

\begin{flushleft}
\textbf{T- and U-channel neutralino:}
\end{flushleft}

\begin{eqnarray}
\tilde{\omega}^{\(\chi^{0}\)}_{Z A} =&& \frac{1}{2 m_{Z}^2}\sum_{k,l=1}^{4} \[m_{\chi_{k}^{0}}m_{\chi_{l}^{0}}I_{k l}^{Z h} + m_{\chi_{k}^{0}} J_{k l}^{Z h} + K_{k l}^{Z h}\]\, ,
\end{eqnarray}

where

\begin{eqnarray}
I_{k l}^{Z A} =&&\(-D_{ZA,\{ij\}}^{+}\mathcal{T}_{2} + G_{Z A}^{I,T\(1\)} \mathcal{T}_{1}+ G_{Z A}^{I,T\(0\)} \mathcal{T}_{0}- C_{ZA,\{ij\}}^{+}  \mathcal{Y}_{2} \r\nn\\
&+&\l\frac{1}{2s} C_{ZA,\[ij\]}^{+}\sigma\delta\(s+m_{Z}^2-m_{A}^2\) \mathcal{Y}_{1}\r\nn\\
&+&\l G_{Z A}^{I,Y\(0\)} \mathcal{Y}_{0}\)\big(s, \sigma,\delta,m_{A}^2,m_{Z}^2,m_{\chi_{k}^0}^2,m_{\chi_{l}^0}^2\big)\, ,\\
J_{k l}^{Z A} =&& \(D_{ZA,\{ij\}}^{+}\(\sigma+\delta\) \mathcal{T}_{2} + G_{Z A}^{J,T\(1\)} \mathcal{T}_{1}+ G_{Z A}^{J,T\(0\)} \mathcal{T}_{0}+ C_{ZA,\{ij\}}^{+}\(\sigma+\delta\) \mathcal{Y}_{2} \r\nn\\
&+&\l G_{Z A}^{J,Y\(1\)}\mathcal{Y}_{1}+G_{Z A}^{J,Y\(0\)} \mathcal{Y}_{0}\)\big(s, \sigma,\delta,m_{A}^2,m_{Z}^2,m_{\chi_{k}^0}^2,m_{\chi_{l}^0}^2\big)\, ,\\
K_{k l}^{Z A} =&& \( G_{Z A}^{K,T\(2\)} \mathcal{T}_{2} + G_{Z A}^{K,T\(1\)} \mathcal{T}_{1}+ G_{Z A}^{K,T\(0\)} \mathcal{T}_{0}\r\nn\\
&+&\l G_{Z A}^{K,Y\(2\)}  \mathcal{Y}_{2} +G_{Z A}^{K,Y\(1\)} \mathcal{Y}_{1}+G_{Z A}^{K,Y\(0\)} \mathcal{Y}_{0}\)\big(s, \sigma,\delta,m_{A}^2,m_{Z}^2,m_{\chi_{k}^0}^2,m_{\chi_{l}^0}^2\big)\, ,\nn\\
\end{eqnarray}

\begin{eqnarray}
G_{Z A}^{I,T\(1\)} =&& \frac{1}{2s}\(2\,D_{ZA,\[ij\]}^{+}\,\left( {m_{A}}^2 - {m_{Z}}^2 - s \right) \,\delta \,\sigma  \,\r\nn\\
&+&\l 
  D_{ZA,\{ij\}}^{+}\,\left( 2\,{m_{Z}}^2\,\left( s - \delta \,\sigma  \right)  + 
     2\,{m_{A}}^2\,\left( s + \delta \,\sigma  \right)  \r\r\nn\\
&+&\l\l 
     s\,\left( -2\,s + {\delta }^2 + {\sigma }^2 \right)  \right)\)\, ,\\
G_{Z A}^{I,T\(0\)} =&-&\frac{1}{16 s^2}\(4\,\left( 6\,D_{ZA,\{ij\}}^{-}\,{m_{Z}}^2\,s^2\,{\left( \delta  + \sigma  \right) }^2 \r\r\nn\\
&+&\l\l 
     D_{ZA,\[ij\]}^{+}\,\left( {m_{A}}^2 - {m_{Z}}^2 - s \right) \,\delta \,
      \sigma \,\left( 2\,{m_{Z}}^2\,\left( s - \delta \,\sigma  \right)  \r\r\r\nn\\
&+&\l\l\l 
        2\,{m_{A}}^2\,\left( s + \delta \,\sigma  \right)  + 
        s\,\left( -2\,s + {\delta }^2 + {\sigma }^2 \right)  \right)  \right)  \r\nn\\
&+&\l
  D_{ZA,\{ij\}}^{+}\,\left( 8\,{m_{Z}}^4\,\delta \,\sigma \,
      \left( -s + \delta \,\sigma  \right)  + 
     8\,{m_{A}}^4\,\delta \,\sigma \,\left( s + \delta \,\sigma  \right)  \r\r\nn\\
&+&\l\l 
     4\,{m_{A}}^2\,\left( s + \delta \,\sigma  \right) \,
      \left( s\,{\left( \delta  - \sigma  \right) }^2 \r\r\r\nn\\
&+&\l\l\l
        4\,{m_{Z}}^2\,\left( s - \delta \,\sigma  \right)  \right)  + 
     s^2\,\left( {\delta }^4 + 6\,{\delta }^2\,{\sigma }^2 + {\sigma }^4 - 
        4\,s\,\left( {\delta }^2 + {\sigma }^2 \right)  \right)  \r\r\nn\\
&-&\l\l 
     4\,{m_{Z}}^2\,s\,\left( 8\,s^2 + 
        \delta \,{\left( \delta  - \sigma  \right) }^2\,\sigma  - 
        s\,\left( 3\,{\delta }^2 + 2\,\delta \,\sigma  + 3\,{\sigma }^2 \right)  \right)  \right)\)\, ,\\
G_{Z A}^{I,Y\(0\)} =&-&\frac{1}{16 s^2}\(24\,C_{ZA,\{ij\}}^{-}\,{m_{Z}}^2\,s^2\,{\left( \delta  + \sigma  \right) }^2 \r\nn\\
&+&\l 
  C_{ZA,\{ij\}}^{+}\,\left( 8\,{m_{Z}}^4\,\delta \,\sigma \,
      \left( -s + \delta \,\sigma  \right)  + 
     8\,{m_{A}}^4\,\delta \,\sigma \,\left( s + \delta \,\sigma  \right)  \r\r\nn\\
&+&\l\l 
     4\,{m_{A}}^2\,\left( s + \delta \,\sigma  \right) \,
      \left( s\,{\left( \delta  - \sigma  \right) }^2 + 
        4\,{m_{Z}}^2\,\left( s - \delta \,\sigma  \right)  \right)  \r\r\nn\\
&+&\l\l 
     s^2\,\left( {\delta }^4 + 6\,{\delta }^2\,{\sigma }^2 + {\sigma }^4 - 4\,s\,\left( {\delta }^2 + {\sigma }^2 \right)  \right)  \r\r\nn\\
&-&\l\l 4\,{m_{Z}}^2\,s\,\left( 8\,s^2 + 
        \delta \,{\left( \delta  - \sigma  \right) }^2\,\sigma  - 
        s\,\left( 3\,{\delta }^2 + 2\,\delta \,\sigma  + 3\,{\sigma }^2 \right)  \right)  \right)\)\, ,
\end{eqnarray}

\begin{eqnarray}
G_{Z A}^{J,T\(1\)} =&-&\frac{1}{2 s}\(\left( \delta  + \sigma  \right) \,\left( -6\,D_{ZA,\{ij\}}^{-}\,{m_{Z}}^2\,s \r\r\nn\\
&+&\l\l 
    D_{ZA,\{ij\}}^{+}\,\left( 2\,{m_{A}}^2\,\delta \,\sigma  - 
       2\,{m_{Z}}^2\,\left( s + \delta \,\sigma  \right)  + 
       s\,\left( {\delta }^2 + {\sigma }^2 \right)  \right)  \right)\)\, ,\\
G_{Z A}^{J,T\(0\)} =&& \frac{1}{16 s}\(\left( \delta  + \sigma  \right) \,\left( 24\,{m_{Z}}^2\,
     \left( -\left( D_{ZA,\[ij\]}^{-}\,
          \left( {m_{A}}^2 - {m_{Z}}^2 + s \right)  \right)  \r\r\r\nn\\
&+&\l\l\l 
       D_{ZA,\[ij\]}^{+}\,\left( -{m_{A}}^2 + {m_{Z}}^2 + s \right)  \right) \,
     \delta \,\sigma  \r\r\nn\\
&+&\l\l 12\,D_{ZA,\{ij\}}^{-}\,{m_{Z}}^2\,
     \left( 2\,\delta \,\sigma \,\left( {m_{Z}}^2 - 2\,\delta \,\sigma  \right)  - 
       2\,{m_{A}}^2\,\left( 2\,s + \delta \,\sigma  \right)  + 
       s\,\left( {\delta }^2 + {\sigma }^2 \right)  \right)  \r\r\nn\\
&+&\l\l 
    D_{ZA,\{ij\}}^{+}\,\left( {m_{Z}}^4\,\left( -32\,s + 8\,\delta \,\sigma  \right)  + 
       {\left( \delta  + \sigma  \right) }^2\,
        \left( s\,{\left( \delta  - \sigma  \right) }^2 + 4\,{m_{A}}^2\,\delta \,\sigma 
          \right)  \r\r\r\nn\\
&+&\l\l\l 4\,{m_{Z}}^2\,
        \left( s\,\left( {\delta }^2 + {\sigma }^2 \right)  - 
          \delta \,\sigma \,\left( 2\,{m_{A}}^2 + {\delta }^2 + 6\,\delta \,\sigma  + 
             {\sigma }^2 \right)  \right)  \right)  \right)\)\, ,\\
G_{Z A}^{J,Y\(1\)} =&-&\frac{1}{2}\(\left( 3\,C_{ZA,\{ij\}}^{-}\,{m_{Z}}^2 + 
    C_{ZA,\{ij\}}^{+}\,\left( {m_{A}}^2 + 2\,{m_{Z}}^2 - s \right)  \right) \,
  \left( \delta  + \sigma  \right)\)\, ,\\
G_{Z A}^{J,Y\(0\)} =&& \frac{1}{16 s}\(\left( \delta  + \sigma  \right) \,\left( 24\,C_{ZA,\{ij\}}^{-}\,{m_{Z}}^2\,
     \left( s\,\left( -{m_{A}}^2 + {m_{Z}}^2 - s + {\delta }^2 \right)  \r\r\r\nn\\
&+&\l\l\l 
       \left( s - 2\,{\delta }^2 \right) \,{\sigma }^2 \right) + C_{ZA,\{ij\}}^{+}\,\left( 8\,{m_{A}}^4\,\left( s + \delta \,\sigma  \right)  - 
       8\,{m_{Z}}^4\,\left( 2\,s + \delta \,\sigma  \right)  \r\r\r\nn\\
&+&\l\l\l 
       s\,\left( 8\,s^2 + {\left( {\delta }^2 - {\sigma }^2 \right) }^2 - 
          4\,s\,\left( {\delta }^2 + {\sigma }^2 \right)  \right)  \r\r\r\nn\\
&-&\l\l\l 
       4\,{m_{Z}}^2\,\left( 6\,s^2 + 
          \delta \,\sigma \,\left( {\delta }^2 + 6\,\delta \,\sigma  + {\sigma }^2 \right)  - 
          s\,\left( 3\,{\delta }^2 + 2\,\delta \,\sigma  + 3\,{\sigma }^2 \right)  \right)  \r\r\r\nn\\
&+&\l\l\l 
       4\,{m_{A}}^2\,\left( 6\,{m_{Z}}^2\,s - 
          \left( s + \delta \,\sigma  \right) \,
           \left( 4\,s - {\left( \delta  + \sigma  \right) }^2 \right)  \right)  \right)  \right)\)\, ,
\end{eqnarray}

\begin{eqnarray}
G_{Z A}^{K,T\(2\)} =&-&\frac{1}{4 s}\(D_{ZA,\{ij\}}^{+}\,s\,\left( 8\,{m_{Z}}^2 - 4\,s + {\delta }^2 \right)  \r\nn\\
&-&\l 
  6\,D_{ZA,\[ij\]}^{+}\,s\,\delta \,\sigma  + 
  D_{ZA,\{ij\}}^{+}\,\left( s - 4\,{\delta }^2 \right) \,{\sigma }^2\)\, ,\\
G_{Z A}^{K,T\(1\)} =&& \frac{1}{8 s^2}\(2\,s\,\left( -6\,D_{ZA,\{ij\}}^{-}\,{m_{Z}}^2\,s\,{\left( \delta  + \sigma  \right) }^2 \r\r\nn\\
&-&\l\l 
     D_{ZA,\[ij\]}^{+}\,\delta \,\sigma \,
      \left( 8\,{m_{Z}}^4 + {m_{Z}}^2\,
         \left( 6\,s - {\delta }^2 - 6\,\delta \,\sigma  - {\sigma }^2 \right)  \r\r\r\nn\\
&+&\l\l\l 2\,s\,\left( -s + {\delta }^2 + {\sigma }^2 \right)  + 
        {m_{A}}^2\,\left( -8\,{m_{Z}}^2 + 2\,s + {\delta }^2 + 6\,\delta \,\sigma  + 
           {\sigma }^2 \right)  \right)  \right)  \r\nn\\
&+&\l
  D_{ZA,\{ij\}}^{+}\,\left( 16\,{m_{Z}}^4\,s\,\left( s - \delta \,\sigma  \right)  + 
     2\,{m_{A}}^2\,\left( s\,\delta \,{\left( \delta  - \sigma  \right) }^2\,\sigma  - 
        4\,{\delta }^3\,{\sigma }^3\r\r\r\nn\\
&+&\l\l\l 8\,{m_{Z}}^2\,s\,\left( s + \delta \,\sigma  \right) -  s^2\,\left( {\delta }^2 + 4\,\delta \,\sigma  + {\sigma }^2 \right)  \right)\r\r\nn\\
&-&\l\l
     2\,{m_{Z}}^2\,\left( 8\,s^3 - 4\,{\delta }^3\,{\sigma }^3 + 
        s\,\delta \,\sigma \,\left( {\delta }^2 - 10\,\delta \,\sigma  + {\sigma }^2 \right)  - 
        s^2\,\left( {\delta }^2 + 4\,\delta \,\sigma  + {\sigma }^2 \right)  \right)  \r\r\nn\\
&+&\l\l 
     s\,\left( -2\,s^2\,\left( {\delta }^2 + {\sigma }^2 \right)-
        4\,{\delta }^2\,{\sigma }^2\,\left( {\delta }^2 + {\sigma }^2 \right)  + 
        s\,\left( {\delta }^4 + 6\,{\delta }^2\,{\sigma }^2 + {\sigma }^4 \right)  \right)  \right)\)\, ,\nn\\
\\
\nn\\
G_{Z A}^{K,T\(0\)} =&& \frac{1}{64 s^2}\(D_{ZA,\{ij\}}^{+}\,\left( 64\,{m_{Z}}^6\,\delta \,\sigma \,
      \left( s - \delta \,\sigma  \right)  \r\r\nn\\
&+&\l\l 
     8\,{m_{A}}^4\,\delta \,\sigma \,\left( s + \delta \,\sigma  \right) \,
      \left( -8\,{m_{Z}}^2 + {\left( \delta  + \sigma  \right) }^2 \right)  \r\r\nn\\
&-&\l\l 
     s\,{\left( {\delta }^2 - {\sigma }^2 \right) }^2\,
      \left( -4\,{\delta }^2\,{\sigma }^2 + s\,\left( {\delta }^2 + {\sigma }^2 \right)  \right)  \r\r\nn\\
&+&\l\l 8\,{m_{Z}}^4\,\left( 4\,s^2\,{\left( \delta  - \sigma  \right) }^2 + 
        s\,\delta \,\sigma \,\left( {\delta }^2 - 14\,\delta \,\sigma  + {\sigma }^2 \right)  \r\r\r\nn\\
&+&\l\l\l
        {\delta }^2\,{\sigma }^2\,\left( {\delta }^2 + 10\,\delta \,\sigma  + {\sigma }^2 \right) 
        \right)  - 4\,{m_{Z}}^2\,
      \left( s^2\,{\left( \delta  - \sigma  \right) }^2\,
         \left( {\delta }^2 + 4\,\delta \,\sigma  + {\sigma }^2 \right)  \r\r\r\nn\\
&+&\l\l\l
        4\,{\delta }^3\,{\sigma }^3\,\left( {\delta }^2 + 6\,\delta \,\sigma  + {\sigma }^2 \right)  - 
        s\,\delta \,\sigma \,\left( {\delta }^4 + 12\,{\delta }^3\,\sigma  + 
           6\,{\delta }^2\,{\sigma }^2 + 12\,\delta \,{\sigma }^3 + {\sigma }^4 \right)  \right)  \r\r\nn\\
&-&\l\l 
     4\,{m_{A}}^2\,\left( 32\,{m_{Z}}^4\,
         \left( s^2 - {\delta }^2\,{\sigma }^2 \right)-
        {\left( \delta  + \sigma  \right) }^2\,
         \left( 4\,{\delta }^3\,{\sigma }^3 - 
           s\,\delta \,\sigma \,{\left( \delta  + \sigma  \right) }^2\r\r\r\r\nn\\
&+&\l\l\l\l 
           s^2\,\left( {\delta }^2 + {\sigma }^2 \right)  \right) - 4\,{m_{Z}}^2\,\left( s^2\,
            \left( {\delta }^2 + 4\,\delta \,\sigma  + {\sigma }^2 \right)  - 
           s\,\delta \,\sigma \,\left( {\delta }^2 + 6\,\delta \,\sigma  + {\sigma }^2 \right)  \r\r\r\r\nn\\
&-&\l\l\l\l  {\delta }^2\,{\sigma }^2\,\left( {\delta }^2 + 6\,\delta \,\sigma  + {\sigma }^2 \right) \right)  \right)  \right)  \r\nn\\
&+&\l 2\,\delta \,\sigma \,
   \left( 24\,\left( D_{ZA,\[ij\]}^{-} - D_{ZA,\{ij\}}^{-} \right) \,{m_{Z}}^2\,
      \left( -{m_{A}}^2 + {m_{Z}}^2 \right) \,s\,{\left( \delta  + \sigma  \right) }^2
      \r\r\nn\\
&+&\l\l D_{ZA,\[ij\]}^{+}\,\left( 32\,{m_{Z}}^6\,\left( s - \delta \,\sigma  \right)  + s^2\,{\left( {\delta }^2 - {\sigma }^2 \right) }^2 \r\r\r\nn\\
&-&\l\l\l 
        2\,{m_{Z}}^2\,s\,{\left( \delta  - \sigma  \right) }^2\,
         \left( 2\,s + {\delta }^2 + 6\,\delta \,\sigma  + {\sigma }^2 \right)  \r\r\r\nn\\
&+&\l\l\l 4\,{m_{A}}^4\,\left( s + \delta \,\sigma  \right) \,
         \left( -8\,{m_{Z}}^2 + {\left( \delta  + \sigma  \right) }^2 \right)  \r\r\r\nn\\
&-&\l\l\l 
        4\,{m_{Z}}^4\,\left( -\left( \delta \,\sigma \,
              \left( {\delta }^2 + 18\,\delta \,\sigma  + {\sigma }^2 \right)  \right) + s\,\left( 7\,{\delta }^2 + 6\,\delta \,\sigma  + 7\,{\sigma }^2 \right)  \right)  \r\r\r\nn\\
&+&\l\l\l 
        2\,{m_{A}}^2\,\left( 32\,{m_{Z}}^4\,\delta \,\sigma  + 
           s\,{\left( \delta  + \sigma  \right) }^2\,
            \left( -2\,s + {\left( \delta  + \sigma  \right) }^2 \right)  \r\r\r\r\nn\\
&+&\l\l\l\l
           4\,{m_{Z}}^2\,\left( 2\,s^2 - 
              \delta \,\sigma \,\left( {\delta }^2 + 10\,\delta \,\sigma  + {\sigma }^2 \right)  + 
              s\,\left( 3\,{\delta }^2 + 2\,\delta \,\sigma  + 3\,{\sigma }^2 \right)  \right)  \right) 
        \right)  \right)\)\, ,\\
G_{Z A}^{K,Y\(2\)} =&& \frac{1}{4 s}\(-\left( C_{ZA,\{ij\}}^{+}\,s\,{\left( \delta  + \sigma  \right) }^2 \right)  + 
  4\,C_{ZA,\{ij\}}^{-}\,\left( 2\,{m_{Z}}^2\,s - s^2 + {\delta }^2\,{\sigma }^2 \right)\)\, ,\\
G_{Z A}^{K,Y\(1\)} =&&\frac{1}{8 s} \(\delta \,\sigma \,\left( 4\,C_{ZA,\[ij\]}^{-}\,{m_{Z}}^2\,
     \left( 2\,{m_{A}}^2 - 2\,{m_{Z}}^2 + s \right)  \r\r\nn\\
&+&\l\l 
   C_{ZA,\[ij\]}^{+}\,\left( {m_{A}}^2 - {m_{Z}}^2 - s \right) \,
     {\left( \delta  + \sigma  \right) }^2 \right)\)\, ,\\
G_{Z A}^{K,Y\(0\)} =&&\frac{1}{64 s^2}\(-\left( C_{ZA,\{ij\}}^{+}\,{\left( \delta  + \sigma  \right) }^2\,
     \left( 8\,{m_{A}}^4\,\delta \,\sigma \,\left( s + \delta \,\sigma  \right)  \r\r\r\nn\\
&-&\l\l\l 
       8\,{m_{Z}}^4\,\left( 6\,s^2 + s\,\delta \,\sigma  - {\delta }^2\,{\sigma }^2 \right)  +  s^2\,\left( {\delta }^4 + 6\,{\delta }^2\,{\sigma }^2 + {\sigma }^4 - 
          4\,s\,\left( {\delta }^2 + {\sigma }^2 \right)  \right)  \r\r\r\nn\\
&+&\l\l\l 
       4\,{m_{Z}}^2\,s\,\left( 4\,s^2 - 
          \delta \,{\left( \delta  - \sigma  \right) }^2\,\sigma +  s\,\left( 3\,{\delta }^2 + 2\,\delta \,\sigma  + 3\,{\sigma }^2 \right)  \right)  \r\r\r\nn\\
&-&\l\l\l 
       4\,{m_{A}}^2\,\left( -\left( s\,{\left( \delta  - \sigma  \right) }^2\,
             \left( s + \delta \,\sigma  \right)  \right)  + 
          4\,{m_{Z}}^2\,\left( 2\,s^2 + {\delta }^2\,{\sigma }^2 \right)  \right)  \right) 
     \right)  \r\nn\\
&+&\l 4\,C_{ZA,\{ij\}}^{-}\,\left( 16\,{m_{Z}}^6\,\delta \,\sigma \,
      \left( -s + \delta \,\sigma  \right)  + 
     s\,{\left( {\delta }^2 - {\sigma }^2 \right) }^2\,
      \left( -s^2 + {\delta }^2\,{\sigma }^2 \right)  \r\r\nn\\
&-&\l\l 
     4\,{m_{Z}}^4\,\left( -\left( {\delta }^2\,{\left( \delta  - \sigma  \right) }^2\,
           {\sigma }^2 \right)  + 2\,s^2\,
         \left( {\delta }^2 - 3\,\delta \,\sigma  + {\sigma }^2 \right)  \r\r\r\nn\\
&+&\l\l\l 
        2\,s\,\delta \,\sigma \,\left( {\delta }^2 - \delta \,\sigma  + {\sigma }^2 \right)  \right)+ 4\,{m_{Z}}^2\,\left( 2\,s^3\,
         \left( {\delta }^2 - \delta \,\sigma  + {\sigma }^2 \right)  \r\r\r\nn\\
&-&\l\l\l 
        {\delta }^3\,{\sigma }^3\,\left( {\delta }^2 + 6\,\delta \,\sigma  + {\sigma }^2 \right)  + s\,{\delta }^2\,{\sigma }^2\,\left( 7\,{\delta }^2 + 2\,\delta \,\sigma  + 7\,{\sigma }^2
           \right)  \r\r\r\nn\\
&-&\l\l\l s^2\,\left( {\delta }^4 - {\delta }^3\,\sigma  + 10\,{\delta }^2\,{\sigma }^2 - 
           \delta \,{\sigma }^3 + {\sigma }^4 \right)  \right)  \r\r\nn\\
&-&\l\l
     4\,{m_{A}}^4\,\left( -4\,{m_{Z}}^2\,s\,\delta \,\sigma +  s^2\,{\left( \delta  + \sigma  \right) }^2 - 
        {\delta }^2\,{\sigma }^2\,\left( 4\,{m_{Z}}^2 + 
           {\left( \delta  + \sigma  \right) }^2 \right)  \right)  \r\r\nn\\
&+&\l\l
     4\,{m_{A}}^2\,\left( {\left( \delta  + \sigma  \right) }^2\,
         {\left( s - \delta \,\sigma  \right) }^2\,\left( s + \delta \,\sigma  \right)  + 
        8\,{m_{Z}}^4\,\left( s^2 - {\delta }^2\,{\sigma }^2 \right)  \r\r\r\nn\\
&+&\l\l\l
        {m_{Z}}^2\,\left( -8\,s^3 - 
           2\,{\delta }^2\,{\sigma }^2\,\left( {\delta }^2 + {\sigma }^2 \right)  \r\r\r\r\nn\\
&+&\l\l\l\l
           2\,s\,\delta \,\sigma \,\left( {\delta }^2 + \delta \,\sigma  + {\sigma }^2 \right)  + 
           s^2\,\left( 5\,{\delta }^2 - 4\,\delta \,\sigma  + 5\,{\sigma }^2 \right)  \right)  \right) 
     \right)\)\, ,
\end{eqnarray}

where we have used the following coupling functions

\begin{eqnarray}
D_{ZA,\{ij\}}^{+} =&& D_{V P}^{\{i j\} k l} + D_{A P}^{\{i j\} k l}\, ,\\
D_{ZA,\{ij\}}^{-} =&& D_{V P}^{\{i j\} k l} - D_{A P}^{\{i j\} k l}\, ,\\
D_{ZA,\[ij\]}^{+} =&& D_{V P}^{\[i j\] k l} + D_{A P}^{\[i j\] k l}\, ,\\
D_{ZA,\[ij\]}^{-} =&& D_{V P}^{\[i j\] k l} - D_{A P}^{\[i j\] k l}\, ,
\end{eqnarray}

\begin{eqnarray}
C_{ZA,\{ij\}}^{+} =&& C_{V P}^{\{i j\} k l} + C_{A P}^{\{i j\} k l}\, ,\\
C_{ZA,\{ij\}}^{-} =&& C_{V P}^{\{i j\} k l} - C_{A P}^{\{i j\} k l}\, ,\\
C_{ZA,\[ij\]}^{+} =&& C_{V P}^{\[i j\] k l} + C_{A P}^{\[i j\] k l}\, ,\\
C_{ZA,\[ij\]}^{-} =&& C_{V P}^{\[i j\] k l} - C_{A P}^{\[i j\] k l}\, ,
\end{eqnarray}

\begin{eqnarray}
D_{V P}^{\{i j\} k l} =&& C_{P}^{\chi_{k}^{0} \chi_{i}^{0} A} \(C_{P}^{\chi_{l}^{0} \chi_{i}^{0} A}\)^{*} C_{V}^{\chi_{j}^{0} \chi_{k}^{0} Z} \(C_{V}^{\chi_{j}^{0} \chi_{l}^{0} Z}\)^{*}\nn\\
&+&C_{V}^{\chi_{k}^{0} \chi_{i}^{0} Z} \(C_{V}^{\chi_{l}^{0} \chi_{i}^{0} Z}\)^{*} C_{P}^{\chi_{j}^{0} \chi_{k}^{0} A} \(C_{P}^{\chi_{j}^{0} \chi_{l}^{0} A}\)^{*}\, ,\\
D_{V P}^{\[i j\] k l} =&&C_{P}^{\chi_{k}^{0} \chi_{i}^{0} A} \(C_{P}^{\chi_{l}^{0} \chi_{i}^{0} A}\)^{*} C_{V}^{\chi_{j}^{0} \chi_{k}^{0} Z} \(C_{V}^{\chi_{j}^{0} \chi_{l}^{0} Z}\)^{*}\nn\\
&-&C_{V}^{\chi_{k}^{0} \chi_{i}^{0} Z} \(C_{V}^{\chi_{l}^{0} \chi_{i}^{0} Z}\)^{*} C_{P}^{\chi_{j}^{0} \chi_{k}^{0} A} \(C_{P}^{\chi_{j}^{0} \chi_{l}^{0} A}\)^{*}\, ,\\
D_{A P}^{\{i j\} k l} =&&C_{P}^{\chi_{k}^{0} \chi_{i}^{0} A} \(C_{P}^{\chi_{l}^{0} \chi_{i}^{0} A}\)^{*} C_{A}^{\chi_{j}^{0} \chi_{k}^{0} Z} \(C_{A}^{\chi_{j}^{0} \chi_{l}^{0} Z}\)^{*}\nn\\
&+&C_{A}^{\chi_{k}^{0} \chi_{i}^{0} Z} \(C_{A}^{\chi_{l}^{0} \chi_{i}^{0} Z}\)^{*} C_{P}^{\chi_{j}^{0} \chi_{k}^{0} A} \(C_{P}^{\chi_{j}^{0} \chi_{l}^{0} A}\)^{*}\, ,\\
D_{A P}^{\[i j\] k l} =&&C_{P}^{\chi_{k}^{0} \chi_{i}^{0} A} \(C_{P}^{\chi_{l}^{0} \chi_{i}^{0} A}\)^{*} C_{A}^{\chi_{j}^{0} \chi_{k}^{0} Z} \(C_{A}^{\chi_{j}^{0} \chi_{l}^{0} Z}\)^{*}\nn\\
&-&C_{A}^{\chi_{k}^{0} \chi_{i}^{0} Z} \(C_{A}^{\chi_{l}^{0} \chi_{i}^{0} Z}\)^{*} C_{P}^{\chi_{j}^{0} \chi_{k}^{0} A} \(C_{P}^{\chi_{j}^{0} \chi_{l}^{0} A}\)^{*}\, ,\,\,\,\,\,\,\,\,\,\,\,\,\,\,\,\,\,\,\,\,
\end{eqnarray}

\begin{eqnarray}
C_{V P}^{\{i j\} k l} =&& C_{V}^{\chi_{k}^{0} \chi_{i}^{0} Z} \(C_{P}^{\chi_{l}^{0} \chi_{i}^{0} A}\)^{*} C_{P}^{\chi_{j}^{0} \chi_{k}^{0} A} \(C_{V}^{\chi_{j}^{0} \chi_{l}^{0} Z}\)^{*}\nn\\
&+&C_{P}^{\chi_{k}^{0} \chi_{i}^{0} A} \(C_{V}^{\chi_{l}^{0} \chi_{i}^{0} Z}\)^{*} C_{V}^{\chi_{j}^{0} \chi_{k}^{0} Z} \(C_{P}^{\chi_{j}^{0} \chi_{l}^{0} A}\)^{*}\, ,\\
C_{V P}^{\[i j\] k l} =&& C_{V}^{\chi_{k}^{0} \chi_{i}^{0} Z} \(C_{P}^{\chi_{l}^{0} \chi_{i}^{0} A}\)^{*} C_{P}^{\chi_{j}^{0} \chi_{k}^{0} A} \(C_{V}^{\chi_{j}^{0} \chi_{l}^{0} Z}\)^{*}\nn\\
&-&C_{P}^{\chi_{k}^{0} \chi_{i}^{0} A} \(C_{V}^{\chi_{l}^{0} \chi_{i}^{0} Z}\)^{*} C_{V}^{\chi_{j}^{0} \chi_{k}^{0} Z} \(C_{P}^{\chi_{j}^{0} \chi_{l}^{0} A}\)^{*}\, ,\\
C_{A P}^{\{i j\} k l} =&& C_{A}^{\chi_{k}^{0} \chi_{i}^{0} Z} \(C_{P}^{\chi_{l}^{0} \chi_{i}^{0} A}\)^{*} C_{P}^{\chi_{j}^{0} \chi_{k}^{0} A} \(C_{A}^{\chi_{j}^{0} \chi_{l}^{0} Z}\)^{*}\nn\\
&+&C_{P}^{\chi_{k}^{0} \chi_{i}^{0} A} \(C_{A}^{\chi_{l}^{0} \chi_{i}^{0} Z}\)^{*} C_{A}^{\chi_{j}^{0} \chi_{k}^{0} Z} \(C_{P}^{\chi_{j}^{0} \chi_{l}^{0} A}\)^{*}\, ,\\
C_{A P}^{\[i j\] k l} =&& C_{A}^{\chi_{k}^{0} \chi_{i}^{0} Z} \(C_{P}^{\chi_{l}^{0} \chi_{i}^{0} A}\)^{*} C_{P}^{\chi_{j}^{0} \chi_{k}^{0} A} \(C_{A}^{\chi_{j}^{0} \chi_{l}^{0} Z}\)^{*}\nn\\
&-&C_{P}^{\chi_{k}^{0} \chi_{i}^{0} A} \(C_{A}^{\chi_{l}^{0} \chi_{i}^{0} Z}\)^{*} C_{A}^{\chi_{j}^{0} \chi_{k}^{0} Z} \(C_{P}^{\chi_{j}^{0} \chi_{l}^{0} A}\)^{*}\, ;
\end{eqnarray}

\begin{flushleft}
\textbf{Higgs (h, H)-neutralino cross term:}
\end{flushleft}

\begin{eqnarray}
\tilde{\omega}^{\(h, H-\chi^{0}\)}_{Z A} =&& \frac{1}{8 m_{Z}^2 s}\sum_{l=1}^{4}\sum_{r=1}^{2}Re\[\(\frac{C^{Z r A} C_{S}^{\chi_{i}^{0} \chi_{j}^{0} r}}{s-m_{r}^{2} + i m_{r} \Gamma_{r}}\)^{*}\(C_{A}^{i j l} G_{r, Z A}^{C_{A}^{i j l}}+C_{P}^{i j l} G_{r, Z A}^{C_{P}^{i j l}}\)\]\, ,\,\,\,\,\,\,\,\,\,\,\,\,\,\,\,\,\,\,\,\,
\end{eqnarray}

\begin{eqnarray}
G_{r, Z A}^{C_{A}^{i j l}} =&& \delta \,\left( 4\,s\,\left( -{m_{A}}^2 + {m_{Z}}^2 + s \right) \,\sigma  + 
    \left( 4\,{m_{\chi_{l}^{0}}}^2\,s\,\left( -{m_{A}}^2 + {m_{Z}}^2 + s \right) \,
        \sigma  \r\r\nn\\
&-&\l\l 4\,m_{\chi_{l}^{0}}\,\left( {m_{A}}^4 + 
          {\left( {m_{Z}}^2 - s \right) }^2- 2\,{m_{A}}^2\,\left( {m_{Z}}^2 + s \right)  \right) \,
        \left( s - {\sigma }^2 \right)  \r\r\nn\\
&+&\l\l 
       \sigma \,\left( 2\,{m_{A}}^4\,\sigma \,\left( \delta  + \sigma  \right)  + 
          {m_{Z}}^2\,s\,\left( 4\,s - {\delta }^2 + 2\,\delta \,\sigma  - 5\,{\sigma }^2
             \right)  \r\r\r\nn\\
&+&\l\l\l s^2\,\left( -{\delta }^2 + {\sigma }^2 \right)  + 
          2\,{m_{Z}}^4\,\left( -2\,s + \sigma \,\left( \delta  + \sigma  \right)  \right)  \r\r\r\nn\\
&+&\l\l\l 
          {m_{A}}^2\,\left( s\,\left( {\delta }^2 - 2\,\delta \,\sigma  - 3\,{\sigma }^2
                \right)  + 4\,{m_{Z}}^2\,
              \left( s - \sigma \,\left( \delta  + \sigma  \right)  \right)  \right)  \right)  \right)\r\nn\\
&\times&\l \mathcal{F}\[s, \sigma,\delta,m_{h}^2,m_{Z}^2,{m_{\chi_{l}^{0}}}^2\] \right)\, ,\\
G_{r, Z A}^{C_{P}^{i j l}} =&-&4\,s\,\left( {m_{A}}^2\,\left( 2\,s + 
        \left( 2\,m_{\chi_{l}^{0}} - \sigma  \right) \,\sigma  \right)  + 
     s\,\left( -2\,s - 2\,m_{\chi_{l}^{0}}\,\sigma  + {\sigma }^2 \right)  \r\nn\\
&+&\l
     {m_{Z}}^2\,\left( 2\,s - 2\,m_{\chi_{l}^{0}}\,\sigma  + {\sigma }^2 \right)  \right)  - \left( -2\,{m_{Z}}^4\,s\,{\delta }^2 + 2\,{m_{Z}}^2\,s^2\,{\delta }^2 \r\nn\\
&+&\l 
     8\,{m_{\chi_{l}^{0}}}^3\,\left( {m_{A}}^2 - {m_{Z}}^2 - s \right) \,s\,\sigma  + 
     4\,{m_{Z}}^4\,s\,\delta \,\sigma  - 4\,{m_{Z}}^2\,s^2\,\delta \,\sigma  \r\nn\\
&-&\l 
     6\,{m_{Z}}^4\,s\,{\sigma }^2 + 6\,{m_{Z}}^2\,s^2\,{\sigma }^2 + 
     2\,{m_{Z}}^4\,{\delta }^2\,{\sigma }^2 - 
     5\,{m_{Z}}^2\,s\,{\delta }^2\,{\sigma }^2 + s^2\,{\delta }^2\,{\sigma }^2 \r\nn\\
&+&\l 
     2\,{m_{Z}}^4\,\delta \,{\sigma }^3 + 2\,{m_{Z}}^2\,s\,\delta \,{\sigma }^3 - {m_{Z}}^2\,s\,{\sigma }^4 - s^2\,{\sigma }^4 \r\nn\\
&-&\l 
     2\,{m_{A}}^4\,\left( \delta  + \sigma  \right) \,
      \left( -\left( \delta \,{\sigma }^2 \right)  + s\,\left( \delta  + \sigma  \right)  \right)  \r\nn\\
&+&\l 
     4\,{m_{\chi_{l}^{0}}}^2\,s\,\left( {m_{A}}^2\,\left( 2\,s - {\sigma }^2 \right)  + 
        s\,\left( -2\,s + {\sigma }^2 \right)  + {m_{Z}}^2\,\left( 2\,s + {\sigma }^2 \right) 
        \right)  \r\nn\\
&-&\l  2\,m_{\chi_{l}^{0}}\,\left( {m_{A}}^2 - {m_{Z}}^2 - s \right) \,
      \sigma \,\left( 2\,{m_{Z}}^2\,\left( s - \delta \,\sigma  \right)  + 
        2\,{m_{A}}^2\,\left( s + \delta \,\sigma  \right)  \r\r\nn\\
&+&\l\l 
        s\,\left( -2\,s + {\delta }^2 + {\sigma }^2 \right)  \right) + {m_{A}}^2\,\left( s\,\left( \delta  + \sigma  \right) \,
         \left( {\sigma }^2\,\left( -3\,\delta  + \sigma  \right)  + 
           2\,s\,\left( \delta  + \sigma  \right)  \right)  \r\r\nn\\
&-&\l\l 
        4\,{m_{Z}}^2\,\left( 4\,s^2 + 
           \delta \,{\sigma }^2\,\left( \delta  + \sigma  \right)  - 
           s\,\left( {\delta }^2 + 2\,{\sigma }^2 \right)  \right)  \right)  \right) \,
   \mathcal{F}\[s, \sigma,\delta,m_{h}^2,m_{Z}^2,{m_{\chi_{l}^{0}}}^2\]\, ,\nn\\
\end{eqnarray}

where we have used the coupling functions

\begin{eqnarray}
C_{P}^{i j l} =&& \(C_{A}^{\chi_{j}^{0} \chi_{l}^{0} Z}\)^{*} \(C_{P}^{\chi_{l}^{0} \chi_{i}^{0} A}\)^{*}+ \(C_{A}^{\chi_{l}^{0} \chi_{i}^{0} Z}\)^{*} \(C_{P}^{\chi_{j}^{0} \chi_{l}^{0} A}\)^{*}\, ,\\
C_{A}^{i j l} =&&\(C_{A}^{\chi_{j}^{0} \chi_{l}^{0} Z}\)^{*} \(C_{P}^{\chi_{l}^{0} \chi_{i}^{0} A}\)^{*}- \(C_{A}^{\chi_{l}^{0} \chi_{i}^{0} Z}\)^{*} \(C_{P}^{\chi_{j}^{0} \chi_{l}^{0} A}\)^{*}\, .
\end{eqnarray}

\subsection*{\mbox{{\Large$\underline{\bf{\chi^{0}_{i} \chi^{0}_{j}\rightarrow W W}}$}}}

Contributions to $\tilde{\omega}$ come from s-channel Z and Higgs boson exchanges, t- and u-channel chargino exchange and cross terms

\begin{equation}
\tilde{\omega}_{\chi^{0}_{i} \chi^{0}_{j}\rightarrow W W}=\tilde{\omega}^{\(h,H\)}_{W W}+\tilde{\omega}^{\(Z\)}_{W W}+\tilde{\omega}^{\(\chi^{\pm}\)}_{W W}+\tilde{\omega}^{\(h,H-\chi_{k}^{\pm}\)}_{W W}+\tilde{\omega}^{\(Z-\chi^{\pm}\)}_{W W}\, :
\end{equation}

\begin{flushleft}
\textbf{S-channel CP-even Higgs boson (h,H):}
\end{flushleft}

\begin{equation}
\tilde{\omega}^{\(h,H\)}_{W W} = \sum_{r=h,H} \left|\frac{C^{W W r} C^{\chi_{i} \chi_{j} r}_{S}}{s-m_{r}^{2} + i m_{r} \Gamma_{r}}\right|^{2} \frac{\(s-\sigma^{2}\)\(s^2-4s\, m_{W}^2+12 m_{W}^4\)}{8 m_{W}^4}\, ;
\end{equation}

\begin{flushleft}
\textbf{S-channel Z-boson:}
\end{flushleft}

\begin{equation}
\tilde{\omega}^{\(Z\)}_{W W} = \left|\frac{C^{W W Z} C^{\chi_{i} \chi_{j} Z}_{A}}{s-m_{Z}^{2} + i m_{Z} \Gamma_{Z}}\right|^{2} \frac{\(s-\sigma^{2}\)\(2 s+\delta^2\)\(s^3+16 s^2 m_{W}^2-68s\, m_{W}^4-48 m_{W}^6\)}{24 s\,m_{W}^4}\, ;
\end{equation}
\begin{flushleft}
\textbf{T- and U-channel chargino:}
\end{flushleft}

\begin{eqnarray}
\tilde{\omega}^{\(\chi^{\pm}\)}_{W W} =&& \frac{1}{m_{W}^4}\sum_{k,l=1}^{2} \[m_{\chi_{k}^{\pm}}m_{\chi_{l}^{\pm}}I_{k l}^{W W} + m_{\chi_{k}^{\pm}} J_{k l}^{W W} + K_{k l}^{W W}\]\, ,
\end{eqnarray}

where

\begin{eqnarray}
I_{k l}^{W W} =&& \(C_{WW,D}^{-+}\(s-4m_{W}^2\)\mathcal{T}_{2} + G_{W W}^{I,T\(1\)} \mathcal{T}_{1}+ G_{W W}^{I,T\(0\)} \mathcal{T}_{0}+C_{WW,D,2}^{--} \(s-4m_{W}^2\)\mathcal{Y}_{2} \r\nn\\
&+&\l G_{W W}^{I,Y\(0\)} \mathcal{Y}_{0}\)\big(s, \sigma,\delta,m_{W}^2,m_{W}^2,m_{\chi_{k}^\pm}^2,m_{\chi_{l}^\pm}^2\big)\, ,\\
J_{k l}^{W W} =&& \(3 D_{WW,C}^{\delta\sigma+} m_{W}^2 \,\mathcal{T}_{2} + G_{W W}^{J,T\(1\)} \mathcal{T}_{1}+ G_{W W}^{J,T\(0\)} \mathcal{T}_{0}\r\nn\\
&+&\l D_{WW,C,2}^{\delta\sigma -}\(s-m_{W}^2\)\mathcal{Y}_{2} +G_{W W}^{J,Y\(1\)}\mathcal{Y}_{1}\r\nn\\
&+&\l G_{W W}^{J,Y\(0\)} \mathcal{Y}_{0}\)\big(s, \sigma,\delta,m_{W}^2,m_{W}^2,m_{\chi_{k}^\pm}^2,m_{\chi_{l}^\pm}^2\big)\, ,\\
K_{k l}^{W W} =&&  \(-C_{WW,D}^{++} \mathcal{T}_{4}-C_{WW,D}^{++}\(s-\sigma^2-\delta^2-2m_{W}^2\)\mathcal{T}_{3}\r\nn\\
&+&\l G_{W W}^{K,T\(2\)} \mathcal{T}_{2} + G_{W W}^{K,T\(1\)} \mathcal{T}_{1}+ G_{W W}^{K,T\(0\)} \mathcal{T}_{0}+ C_{WW,D,2}^{+-}\,\mathcal{Y}_{4} \r\nn\\
&+&\l G_{W W}^{K,Y\(2\)}  \mathcal{Y}_{2} +G_{W W}^{K,Y\(0\)} \mathcal{Y}_{0}\)\big(s, \sigma,\delta,m_{W}^2,m_{W}^2,m_{\chi_{k}^\pm}^2,m_{\chi_{l}^\pm}^2\big)\, ,
\end{eqnarray}

\begin{eqnarray}
G_{W W}^{I,T\(1\)} =&& \frac{1}{2}C_{WW,D}^{-+}\left( 16\,{m_{W}}^4 - s\,\left( {\delta }^2 + {\sigma }^2 \right)  + 
    4\,{m_{W}}^2\,\left( -2\,s + {\delta }^2 + {\sigma }^2 \right)  \right)\, ,\\
G_{W W}^{I,T\(0\)} =&& \frac{1}{16}\(C_{WW,D}^{-+}\(-64\,{m_{W}}^6 + 64\,{m_{W}}^4\,s - 40\,{m_{W}}^4\,{\delta }^2 + 
  16\,{m_{W}}^2\,s\,{\delta }^2 \r\r\nn\\
&-&\l\l 4\,{m_{W}}^2\,{\delta }^4+ s\,{\delta }^4 - 
  40\,{m_{W}}^4\,{\sigma }^2 + 16\,{m_{W}}^2\,s\,{\sigma }^2 - 
  8\,{m_{W}}^2\,{\delta }^2\,{\sigma }^2 \r\r\nn\\
&-&\l\l 2\,s\,{\delta }^2\,{\sigma }^2 - 4\,{m_{W}}^2\,{\sigma }^4 + 
  s\,{\sigma }^4\)\r\nn\\
&+&\l C_{WW,D}^{--}\(72\,{m_{W}}^4\,\left( {\delta }^2 - {\sigma }^2 \right)\)\)\, ,\\
G_{W W}^{I,Y\(0\)} =&&\frac{1}{16}\(C_{WW,D,2}^{--}\(-64\,{m_{W}}^6 + s\,{\left( {\delta }^2 - {\sigma }^2 \right) }^2 + 
  8\,{m_{W}}^4\,\left( {\delta }^2 + {\sigma }^2 \right)  \r\r\nn\\
&+&\l\l 
  4\,{m_{W}}^2\,\left( 2\,s - {\delta }^2 - {\sigma }^2 \right) \,\left( {\delta }^2 + {\sigma }^2 \right)\r\r\nn\\
&+&\l\l 4 C_{WW,D,2}^{-+}\left( 6\,{m_{W}}^4 + 4\,{m_{W}}^2\,s - s^2 \right) \,\left( {\delta }^2 - {\sigma }^2 \right)\)\)\, ,\,\,\,\,\,\,\,\,\,\,\,\,\,\,\,\,\,\,\,\,
\end{eqnarray}
\begin{eqnarray}
G_{W W}^{J,T\(1\)} =&-&\frac{3}{2} m_{W}^2 \(D_{WW}^{\{-+\}}\delta\(\delta^2-2m_{W}^2\)+C_{WW}^{\{-+\}}\sigma\(\sigma^2-2m_{W}^2\)\)\, ,\\
G_{W W}^{J,T\(0\)} =&& \frac{3 m_{W}^2}{16 s}\(D_{WW}^{\{-+\}}\(-32\,{m_{W}}^4\,s\,\delta  + 4\,{m_{W}}^2\,s\,{\delta }^3 + s\,{\delta }^5 \r\r\nn\\
&+&\l\l 
  12\,{m_{W}}^2\,s\,\delta \,{\sigma }^2 - 8\,{m_{W}}^2\,{\delta }^3\,{\sigma }^2 - s\,\delta \,{\sigma }^4\)\r\nn\\
&+&\l C_{WW}^{\{-+\}}\(-32\,{m_{W}}^4\,s\,\sigma  + 12\,{m_{W}}^2\,s\,{\delta }^2\,\sigma  - s\,{\delta }^4\,\sigma  + 
  4\,{m_{W}}^2\,s\,{\sigma }^3 \r\r\nn\\
&-&\l\l 8\,{m_{W}}^2\,{\delta }^2\,{\sigma }^3 + s\,{\sigma }^5\)\)\, ,\\
G_{W W}^{J,Y\(1\)} =&& \frac{m_{W}^2}{8}\(D_{WW,C,2}^{\delta\sigma-}\(12 m_{W}^2-8s+3\(\delta^2+\sigma^2\)\)\r\nn\\
&+&\l 3 D_{WW,C,2}^{\delta\sigma+}\(\sigma^2-\delta^2\)\)\, ,\\
G_{W W}^{J,Y\(0\)} =&&\frac{1}{16 s}\(-D_{WW,C,2}^{\delta\sigma-}\(16\,{m_{W}}^6\,s - s^2\,{\left( {\delta }^2 - {\sigma }^2 \right) }^2 \r\r\nn\\
&-&\l\l 
  8\,{m_{W}}^4\,\left( 11\,s^2 + {\delta }^2\,{\sigma }^2 - 5\,s\,\left( {\delta }^2 + {\sigma }^2 \right)  \right)
      \r\r\nn\\
&+&\l\l {m_{W}}^2\,s\,\left( 16\,s^2 + {\delta }^4 - 14\,{\delta }^2\,{\sigma }^2 + {\sigma }^4 - 
     2\,s\,\left( {\delta }^2 + {\sigma }^2 \right)  \right)\)\r\nn\\
&+&\l 6 s m_{W}^2 D_{WW,C,2}^{\delta\sigma+}\left( 4\,{m_{W}}^2 - s \right) \,\left( {\delta }^2 - {\sigma }^2 \right)\)\, ,
\end{eqnarray}

\begin{eqnarray}
G_{W W}^{K,T\(2\)} =&&\frac{1}{8 s}\(C_{WW,D}^{++}\(-40\,{m_{W}}^4\,s + 32\,{m_{W}}^2\,s^2 - 12\,{m_{W}}^2\,s\,{\delta }^2 + 4\,s^2\,{\delta }^2 - 
  3\,s\,{\delta }^4 \r\r\nn\\
&-&\l\l 12\,{m_{W}}^2\,s\,{\sigma }^2 + 4\,s^2\,{\sigma }^2 +  8\,{m_{W}}^2\,{\delta }^2\,{\sigma }^2 - 2\,s\,{\delta }^2\,{\sigma }^2 - 3\,s\,{\sigma }^4\)\r\nn\\
&+&\l 8 m_{W}^2 \delta^2\sigma^2\)\, ,\\
G_{W W}^{K,T\(1\)} =&&\frac{1}{16 s} \(C_{WW,D}^{++}\(128\,{m_{W}}^6\,s - s\,\left( s - {\delta }^2 - {\sigma }^2 \right) \,{\left( {\delta }^2 - {\sigma }^2 \right) }^2 \r\r\nn\\
&-&\l\l 
  16\,{m_{W}}^4\,\left( 4\,s^2 - {\delta }^2\,{\sigma }^2 \right) - 2\,{m_{W}}^2\,\left( 8\,s^2\,\left( {\delta }^2 + {\sigma }^2 \right)  + 
     4\,{\delta }^2\,{\sigma }^2\,\left( {\delta }^2 + {\sigma }^2 \right)  \r\r\r\nn\\
&-&\l\l\l 
     s\,\left( 3\,{\delta }^4 + 2\,{\delta }^2\,{\sigma }^2 + 3\,{\sigma }^4 \right)  \right)\)+ 72 m_{W}^4 s \(\delta^2-\sigma^2\)C_{WW,D}^{+-}\)\, ,\\
G_{W W}^{K,T\(0\)} =&-&\frac{1}{256 s^2}\(C_{WW,D}^{++} \(1024\,{m_{W}}^8\,s^2 - 256\,{m_{W}}^6\,s^2\,{\delta }^2 - 48\,{m_{W}}^4\,s^2\,{\delta }^4 \r\r\nn\\
&+&\l\l 
  8\,{m_{W}}^2\,s^2\,{\delta }^6 + s^2\,{\delta }^8- 256\,{m_{W}}^6\,s^2\,{\sigma }^2 \r\r\nn\\
&+&\l\l 
  512\,{m_{W}}^6\,s\,{\delta }^2\,{\sigma }^2 - 416\,{m_{W}}^4\,s^2\,{\delta }^2\,{\sigma }^2 - 
  64\,{m_{W}}^4\,s\,{\delta }^4\,{\sigma }^2 - 8\,{m_{W}}^2\,s^2\,{\delta }^4\,{\sigma }^2 \r\r\nn\\
&-&\l\l 
  16\,{m_{W}}^2\,s\,{\delta }^6\,{\sigma }^2 - 4\,s^2\,{\delta }^6\,{\sigma }^2 - 
  48\,{m_{W}}^4\,s^2\,{\sigma }^4 - 64\,{m_{W}}^4\,s\,{\delta }^2\,{\sigma }^4 \r\r\nn\\
&-&\l\l 
  8\,{m_{W}}^2\,s^2\,{\delta }^2\,{\sigma }^4 + 128\,{m_{W}}^4\,{\delta }^4\,{\sigma }^4 \r\r\nn\\
&+&\l\l 
  32\,{m_{W}}^2\,s\,{\delta }^4\,{\sigma }^4 + 6\,s^2\,{\delta }^4\,{\sigma }^4 \r\r\nn\\
&+&\l\l 
  8\,{m_{W}}^2\,s^2\,{\sigma }^6 - 16\,{m_{W}}^2\,s\,{\delta }^2\,{\sigma }^6 - 
  4\,s^2\,{\delta }^2\,{\sigma }^6 + s^2\,{\sigma }^8\)\)\, ,\\
G_{W W}^{K,Y\(2\)} =&& \frac{1}{8 s} \(C_{WW,D,2}^{+-}\(8\,{m_{W}}^4\,s - 32\,{m_{W}}^2\,s^2 + s\,{\delta }^4 + 8\,{m_{W}}^2\,{\delta }^2\,{\sigma }^2 - 
  2\,s\,{\delta }^2\,{\sigma }^2 + s\,{\sigma }^4\)\r\nn\\
&+&\l C_{WW,D,2}^{++}\(8\,{m_{W}}^2\,s\,{\delta }^2 - 2\,s^2\,{\delta }^2 - 8\,{m_{W}}^2\,s\,{\sigma }^2 + 2\,s^2\,{\sigma }^2\)\)\, ,\\
G_{W W}^{K,Y\(0\)} =&&\frac{1}{256 s}\(-4 C_{WW,D,2}^{++}\left( {\delta }^2 - {\sigma }^2 \right) \,\left( 32\,{m_{W}}^6\,s + 
    s^2\,{\left( {\delta }^2 - {\sigma }^2 \right) }^2 \r\r\nn\\
&-&\l\l 
    8\,{m_{W}}^4\,\left( 6\,s^2 + 10\,{\delta }^2\,{\sigma }^2 - s\,\left( {\delta }^2 + {\sigma }^2 \right)  \right)
        \r\r\nn\\
&+&\l\l 4\,{m_{W}}^2\,s\,\left( -{\delta }^4 + 6\,{\delta }^2\,{\sigma }^2 - {\sigma }^4 + 
       2\,s\,\left( {\delta }^2 + {\sigma }^2 \right)  \right)  \right)+C_{WW,D,2}^{+-}\(s\,{\left( {\delta }^2 - {\sigma }^2 \right) }^4 \r\r\nn\\
&+&\l\l 
  16\,{m_{W}}^2\,{\left( {\delta }^2 - {\sigma }^2 \right) }^2\,\left( -4\,s^2 + {\delta }^2\,{\sigma }^2 \right)  \r\r\nn\\
&-&\l\l 
  256\,{m_{W}}^6\,\left( 8\,s^2 - {\delta }^2\,{\sigma }^2 - 2\,s\,\left( {\delta }^2 + {\sigma }^2 \right)  \right)+ 16\,{m_{W}}^4\,\left( 48\,s^2\,\left( {\delta }^2 + {\sigma }^2 \right)  \r\r\r\nn\\
&+&\l\l\l 
     20\,{\delta }^2\,{\sigma }^2\,\left( {\delta }^2 + {\sigma }^2 \right)  - 
     s\,\left( 9\,{\delta }^4 + 118\,{\delta }^2\,{\sigma }^2 + 9\,{\sigma }^4 \right)  \right)\)\)\, ,
\end{eqnarray}

where we have used the following coupling functions:

\begin{eqnarray}
D_{WW,C}^{\delta\sigma+} =&& \delta D_{WW}^{\{-+\}}+\sigma C_{WW}^{\{-+\}}\, ,\\
D_{WW,C}^{\delta\sigma-} =&& \delta D_{WW}^{\{-+\}}-\sigma C_{WW}^{\{-+\}}\, ,\\
D_{WW,C,2}^{\delta\sigma+} =&& \delta D_{WW,2}^{\{-+\}}+\sigma C_{WW,2}^{\{-+\}}\, ,\\
D_{WW,C,2}^{\delta\sigma-} =&& \delta D_{WW,2}^{\{-+\}}-\sigma C_{WW,2}^{\{-+\}}\, ,
\end{eqnarray}

\begin{eqnarray}
C_{WW,D}^{-\pm} =&& C_{WW}^{-}\pm D_{WW}^{-}\, ,\\
C_{WW,D}^{+\pm} =&& C_{WW}^{+}\pm D_{WW}^{+}\, ,\\
C_{WW,D,2}^{-\pm} =&& C_{WW,2}^{-}\pm D_{WW,2}^{-}\, ,\\
C_{WW,D,2}^{+\pm} =&& C_{WW,2}^{+}\pm D_{WW,2}^{+}\, ,
\end{eqnarray}

\begin{eqnarray}
C_{WW}^{\{-+\}} =&& C_{WW}^{-+}+C_{WW}^{+-}\, ,\\
C_{WW}^{\[-+\]} =&& C_{WW}^{-+}-C_{WW}^{+-}\, ,\\
D_{WW}^{\{-+\}} =&& D_{WW}^{-+}+D_{WW}^{+-}\, ,\\
D_{WW}^{\[-+\]} =&& D_{WW}^{-+}-D_{WW}^{+-}\, ,\\
C_{WW,2}^{\{-+\}} =&& C_{WW,2}^{-+}+C_{WW,2}^{+-}\, ,\\
C_{WW,2}^{\[-+\]} =&& C_{WW,2}^{-+}-C_{WW,2}^{+-}\, ,\\
D_{WW,2}^{\{-+\}} =&& D_{WW,2}^{-+}+D_{WW,2}^{+-}\, ,\\
D_{WW,2}^{\[-+\]} =&& D_{WW,2}^{-+}-D_{WW,2}^{+-}\, ,
\end{eqnarray}

\begin{eqnarray}
C_{WW}^{\pm} =&& Re\(C_{WW,k}^{\pm} \(C_{WW,l}^{\pm}\)^{*}\)\, ,\\
D_{WW}^{\pm} =&& Re\(D_{WW,k}^{\pm} \(D_{WW,l}^{\pm}\)^{*}\)\, ,\\
C_{WW,2}^{\pm} =&& Re\(C_{WW,k}^{\pm} C_{WW,l}^{\pm}\)\, ,\\
D_{WW,2}^{\pm} =&& Re\(D_{WW,k}^{\pm} D_{WW,l}^{\pm}\)\, ,
\end{eqnarray}

\begin{eqnarray}
C_{WW}^{-+} =&& Re\(C_{WW,k}^{-} \(C_{WW,l}^{+}\)^{*}\)\, ,\\
C_{WW}^{+-} =&& Re\(C_{WW,k}^{+} \(C_{WW,l}^{-}\)^{*}\)\, ,\\
D_{WW}^{-+} =&& Re\(D_{WW,k}^{-} \(D_{WW,l}^{+}\)^{*}\)\, ,\\
D_{WW}^{+-} =&& Re\(D_{WW,k}^{+} \(D_{WW,l}^{-}\)^{*}\)\, ,\\
C_{WW,2}^{-+} =&& Re\(C_{WW,k}^{-} C_{WW,l}^{+}\)\, ,\\
C_{WW,2}^{+-} =&& Re\(C_{WW,k}^{+} C_{WW,l}^{-}\)\, ,\\
D_{WW,2}^{-+} =&& Re\(D_{WW,k}^{-} D_{WW,l}^{+}\)\, ,\\
D_{WW,2}^{+-} =&& Re\(D_{WW,k}^{+} D_{WW,l}^{-}\)\, ,
\end{eqnarray}

\begin{eqnarray}
C_{WW,k}^{\pm} =&& C_{V}^{\chi_{k}^{+} \chi_{j}^{0} W^{-}} \(C_{V}^{\chi_{k}^{+} \chi_{i}^{0} W^{-}}\)^{*}\pm C_{A}^{\chi_{k}^{+} \chi_{j}^{0} W^{-}} \(C_{A}^{\chi_{k}^{+} \chi_{i}^{0} W^{-}}\)^{*}\, ,\\
D_{WW,l}^{\pm} =&& C_{A}^{\chi_{l}^{+} \chi_{j}^{0} W^{-}} \(C_{V}^{\chi_{l}^{+} \chi_{i}^{0} W^{-}}\)^{*}\pm C_{V}^{\chi_{l}^{+} \chi_{j}^{0} W^{-}} \(C_{A}^{\chi_{l}^{+} \chi_{i}^{0} W^{-}}\)^{*}\, ;
\end{eqnarray}

\begin{flushleft}
\textbf{Higgs (h, H)-chargino cross term:}
\end{flushleft}

\begin{eqnarray}
\tilde{\omega}^{\(h,H-\chi_{k}^{\pm}\)}_{W W} =&& \sum_{k=1}^{2} Re \[\(\sum_{r=h,H}\frac{C^{W W r} C_{S}^{\chi_{i}^{0} \chi_{j}^{0} r}}{s-m_{r}^{2} + i m_{r} \Gamma_{r}}\)\frac{1}{16  m_{W}^4 s} \r\nn\\
&\times&\l \(C_{WW,k}^{-} G_{r, WW}^{C_{WW,k}^{-}}+C_{WW,k}^{+} G_{r, WW}^{C_{WW,k}^{+}}\)\]\, ,\,\,\,\,\,\,\,\,\,\,\,\,\,\,\,\,\,\,\,\,
\end{eqnarray}

\begin{eqnarray}
G_{r, WW}^{C_{WW,k}^{-}} =&& 4\,m_{\chi_{k}^{+}}\,s\,\left( 4\,\left( 2\,{m_{W}}^2 - s \right) \,s + 
    \left( 4\,{m_{\chi_{k}^{+}}}^2\,\left( 2\,{m_{W}}^2 - s \right) \,s + 
       8\,{m_{W}}^4\,\left( 2\,s - 3\,{\sigma }^2 \right)  \r\r\nn\\
&-&\l\l
       2\,{m_{W}}^2\,s\,\left( {\delta }^2 + 2\,\delta \,\sigma  - 3\,{\sigma }^2 \right)  + 
       s^2\,\left( {\delta }^2 + 2\,\delta \,\sigma  - {\sigma }^2 \right)  \right) \r\nn\\
&\times&\l\mathcal{F}\[s, \sigma,\delta,m_{W}^2,m_{W}^2,{m_{\chi_{k}^{+}}}^2\]
    \right)\, ,\\
G_{r, WW}^{C_{WW,k}^{+}} =&& \sigma \,\left( 4\,s\,\left( 16\,{m_{W}}^4 + 4\,{m_{\chi_{k}^{+}}}^2\,\left( 2\,{m_{W}}^2 + s \right)  + 
       s\,\left( 2\,s - {\left( \delta  + \sigma  \right) }^2 \right)  \r\r\nn\\
&-&\l\l 
       2\,{m_{W}}^2\,\left( 2\,s + {\left( \delta  + \sigma  \right) }^2 \right)  \right) + \left( -64\,{m_{W}}^6\,s + 16\,{m_{\chi_{k}^{+}}}^4\,s\,\left( 2\,{m_{W}}^2 + s \right)  \r\r\nn\\
&+&\l\l 
       8\,{m_{W}}^4\,\left( 4\,s^2 + s\,{\left( \delta  - \sigma  \right) }^2 - 2\,{\delta }^2\,{\sigma }^2 \right)
           \r\r\nn\\
&+&\l\l s^2\,\left( {\delta }^4 - 8\,s\,\delta \,\sigma  + 4\,{\delta }^3\,\sigma  + 2\,{\delta }^2\,{\sigma }^2 + 
          4\,\delta \,{\sigma }^3 + {\sigma }^4 \right)  \r\r\nn\\
&+&\l\l 
       2\,{m_{W}}^2\,s\,\left( {\delta }^4 + 4\,{\delta }^3\,\sigma  + 14\,{\delta }^2\,{\sigma }^2 + 
          4\,\delta \,{\sigma }^3 + {\sigma }^4 \r\r\r\nn\\
&+&\l\l\l s\,\left( -6\,{\delta }^2 + 4\,\delta \,\sigma  - 6\,{\sigma }^2 \right) 
          \right)  + 8\,{m_{\chi_{k}^{+}}}^2\,s\,\left( 4\,{m_{W}}^4 + 
          s\,\left( 2\,s - {\left( \delta  + \sigma  \right) }^2 \right)  \r\r\r\nn\\
&-&\l\l\l 
          2\,{m_{W}}^2\,\left( s + {\left( \delta  + \sigma  \right) }^2 \right)  \right)  \right) \,
     \mathcal{F}\[s, \sigma,\delta,m_{W}^2,m_{W}^2,{m_{\chi_{k}^{+}}}^2\] \right)\, ,
\end{eqnarray}

where we have used the following coupling functions

\begin{eqnarray}
C_{WW,k}^{\pm} =&& C_{V}^{\chi_{k}^{+} \chi_{j}^{0} W^{-}} \(C_{V}^{\chi_{k}^{+} \chi_{i}^{0} W^{-}}\)^{*}\pm C_{A}^{\chi_{k}^{+} \chi_{j}^{0} W^{-}} \(C_{A}^{\chi_{k}^{+} \chi_{i}^{0} W^{-}}\)^{*}\, ;
\end{eqnarray}

\begin{flushleft}
\textbf{Z-chargino cross term:}
\end{flushleft}

\begin{eqnarray}
\tilde{\omega}^{\(Z-\chi^{\pm}\)}_{W W} =&-& \sum_{k=1}^{2} Re \[\(\frac{C^{W W Z} C_{Z}^{\chi_{i}^{0} \chi_{j}^{0} Z}}{s-m_{Z}^{2} + i m_{Z} \Gamma_{Z}}\)\frac{1}{96 m_{W}^4 s} \r\nn\\
&\times&\l \(D_{WW,k}^{-}G_{Z, W W}^{D_{WW,k}^{-}} +D_{WW,k}^{+}G_{Z, W W}^{D_{WW,k}^{+}}\)\]\, ,\,\,\,\,\,\,\,\,\,\,\,\,\,\,\,\,\,\,\,\,
\end{eqnarray}

\begin{eqnarray}
G_{Z, W W}^{D_{WW,k}^{-}} =&& 24\,m_{\chi_{k}^{+}}\,s\,\delta \,\left( 48\,{m_{W}}^4 - 4\,s^2 + 
    \left( -48\,{m_{W}}^6 + {m_{\chi_{k}^{+}}}^2\,\left( 48\,{m_{W}}^4 - 4\,s^2 \right)  \r\r\nn\\
&-&\l\l 
       4\,{m_{W}}^4\,\left( 14\,s + 3\,{\delta }^2 + 6\,\delta \,\sigma- 17\,{\sigma }^2 \right)  + 
       4\,{m_{W}}^2\,s\,\left( 4\,s - 3\,{\sigma }^2 \right)  \r\r\nn\\
&+&\l\l 
       s^2\,\left( {\delta }^2 + 2\,\delta \,\sigma  - {\sigma }^2 \right)  \right) \,\mathcal{F}\[s, \sigma,\delta,m_{W}^2,m_{W}^2,{m_{\chi_{k}^{+}}}^2\]
    \right)\, ,\\
G_{Z, W W}^{D_{WW,k}^{+}} =&-&4\,\left( -192\,{m_{W}}^6\,s + 48\,{m_{\chi_{k}^{+}}}^4\,\left( 2\,{m_{W}}^2 - s \right) \,s \r\nn\\
&+&\l 
     6\,{m_{W}}^2\,s\,\left( 24\,s^2 + {\delta }^4 + 4\,{\delta }^3\,\sigma  - 6\,{\delta }^2\,{\sigma }^2 + 
        4\,\delta \,{\sigma }^3 + {\sigma }^4 \r\r\nn\\
&+&\l\l 4\,s\,\left( 2\,{\delta }^2 - 2\,\delta \,\sigma  - 3\,{\sigma }^2 \right) 
        \right) - s^2\,\left( -8\,s^2 + 3\,{\delta }^4 + 12\,{\delta }^3\,\sigma  + 10\,{\delta }^2\,{\sigma }^2 \r\r\nn\\
&+&\l\l 12\,\delta \,{\sigma }^3 + 3\,{\sigma }^4+ 2\,s\,\left( {\delta }^2 - 6\,\delta \,\sigma  + {\sigma }^2 \right) 
        \right)  \r\nn\\
&-&\l 8\,{m_{W}}^4\,\left( 28\,s^2 + 10\,{\delta }^2\,{\sigma }^2 + 
        s\,\left( -7\,{\delta }^2 + 6\,\delta \,\sigma  + 11\,{\sigma }^2 \right)  \right)  \r\nn\\
&+&\l 
     24\,{m_{\chi_{k}^{+}}}^2\,s\,\left( 4\,{m_{W}}^4 + 
        {m_{W}}^2\,\left( 4\,s - 2\,{\left( \delta  + \sigma  \right) }^2 \right)  + 
        s\,\left( -s + {\left( \delta  + \sigma  \right) }^2 \right)  \right)  \right)  \nn\\
&-& 
  3\left( 256\,{m_{W}}^8 s + 64{m_{\chi_{k}^{+}}}^6\left( 2{m_{W}}^2 - s \right) s \r\nn\\
&+&\l 
     s^2{\left( \delta  + \sigma  \right) }^2\left( {\delta }^4 - 8\,s\,\delta \,\sigma  + 4{\delta }^3\,\sigma  + 
        2{\delta }^2{\sigma }^2 + 4\delta\, {\sigma }^3 + {\sigma }^4 \right)  \r\nn\\
&+&\l 
     32{m_{W}}^6\left( 16s^2 + 2\,{\delta }^2\,{\sigma }^2 + 
        s\left( {\delta }^2 + 6\,\delta \,\sigma  - 5{\sigma }^2 \right)  \right)  \r\nn\\
&-&\l 
     2{m_{W}}^2s\,\left( 64s^2\,\delta \,\sigma  + 
        {\left( {\delta }^2 - {\sigma }^2 \right) }^2\left( {\delta }^2 + 6\,\delta \,\sigma  + {\sigma }^2 \right)  \r\r\nn\\
&+&\l\l 
        4\,s\,\left( 2\,{\delta }^4 - {\delta }^2\,{\sigma }^2 - 10\,\delta \,{\sigma }^3 - 3\,{\sigma }^4 \right)  \right)  -
      16\,{m_{W}}^4\left( -\left( {\delta }^2{\sigma }^2\,{\left( \delta  + \sigma  \right) }^2 \right)  \r\r\nn\\
&+&\l\l 
        5s^2\left( {\delta }^2 - 2\,\delta \,\sigma  + 3\,{\sigma }^2 \right)  +  s\,\left( {\delta }^4 + 2\,{\delta }^3\,\sigma  - 11\,{\delta }^2\,{\sigma }^2 - 4\,\delta \,{\sigma }^3 - 
           2\,{\sigma }^4 \right)  \right)  \r\nn\\
&+&\l 16\,{m_{\chi_{k}^{+}}}^4\,s\,
      \left( 2\,{m_{W}}^2\,\left( 8\,s - 3\,{\left( \delta  + \sigma  \right) }^2 \right)  +  s\,\left( -4\,s + 3\,{\left( \delta  + \sigma  \right) }^2 \right)  \right)  \r\nn\\
&-&\l 
     4\,{m_{\chi_{k}^{+}}}^2\,\left( 96\,{m_{W}}^6\,s + 
        16\,{m_{W}}^4\,\left( 5\,s^2 - s\,{\delta }^2 + 2\,s\,{\sigma }^2 + {\delta }^2\,{\sigma }^2 \right)  \r\r\nn\\
&-&\l\l
        2\,{m_{W}}^2\,s\,\left( 32\,s^2 + 3\,{\delta }^4 + 12\,{\delta }^3\,\sigma  + 2\,{\delta }^2\,{\sigma }^2 + 
           12\,\delta \,{\sigma }^3 + 3\,{\sigma }^4 \r\r\r\nn\\
&-&\l\l\l 4\,s\,\sigma \,\left( 8\,\delta  + 5\,\sigma  \right)  \right)  +  s^2\,\left( 3\,{\delta }^4 + 12\,{\delta }^3\,\sigma  + 14\,{\delta }^2\,{\sigma }^2 + 12\,\delta \,{\sigma }^3 + 
           3\,{\sigma }^4 \r\r\r\nn\\
&-&\l\l\l 4\,s\,\left( {\delta }^2 + 4\,\delta \,\sigma  + {\sigma }^2 \right)  \right)  \right)  \right) \,
  \mathcal{F}\[s, \sigma,\delta,m_{W}^2,m_{W}^2,{m_{\chi_{k}^{+}}}^2\]\, ,
\end{eqnarray}

where we have used the following coupling functions

\begin{eqnarray}
D_{WW,k}^{\pm} =&& C_{A}^{\chi_{k}^{+} \chi_{j}^{0} W^{-}} \(C_{V}^{\chi_{k}^{+} \chi_{i}^{0} W^{-}}\)^{*}\pm C_{V}^{\chi_{k}^{+} \chi_{j}^{0} W^{-}} \(C_{A}^{\chi_{k}^{+} \chi_{i}^{0} W^{-}}\)^{*}\, .
\end{eqnarray}
\subsection*{\mbox{{\Large$\underline{\bf{\chi^{0}_{i} \chi^{0}_{j}\rightarrow Z Z}}$}}}

Contributions to $\tilde{\omega}$ come from s-channel Higgs boson exchange, t- and u-channel neutralino exchange and cross terms

\begin{equation}
\tilde{\omega}_{\chi^{0}_{i} \chi^{0}_{j}\rightarrow Z Z}= \tilde{\omega}^{\(h,H\)}_{Z Z}+\tilde{\omega}^{\(\chi^{0}\)}_{Z Z}+\tilde{\omega}^{\(h,H-\chi_{k}^{0}\)}_{Z Z}\, :
\end{equation}
\begin{flushleft}
\textbf{S-channel CP-even Higgs boson (h,H):}
\end{flushleft}

\begin{equation}
\tilde{\omega}^{\(h,H\)}_{Z Z} = \sum_{r=h,H}\left|\frac{C^{Z Z r} C^{\chi_{i} \chi_{j} r}_{S}}{s-m_{r}^{2} + i m_{r} \Gamma_{r}}\right|^{2} \frac{\(s-\sigma^{2}\)\(s^2-4s\, m_{Z}^2+12 m_{Z}^4\)}{16 m_{Z}^4}\, ;
\end{equation}

\begin{flushleft}
\textbf{T- and U-channel neutralino:}
\end{flushleft} 
\begin{eqnarray}
\tilde{\omega}^{\(\chi^{0}\)}_{Z Z} =&& \frac{1}{2m_{Z}^4}\sum_{k,l=1}^{4} \[m_{\chi_{k}^{0}}m_{\chi_{l}^{0}}I_{k l}^{Z Z} + m_{\chi_{k}^{0}} J_{k l}^{Z Z} + K_{k l}^{Z Z}\]\, ,
\end{eqnarray}

where

\begin{eqnarray}
I_{k l}^{Z Z} =&& \(C_{ZZ,P}^{D,-}\(s-4m_{Z}^2\)\mathcal{T}_{2} + G_{Z Z}^{I,T\(1\)} \mathcal{T}_{1}+ G_{Z Z}^{I,T\(0\)} \mathcal{T}_{0}\r\nn\\
&+&\l C_{ZZ,P}^{D,-} \(s-4m_{Z}^2\)\mathcal{Y}_{2} + G_{Z Z}^{I,Y\(0\)} \mathcal{Y}_{0}\)\big(s, \sigma,\delta,m_{Z}^2,m_{Z}^2,m_{\chi_{k}^0}^2,m_{\chi_{l}^0}^2\big)\, ,\\
J_{k l}^{Z Z} =&& \(3 m_{Z}^2 \(C_{ZZ}^{+-}\sigma +D_{ZZ}^{+-} \delta\)\mathcal{T}_{2} + G_{Z Z}^{J,T\(1\)} \mathcal{T}_{1}+ G_{Z Z}^{J,T\(0\)} \mathcal{T}_{0}\r\nn\\
&-& \l\(C_{ZZ}^{+-} \sigma+D_{ZZ}^{+-} \delta\) \(s-m_{Z}^2\)\mathcal{Y}_{2} +G_{Z Z}^{J,Y\(1\)}\mathcal{Y}_{1}\r\nn\\
&+&\l G_{Z Z}^{J,Y\(0\)} \mathcal{Y}_{0}\)\big(s, \sigma,\delta,m_{Z}^2,m_{Z}^2,m_{\chi_{k}^0}^2,m_{\chi_{l}^0}^2\big)\, ,\\
K_{k l}^{Z Z} =&&  \(-C_{ZZ,P}^{D,+} \mathcal{T}_{4}-C_{ZZ,P}^{D,+}\(s-\sigma^2-\delta^2-2m_{Z}^2\)\mathcal{T}_{3}\r\nn\\
&+&\l G_{Z Z}^{K,T\(2\)} \mathcal{T}_{2} + G_{Z Z}^{K,T\(1\)} \mathcal{T}_{1}+ G_{Z Z}^{K,T\(0\)} \mathcal{T}_{0}\r\nn\\
&+&\l C_{ZZ,M}^{D,+}\,\mathcal{Y}_{4} +G_{Z Z}^{K,Y\(2\)}  \mathcal{Y}_{2} +G_{Z Z}^{K,Y\(0\)} \mathcal{Y}_{0}\)\big(s, \sigma,\delta,m_{Z}^2,m_{Z}^2,m_{\chi_{k}^0}^2,m_{\chi_{l}^0}^2\big)\, ,\nn\\
\end{eqnarray}

\begin{eqnarray}
G_{Z Z}^{I,T\(1\)} =&& \frac{1}{2}\(C_{ZZ,P}^{D,-}\,\left( 16\,{m_{Z}}^4 - s\,\left( {\delta }^2 + {\sigma }^2 \right)  + 
    4\,{m_{Z}}^2\,\left( -2\,s + {\delta }^2 + {\sigma }^2 \right)  \right)\)\, ,\\
G_{Z Z}^{I,T\(0\)} =&& \frac{1}{16}\(C_{ZZ,P}^{D,-}\(-64\,{m_{Z}}^6 + s\,{\left( {\delta }^2 - {\sigma }^2 \right) }^2 + 
  4\,{m_{Z}}^2\,\left( 4\,s - {\delta }^2 - {\sigma }^2 \right) \,\left( {\delta }^2 + {\sigma }^2 \right)  \r\r\nn\\
&+&\l\l 
  8\,{m_{Z}}^4\,\left( 8\,s - 5\,\left( {\delta }^2 + {\sigma }^2 \right)  \right)\)+72\,{m_{Z}}^4\,\left( {\delta }^2 - {\sigma }^2 \right)C_{ZZ,M}^{D,-} \)\, ,\\
G_{Z Z}^{I,Y\(0\)} =&&\frac{1}{16}\(C_{ZZ,P}^{D,-}\(-64\,{m_{Z}}^6 + s\,{\left( {\delta }^2 - {\sigma }^2 \right) }^2 + 
  8\,{m_{Z}}^4\,\left( {\delta }^2 + {\sigma }^2 \right)  \r\r\nn\\
&+&\l\l 
  4\,{m_{Z}}^2\,\left( 2\,s - {\delta }^2 - {\sigma }^2 \right) \,\left( {\delta }^2 + {\sigma }^2 \right)\)\r\nn\\
&+&\l 4\,\left( 6\,{m_{Z}}^4 + 4\,{m_{Z}}^2\,s - s^2 \right) \,\left( {\delta }^2 - {\sigma }^2 \right)C_{ZZ,M}^{D,-}\right)\, ,
\end{eqnarray}

\begin{eqnarray}
G_{Z Z}^{J,T\(1\)} =&-&\frac{3}{2} m_{Z}^2\left( D_{ZZ}^{+-}\,\delta   + 
    C_{ZZ}^{+-}\,\sigma\right)\,\left( -2\,{m_{Z}}^2 + {\sigma }^2 \right)\, ,\\
G_{Z Z}^{J,T\(0\)} =&& \frac{3 m_{Z}^2}{16 s} \,\left( D_{ZZ}^{+-}\,\delta \,
     \left( -32\,{m_{Z}}^4\,s + 
       4\,{m_{Z}}^2\,\left( s\,{\delta }^2 +7\,s\,{\sigma }^2 - 4\,{\delta }^2\,{\sigma }^2 \right)  \r\r\nn\\
&+&\l\l 
       s\,\left( -{\delta }^2 + {\sigma }^2 \right)^2\right)+ C_{ZZ}^{+-}\,\sigma\,\left( -32\,{m_{Z}}^4\,s \r\r\nn\\
&+&\l\l 
       4\,{m_{Z}}^2\,\left(7\, s\,{\delta }^2 +s\,{\sigma }^2 - 4\,{\delta }^2\,{\sigma }^2 \right)  + 
       s\,\left( -{\delta }^2 + {\sigma }^2 \right)^2  \right)  \right) \, ,\\
G_{Z Z}^{J,Y\(1\)} =&-&\frac{1}{2}\(3 m_{Z}^2-2s\)\(D_{ZZ}^{+-} \delta+C_{ZZ}^{+-} \sigma\)\\
G_{Z Z}^{J,Y\(0\)} =&& \frac{1}{16 s}\left(C_{ZZ}^{+-}\,\sigma \,
   \left( 16\,{m_{Z}}^6\,s - 
     s^2\,{\left( {\delta }^2 - {\sigma }^2 \right) }^
       2 \r\r\nn\\
&-&\l\l 8\,{m_{Z}}^4\,
      \left( s\,\left( 11\,s - 9\,{\delta }^2 \right)  - 
        2\,\left( s - {\delta }^2 \right) \,{\sigma }^2
        \right)  \r\r\nn\\
&+&\l\l {m_{Z}}^2\,s\,
      \left( 16\,s^2 + {\delta }^4 - 
        18\,{\delta }^2\,{\sigma }^2 + {\sigma }^4 - 
        4\,s\,\left( \delta  - \sigma  \right) \,
         \left( \delta  + \sigma  \right)  \right) 
     \right)  \r\nn\\
&+&\l D_{ZZ}^{+-}\,\delta \,
   \left( 16\,{m_{Z}}^6\,s - 
     s^2\,{\left( {\delta }^2 - {\sigma }^2 \right) }^
       2 \r\r\nn\\
&-&\l\l 8\,{m_{Z}}^4\,
      \left( 11\,s^2 - 2\,s\,{\delta }^2 - 
        9\,s\,{\sigma }^2 + 2\,{\delta }^2\,{\sigma }^2
        \right)  \r\r\nn\\
&+&\l\l {m_{Z}}^2\,s\,
      \left( 16\,s^2 + {\delta }^4 - 
        18\,{\delta }^2\,{\sigma }^2 + {\sigma }^4 + 
        4\,s\,\left( \delta  - \sigma  \right) \,
         \left( \delta  + \sigma  \right)  \right) 
     \right)\right)\, ,
\end{eqnarray}

\begin{eqnarray}
G_{Z Z}^{K,T\(2\)} =&-&\frac{1}{8 s}C_{ZZ,P}^{D,+}\,\left( 40\,{m_{Z}}^4\,s + 
    s\,\left( 3\,{\delta }^4 + 2\,{\delta }^2\,{\sigma }^2 + 3\,{\sigma }^4 - 
       4\,s\,\left( {\delta }^2 + {\sigma }^2 \right)  \right)  \r\nn\\
&-&\l 
    4\,{m_{Z}}^2\,\left( 8\,s^2 + 4\,{\delta }^2\,{\sigma }^2 - 
       3\,s\,\left( {\delta }^2 + {\sigma }^2 \right)  \right)  \right)\, ,\\
G_{Z Z}^{K,T\(1\)} =&&\frac{1}{16 s} \(C_{ZZ,P}^{D,+} \(128\,{m_{Z}}^6\,s - s\,\left( s - {\delta }^2 - {\sigma }^2 \right) \,
   {\left( {\delta }^2 - {\sigma }^2 \right) }^2 \r\r\nn\\
&-&\l\l 
  32\,{m_{Z}}^4\,\left( 2\,s^2 - {\delta }^2\,{\sigma }^2 \right)-  2\,{m_{Z}}^2\,\left( 8\,s^2\,\left( {\delta }^2 + {\sigma }^2 \right)  +8 \,{\delta }^2\,{\sigma }^2\,\left( {\delta }^2 + {\sigma }^2 \right)  \r\r\r\nn\\
&-&\l\l\l 
     s\,\left( 3\,{\delta }^4 + 10\,{\delta }^2\,{\sigma }^2 + 3\,{\sigma }^4 \right)  \right)\)+ 72\,{m_{Z}}^4\,s\,\left( {\delta }^2 - {\sigma }^2 \right) C_{ZZ,M}^{D,+}\)\, ,\\
G_{Z Z}^{K,T\(0\)} =&-&\frac{1}{256 s^2}C_{ZZ,P}^{D,+}\,\left( 1024\,{m_{Z}}^8\,s^2 + s^2\,{\left( {\delta }^2 - {\sigma }^2 \right) }^4 \r\nn\\
&-&\l 
    256\,{m_{Z}}^6\,s\,\left( -4\,{\delta }^2\,{\sigma }^2 + s\,\left( {\delta }^2 + {\sigma }^2 \right)  \right)
        \r\nn\\
&+&\l 8\,{m_{Z}}^2\,s\,{\left( {\delta }^2 - {\sigma }^2 \right) }^2\,
     \left( -4\,{\delta }^2\,{\sigma }^2 + s\,\left( {\delta }^2 + {\sigma }^2 \right)  \right)  \r\nn\\
&-&\l 
    16\,{m_{Z}}^4\,\left( -16\,{\delta }^4\,{\sigma }^4 + 
       8\,s\,{\delta }^2\,{\sigma }^2\,\left( {\delta }^2 + {\sigma }^2 \right)  \r\r\nn\\
&+&\l\l s^2\,\left( 3\,{\delta }^4 + 26\,{\delta }^2\,{\sigma }^2 + 3\,{\sigma }^4 \right)  \right)  \right)\, ,\\
G_{Z Z}^{K,Y\(2\)} =&& \frac{1}{8 s} \(C_{ZZ,P}^{D,+}\,\left( 8\,{m_{Z}}^4 \, s +s\,\left( {\delta }^2 - {\sigma }^2 \right)^2+m_{Z}^2\(-32 s^2+16 \delta^2\sigma^2\)  \r\r\nn\\
&+&\l\l 
  2\, C_{ZZ,M}^{D,+}\,s\,\left(4m_{Z}^2-s\right)\(\delta^2-\sigma^2\)\)\)\, ,\\
G_{Z Z}^{K,Y\(0\)} =&&\frac{1}{256 s^2}\(C_{ZZ,P}^{D,+}\(s^2\,{\left( {\delta }^2 - {\sigma }^2 \right) }^4 - 
  32\,{m_{Z}}^2\,s\,{\left( {\delta }^2 - {\sigma }^2 \right) }^2\,
   \left( 2\,s^2 - {\delta }^2\,{\sigma }^2 \right)  \r\r\nn\\
&-&\l\l 
  512\,{m_{Z}}^6\,s\,\left( 4\,s^2 - {\delta }^2\,{\sigma }^2 - 
     s\,\left( {\delta }^2 + {\sigma }^2 \right)  \right)  \r\r\nn\\
&+&\l\l 
  16\,{m_{Z}}^4\,\left( 16\,{\delta }^4\,{\sigma }^4 + 
     48\,s^3\,\left( {\delta }^2 + {\sigma }^2 \right)  - s^2\,\left( 9\,{\delta }^4 + 94\,{\delta }^2\,{\sigma }^2 + 9\,{\sigma }^4 \right) 
     \right)\)\r\nn\\
&+&\l C_{ZZ,M}^{D,+}\(-4\,s^2\,\left( {\delta }^2 - {\sigma }^2 \right) \,
  \left( 32\,{m_{Z}}^6 + s\,{\left( {\delta }^2 - {\sigma }^2 \right) }^2 \r\r\r\nn\\
&+&\l\l\l 4\,{m_{Z}}^2\,\left( 2\,s - {\delta }^2 - {\sigma }^2 \right) \,
     \left( {\delta }^2 + {\sigma }^2 \right)  + 
    8\,{m_{Z}}^4\,\left( -6\,s + {\delta }^2 + {\sigma }^2 \right)  \right)\)\)\, ,
\end{eqnarray}

where we have used the following coupling functions

\begin{eqnarray}
C_{ZZ,P}^{D,\pm} =&& C_{ZZ,k}^{\pm} \(C_{ZZ,l}^{\pm}\)^{*} + D_{ZZ,k}^{\pm} \(D_{ZZ, \, l}^{\pm}\)^{*}\, ,\\
C_{ZZ,M}^{D,\pm} =&& C_{ZZ,k}^{\pm} \(C_{ZZ,l}^{\pm}\)^{*} - D_{ZZ,k}^{\pm} \(D_{ZZ, \, l}^{\pm}\)^{*}\, ,
\end{eqnarray}

\begin{eqnarray}
C_{ZZ}^{+-} =&& C_{ZZ,k}^{+} \(C_{ZZ,l}^{-}\)^{*} + C_{ZZ,k}^{-} \(C_{ZZ,l}^{+}\)^{*}\, ,\\
C_{ZZ}^{-+} =&& C_{ZZ,k}^{+} \(C_{ZZ,l}^{-}\)^{*} - C_{ZZ,k}^{-} \(C_{ZZ,l}^{+}\)^{*}\, ,\\
D_{ZZ}^{+-} =&& D_{ZZ,k}^{+} \(D_{ZZ, \, l}^{-}\)^{*} + D_{ZZ,k}^{-} \(D_{ZZ, \, l}^{+}\)^{*}\, ,\\
D_{ZZ}^{-+} =&& D_{ZZ,k}^{+} \(D_{ZZ, \, l}^{-}\)^{*} - D_{ZZ,k}^{-} \(D_{ZZ, \, l}^{+}\)^{*}\, ,
\end{eqnarray}

\begin{eqnarray}
C_{ZZ,k}^{\pm} =&& C_{V}^{\chi_{k}^{0} \chi_{j}^{0} Z} \(C_{V}^{\chi_{k}^{0} \chi_{i}^{0} Z}\)^{*}\pm C_{A}^{\chi_{k}^{0} \chi_{j}^{0} Z} \(C_{A}^{\chi_{k}^{0} \chi_{i}^{0} Z}\)^{*}\, ,\\
C_{ZZ,l}^{\pm} =&& C_{V}^{\chi_{l}^{0} \chi_{j}^{0} Z} \(C_{V}^{\chi_{l}^{0} \chi_{i}^{0} Z}\)^{*}\pm C_{A}^{\chi_{l}^{0} \chi_{j}^{0} Z} \(C_{A}^{\chi_{l}^{0} \chi_{i}^{0} Z}\)^{*}\, ,\\
D_{ZZ,k}^{\pm} =&& C_{A}^{\chi_{k}^{0} \chi_{j}^{0} Z} \(C_{V}^{\chi_{k}^{0} \chi_{i}^{0} Z}\)^{*}\pm C_{V}^{\chi_{k}^{0} \chi_{j}^{0} Z} \(C_{A}^{\chi_{k}^{0} \chi_{i}^{0} Z}\)^{*}\, ,\\
D_{ZZ, \, l}^{\pm} =&& C_{A}^{\chi_{l}^{0} \chi_{j}^{0} Z} \(C_{V}^{\chi_{l}^{0} \chi_{i}^{0} Z}\)^{*}\pm C_{V}^{\chi_{l}^{0} \chi_{j}^{0} Z} \(C_{A}^{\chi_{l}^{0} \chi_{i}^{0} Z}\)^{*}\, ;
\end{eqnarray}
\\
\begin{flushleft}
\textbf{Higgs (h, H)-neutralino cross term:}
\end{flushleft}

\begin{eqnarray}
\tilde{\omega}^{\(h,H-\chi_{k}^{0}\)}_{Z Z} =&-&\sum_{r=h,H}\sum_{k=1}^{4} Re \[\(\sum_{r=h,H}\frac{C^{Z Z r} C_{S}^{\chi_{i}^{0} \chi_{j}^{0} r}}{s-m_{r}^{2} + i m_{r} \Gamma_{r}}\)\frac{1}{32 m_{Z}^4 s}\r\nn\\
&\times&\l \(C_{ZZ,k}^{-} G_{r,ZZ}^{C_{ZZ,k}^{-}}+C_{ZZ,k}^{+} G_{r,ZZ}^{C_{ZZ,k}^{+}}\)\]\, ,
\end{eqnarray}

\begin{eqnarray}
G_{r,ZZ}^{C_{ZZ,k}^{-}} =&& 4\,m_{\chi_{k}^{0}}\,s\,\left( 4\,s\,\left( -2\,{m_{Z}}^2 + s \right)  + 
    \left( 4\,{m_{\chi_{k}^{0}}}^2\,s\,\left( -2\,{m_{Z}}^2 + s \right)  - 
       8\,{m_{Z}}^4\,\left( 2\,s - 3\,{\sigma }^2 \right)  \r\r\nn\\
&+&\l\l 
       2\,{m_{Z}}^2\,s\,\left( {\delta }^2 - 3\,{\sigma }^2 \right)  + s^2\,\left( -{\delta }^2 + {\sigma }^2 \right)  \right) \,
     \mathcal{F}\[s, \sigma,\delta,m_{Z}^2,m_{Z}^2,{m_{\chi_{k}^{0}}}^2\] \right)\, ,\\
G_{r,ZZ}^{C_{ZZ,k}^{+}} =&-&\sigma \,\left( 4\,s\,\left( 16\,{m_{Z}}^4 + 
         4\,{m_{\chi_{k}^{0}}}^2\,\left( 2\,{m_{Z}}^2 + s \right)  \r\r\nn\\
&+&\l\l 
         s\,\left( 2\,s - {\delta }^2 - {\sigma }^2 \right)  - 
         2\,{m_{Z}}^2\,\left( 2\,s + {\delta }^2 + {\sigma }^2 \right)  \right)  \r\nn\\
&+&\l 
      \left( -64\,{m_{Z}}^6\,s + 
         16\,{m_{\chi_{k}^{0}}}^4\,s\,\left( 2\,{m_{Z}}^2 + s \right)  + 
         s^2\,{\left( {\delta }^2 - {\sigma }^2 \right) }^2 \r\r\nn\\
&+&\l\l 
         8\,{m_{\chi_{k}^{0}}}^2\,s\,\left( 4\,{m_{Z}}^4 + 
            s\,\left( 2\,s - {\delta }^2 - {\sigma }^2 \right)- 2\,{m_{Z}}^2\,\left( s + {\delta }^2 + {\sigma }^2 \right)  \right)  \r\r\nn\\
&+&\l\l 
         8\,{m_{Z}}^4\,\left( 4\,s^2 - 4\,{\delta }^2\,{\sigma }^2 + 
            s\,\left( 3\,{\delta }^2 + {\sigma }^2 \right)  \right)  \r\r\nn\\
&+&\l\l
         2\,{m_{Z}}^2\,s\,\left( {\delta }^4 + 6\,{\delta }^2\,{\sigma }^2 + 
            {\sigma }^4 - 2\,s\,\left( {\delta }^2 + 3\,{\sigma }^2 \right)  \right)  \right)
         \r\nn\\
&\times&\l\mathcal{F}\[s, \sigma,\delta,m_{Z}^2,m_{Z}^2,{m_{\chi_{k}^{0}}}^2\] \right)\, ,
\end{eqnarray}

where we have used the following coupling function

\begin{eqnarray}
C_{ZZ,k}^{\pm} =&& C_{V}^{\chi_{k}^{0} \chi_{j}^{0} Z} \(C_{V}^{\chi_{k}^{0} \chi_{i}^{0} Z}\)^{*}\pm C_{A}^{\chi_{k}^{0} \chi_{j}^{0} Z} \(C_{A}^{\chi_{k}^{0} \chi_{i}^{0} Z}\)^{*}\, .
\end{eqnarray}

\subsection*{\mbox{{\Large$\underline{\bf{\chi^{0}_{i} \chi^{0}_{j}\rightarrow f \bar{f}}}$}}}

Contributions to $\tilde{\omega}$ come from s-channel Z and Higgs boson (both CP-even and CP-odd) exchanges, t- and u-channel sfermion exchange and cross terms

\begin{equation}
\tilde{\omega}_{\chi^{0}_{i} \chi^{0}_{j}\rightarrow f \bar{f}}=\tilde{\omega}^{\(h,H\)}_{f \bar{f}}+\tilde{\omega}^{\(A\)}_{f \bar{f}}+\tilde{\omega}^{\(Z\)}_{f \bar{f}}+\tilde{\omega}^{\(\tilde{f}\)}_{f \bar{f}}+\tilde{\omega}^{\(h,H-\tilde{f}\)}_{f \bar{f}}+\tilde{\omega}^{\(A-Z\)}_{f \bar{f}}+\tilde{\omega}^{\(A-\tilde{f}\)}_{f \bar{f}}+\tilde{\omega}^{\(Z-\tilde{f}\)}_{f \bar{f}}\, :
\end{equation}

\begin{flushleft}
\textbf{S-channel CP-even Higgs boson (h,H):}
\end{flushleft}

\begin{equation}
\tilde{\omega}^{\(h,H\)}_{f \bar{f}} = \sum_{r=h,H}\left|\frac{C^{f f r}_{S} C^{\chi_{i} \chi_{j} r}_{S}}{s-m_{r}^{2} + i m_{r} \Gamma_{r}}\right|^{2} \(s-\sigma^{2}\)\(s-4 m_{f}^2\)\, ;
\end{equation}
\\
\\
\begin{flushleft}
\textbf{S-channel CP-odd Higgs boson A:}
\end{flushleft}

\begin{equation}
\tilde{\omega}^{\(A\)}_{f \bar{f}} = \left|\frac{C^{f f A}_{P} C^{\chi_{i} \chi_{j} A}_{P}}{s-m_{A}^{2} + i m_{A} \Gamma_{A}}\right|^{2} s\(s-\delta^{2}\)\, ;
\end{equation}

\begin{flushleft}
\textbf{S-channel Z boson:}
\end{flushleft}

\begin{eqnarray}
\tilde{\omega}^{\(Z\)}_{f \bar{f}} =&& \left|\frac{1}{s-m_{Z}^{2} + i m_{Z} \Gamma_{Z}}\right|^{2} \frac{2}{3 m_{Z}^4}\nn\\
&\times&\(2 \left|C^{\chi_{i} \chi_{j} Z}_{V}\right|^2 m_{Z}^4 \(\left|C^{f f Z}_{A}\right|^2 \(s-4 m_{f}^2\)+\left|C^{f f Z}_{V}\right|^2 \(s+2 m_{f}^2\)\)\(2s + \sigma^2\)\r\nn\\
&+&\l 2\left|C^{\chi_{i} \chi_{j} Z}_{A}\right|^2 \(\left|C^{f f Z}_{V}\right|^2 m_{Z}^4 \(s+2 m_{f}^2\)\(s-\sigma^2\)\r\r\nn\\
&+&\l\l\left|C^{f f Z}_{A}\right|^2 \(s\, m_{Z}^4 \(s-\sigma^2\)+m_{f}^2 \(3 s^2 \sigma^2 -6 s\, m_{Z}^2 \sigma^2 - m_{Z}^4 \(4 s - 7 \sigma^2\)\)\)\)\r\nn\\
&-&\l\delta\(\left|C^{\chi_{i} \chi_{j} Z}_{A}\right|^2\(\left|C^{f f Z}_{V}\right|^2 m_{Z}^4 \(s+2 m_{f}^2\)\(5 \sigma - 2 \delta\)\r\r\r\nn\\
&+&\l\l\l\left|C^{f f Z}_{A}\right|^2 \(s \, m_{Z}^4 \(5 \sigma - 2 \delta\)+2 m_{f}^2\(m_{Z}^4\(4 \delta - 13 \sigma\)+ 6 s\, m_{Z}^2 \sigma - 3 s^2 \sigma\)\)\)\r\r\nn\\
&+&\l\l\left|C^{\chi_{i} \chi_{j} Z}_{V}\right|^2\(\left|C^{f f Z}_{V}\right|^2 m_{Z}^4 \(s+2 m_{f}^2\)\(5 \sigma + 4 \delta\)+\left|C^{f f Z}_{A}\right|^2 \(s\, m_{Z}^4 \(5 \sigma+4 \delta\)\r\r\r\r\nn\\
&-&\l\l\l\l 2 m_{f}^2 \(\(3 s^2 - 6 s\, m_{Z}^2+7 m_{Z}^4\)\(\sigma+2\delta\)+6 m_{Z}^4 \sigma\)\)\)\)\)\, ;
\end{eqnarray}

\begin{flushleft}
\textbf{T- and U-channel sfermion:}
\end{flushleft}

\begin{eqnarray}
\tilde{\omega}^{\(\tilde{f}\)}_{f \bar{f}} =&& \frac{1}{8}\sum_{a,b} \[\(C_{+\left\{ij\right\}}^{a b}+D_{-\left\{ij\right\}}^{a b}\)\mathcal{T}_{2}\r\nn\\
&-&\l\frac{1}{2}\(\(C_{+\left\{ij\right\}}^{a b}+D_{-\left\{ij\right\}}^{a b}\)\(4m_{f}^2+\sigma^2+\delta^2\)+4 \tilde{C}_{+\left\{ij\right\}}^{a b} m_{f} \sigma-4 \tilde{D}_{-\left[ij\right]}^{a b} m_{f} \delta\)\mathcal{T}_{1}\r\nn\\
&+&\l\frac{1}{16} G_{\bar{f} f}^{T \(0\)} \mathcal{T}_{0}+ B_{-\left\{ij\right\}}^{a b} \mathcal{Y}_{2}+\frac{1}{2} m_{f} \left(\tilde{B}_{-\left[ij\right]}^{a b} \delta - \tilde{A}_{-\left\{ij\right\}}^{a b} \sigma \right)\mathcal{Y}_{1}\r\nn\\
&+&\l\frac{1}{16} G_{\bar{f} f}^{Y \(0\)}\mathcal{Y}_{0}\]\big(s, \sigma,\delta,m_{f}^2,m_{f}^2,m_{\tilde{f}_{a}}^2,m_{\tilde{f}_{b}}^2\big)\, ,\nn\\
\end{eqnarray}

\begin{eqnarray}
G_{\bar{f} f}^{T \(0\)} =&-&8\,m_{f}\,\left( 2\,C_{-\left\{ij\right\}}^{a b}\,m_{f}\,\left( {\delta }^2 - {\sigma }^2 \right)  + 
     2\,D_{+\left\{ij\right\}}^{a b}\,m_{f}\,\left( {\delta }^2 - {\sigma }^2 \right)  \r\nn\\
&-&\l \tilde{C}_{+\left\{ij\right\}}^{a b}\,\sigma \,\left( 4\,{m_{f}}^2 - {\delta }^2 + {\sigma }^2 \right)  + \tilde{D}_{+\left[ij\right]}^{a b}\,\delta \,\left( 4\,{m_{f}}^2 + {\delta }^2 - {\sigma }^2 \right)\right)  \nn\\
&+& C_{+\left\{ij\right\}}^{a b}\,\left( 16\,{m_{f}}^4 + {\left( {\delta }^2 - {\sigma }^2 \right) }^2 + 
     8\,{m_{f}}^2\,\left( {\delta }^2 + {\sigma }^2 \right)  \right)  \nn\\
&+& 
  D_{-\left\{ij\right\}}^{a b}\,\left( 16\,{m_{f}}^4 + {\left( {\delta }^2 - {\sigma }^2 \right) }^2 + 
     8\,{m_{f}}^2\,\left( {\delta }^2 + {\sigma }^2 \right)  \right)\, ,\\
G_{\bar{f} f}^{Y \(0\)} =&& B_{-\left\{ij\right\}}^{a b} \(16 m_{f}^4+\(\sigma^2-\delta^2\)^2\)\nn\\
&+&4\(2 \, A_{-\left\{ij\right\}}^{a b} m_{f}^2 \(2s-\sigma^2-\delta^2\)+A_{+\left\{ij\right\}}^{a b} \(s-2m_{f}^2\)\(\sigma^2-\delta^2\)\r\nn\\
&+&\l2m_{f} \(\tilde{A}_{+\left\{ij\right\}}^{a b}\sigma\(s-\delta^2\)-\tilde{B}_{+\left[ij\right]}^{a b}\delta\(s-\sigma^2\)+B_{+\left\{ij\right\}}^{a b} m_{f}\(\sigma^2-\delta^2\)\)\)\, ;\nn\\
\end{eqnarray}

\begin{flushleft}
\textbf{Higgs (h, H)-sfermion cross term:}
\end{flushleft}

\begin{eqnarray}
\tilde{\omega}^{\(h,H-\tilde{f}\)}_{f \bar{f}} =&& \frac{1}{8}\sum_{b} Re \[\sum_{r=h,H}\frac{C_{S}^{f f r} C_{S}^{\chi_{i}^{0} \chi_{j}^{0} r}}{s-m_{r}^{2} + i m_{r} \Gamma_{r}}\(2\,m_{f}\,\sigma\, C_{+\left\{ij\right\}}^{b}\left( 4 + \left( 4{m_{\tilde{f}_{b}}}^2 - 4{m_{f}}^2 \r\r\r\r\nn\\
&+&\l\l\l\l 2\,s - {\delta }^2 - {\sigma }^2 \right) \mathcal{F}\[s, \sigma,\delta,m_{f}^2,m_{f}^2,{m_{\tilde{f}_{b}}}^2\]\)\r\r\nn\\
&+&\l\l D_{+\left\{ij\right\}}^{b}\(4\,s + \left( 4\,{m_{f}}^2\,\left( s - 2{\sigma }^2 \right)  + 
     s\,\left( 4\,{m_{\tilde{f}_{b}}}^2- {\delta }^2 + {\sigma }^2 \right)
         \right) \r\r\r\nn\\
&\times&\l\l\l\mathcal{F}\[s, \sigma,\delta,m_{f}^2,m_{f}^2,{m_{\tilde{f}_{b}}}^2\]\)\)\]\, ;\nn\\
\end{eqnarray}

\begin{flushleft}
\textbf{Higgs (A)-Z cross term:}
\end{flushleft}

\begin{equation}
\tilde{\omega}^{\(A-Z\)}_{f \bar{f}} = Re\[ \(\frac{C_{P}^{f f A} C_{P}^{\chi_{i}^{0} \chi_{j}^{0} A}}{s-m_{A}^{2} + i m_{A} \Gamma_{A}}\)^{*}\(\frac{C_{A}^{f f Z} C_{A}^{\chi_{i}^{0} \chi_{j}^{0} Z}}{s-m_{Z}^{2} + i m_{Z} \Gamma_{Z}}\)\] \frac{4\sigma m_{f}}{m_{Z}^2}\(s-\delta^2\)\(m_{Z}^2-s\)\, ;
\end{equation}

\begin{flushleft}
\textbf{Higgs (A)-sfermion cross term:}
\end{flushleft}

\begin{eqnarray}
\tilde{\omega}^{\(A-\tilde{f}\)}_{f \bar{f}} =&-&\frac{1}{8}\sum_{a} Re \[\frac{C_{P}^{f f A} C_{P}^{\chi_{i}^{0} \chi_{j}^{0} A}}{s-m_{A}^{2} + i m_{A} \Gamma_{A}}\r\nn\\
&\times&\l\(C_{+\left\{ij\right\}}^{a}\(4\,m_{f}\,\left( s - {\delta }^2 \right) \,\sigma \,
  \mathcal{F}\[s, \sigma,\delta,m_{f}^2,m_{f}^2,{m_{\tilde{f}_{a}}}^2\]\)\r\r\nn\\
&+&\l\l D_{+\left\{ij\right\}}^{a}\(s\,\left( -4 + \left( -4\,{m_{\tilde{f}_{a}}}^2 + 4\,{m_{f}}^2 - {\delta }^2 + 
       {\sigma }^2 \right) \r\r\r\r\nn\\
&\times&\l\l\l\l\mathcal{F}\[s, \sigma,\delta,m_{f}^2,m_{f}^2,{m_{\tilde{f}_{a}}}^2\] \right)\)\)\]\, ;
\end{eqnarray}
\\
\\
\begin{flushleft}
\textbf{Z-sfermion cross term:}
\end{flushleft}

\begin{equation}
\tilde{\omega}^{\(Z-\tilde{f}\)}_{f \bar{f}} = \frac{1}{4m_{Z}^2}\sum_{a} Re \[\frac{1}{s-m_{Z}^{2} + i m_{Z} \Gamma_{Z}}\(C_{A}^{f f Z} C_{A}^{\chi_{i}^{0} \chi_{j}^{0} Z} Z_{a}^{\(+\) \bar{f} f}+C_{V}^{f f Z} C_{A}^{\chi_{i}^{0} \chi_{j}^{0} Z} Z_{a}^{\(-\) \bar{f} f}\)\]\, ,
\end{equation}

where

\begin{eqnarray}
Z_{a}^{\(+\) \bar{f} f} =&-&4\,D_{+\{ij\}}^{a}\,m_{f}\,\left( -3\,{m_{Z}}^2 + s \right) \,\sigma  +
  C_{+\{ij\}}^{a}\,{m_{Z}}^2\,\left( -4\,{m_{\tilde{f}_{a}}}^2 + 4\,{m_{f}}^2 + 
     2\,s + {\delta }^2 + {\sigma }^2 \right)\nn\\
&-&\frac{1}{4}\(4D_{+\{ij\}}^{a}\,m_{f}\,\sigma\(4\,m_{f}^2\left( 3\,{m_{Z}}^2 - s \right) -4m_{Z}^2\, s +4{m_{\tilde{f}_{a}}}^2\(-3m_{Z}^2+s\)\r\r\nn\\
&+&\l\l \delta^2\,m_{Z}^2+\delta^2 \,s+3m_{Z}^2 \sigma-\sigma^2\,s\)\r\nn\\
&-&\l C_{+\{ij\}}^{a}\(16m_{f}^4\,m_{Z}^2-8m_{f}^2\(4{m_{\tilde{f}_{a}}}^2\,m_{Z}^2+2\sigma^2\(s-\delta^2\)+m_{Z}^2\(2s+\delta^2-5\sigma^2\)\)\r\r\nn\\
&+&\l\l m_{Z}^2\(16 {m_{\tilde{f}_{a}}}^4+\(\delta^2-\sigma^2\)\(4s-\delta^2-\sigma^2\)-8{m_{\tilde{f}_{a}}}^2\(\delta^2+\sigma^2\)\)\)\right)\nn\\
&\times&\mathcal{F}\[s, \sigma,\delta,m_{f}^2,m_{f}^2,m_{\tilde{f}_{a}}^2\]\, ,\\
Z_{a}^{\(-\) \bar{f} f} =&& 4\, D_{-\[ij\]}^{a} m_{f} \delta+ C_{-\{ij\}}^{a}\,{m_{Z}}^2\,\left( -4\,{m_{\tilde{f}_{a}}}^2 + 4\,{m_{f}}^2 + 
    2\,s + {\delta }^2 + {\sigma }^2 \right) \nn\\
&-&\frac{1}{4} \(C_{-\{ij\}}^{a}\,{m_{Z}}^2\,\left( 16\,{m_{\tilde{f}_{a}}}^4 + 16\,{m_{f}}^4 + 16\,{m_{f}}^2\,s + 4\,s\,{\delta }^2+{\delta }^4 - 4\,s\,{\sigma }^2 \r\r\nn\\
&-&\l\l 2\,{\delta }^2\,{\sigma }^2 + 
    {\sigma }^4 + 8\,{m_{f}}^2\,\left( {\delta }^2 - {\sigma }^2 \right)  - 
    8\,{m_{\tilde{f}_{a}}}^2\,\left( 4\,{m_{f}}^2 + {\delta }^2 + {\sigma }^2 \right) \right)\r\nn\\ 
&+&\l 4\, D_{-\[ij\]}^{a} m_{f} \delta \(4\, m_{f}^2 - 4\,{m_{\tilde{f}_{a}}}^2+4\,s+\delta^2-5\,\sigma^2\)   \)\nn\\
&\times&\mathcal{F}\[s, \sigma,\delta,m_{f}^2,m_{f}^2,m_{\tilde{f}_{a}}^2\]\, ,\nn\\
\end{eqnarray}

Any symmetric or antisymmetric coupling function is defined $F_{\left\{ij\right\}}=F_{ij}+F_{ji}$ or $F_{\left[ij\right]}=F_{ij}-F_{ji}$.  In the above expressions we have used

\begin{eqnarray}
A_{\pm ij}^{ab} = C_{+ij}^{a} C_{+ij}^{b}\pm C_{-ij}^{a} C_{-ij}^{b}\, ,\\
\tilde{A}_{\pm ij}^{ab} = C_{+ij}^{a} D_{+ij}^{b}\pm D_{+ij}^{a} C_{+ij}^{b}\, ,\\
B_{\pm ij}^{ab} = D_{+ij}^{a} D_{+ij}^{b}\pm D_{-ij}^{a} D_{-ij}^{b}\, ,\\
\tilde{B}_{\pm ij}^{ab} = C_{-ij}^{a} D_{-ij}^{b}\pm D_{-ij}^{a} C_{-ij}^{b}\, ,\\
C_{\pm ij}^{ab} = C_{+ij}^{a} C_{+ji}^{b}\pm C_{-ij}^{a} C_{-ji}^{b}\, ,\\
\tilde{C}_{-ij}^{ab} = C_{+ij}^{a} D_{+ji}^{b}\pm D_{+ij}^{a} C_{+ji}^{b}\, ,\\
D_{\pm ij}^{ab} = D_{+ij}^{a} D_{+ji}^{b}\pm D_{-ij}^{a} D_{-ji}^{b}\, ,\\
\tilde{D}_{\pm ij}^{ab} = C_{-ij}^{a} D_{-ji}^{b}\pm D_{-ij}^{a} C_{-ji}^{b}\, ,
\end{eqnarray}

\begin{eqnarray}
C^{a}_{\pm ij} = \Lambda^{\(f\) L}_{n a i} \(\Lambda^{\(f\) L}_{n a j}\)^{*} \pm \Lambda^{\(f\) R}_{n a i} \(\Lambda^{\(f\) R}_{n a j}\)^{*}\, ,\\
D^{a}_{\pm ij} = \Lambda^{\(f\) L}_{n a i} \(\Lambda^{\(f\) R}_{n a j}\)^{*} \pm \Lambda^{\(f\) R}_{n a i} \(\Lambda^{\(f\) L}_{n a j}\)^{*}\, .
\end{eqnarray}

\section{Conclusions}

We have presented exact cross section expressions for all tree level two body final state processes $\chi_{i} \chi_{j} \rightarrow X$.  No approximations have been made in these calculations except for ignoring most CP-violating couplings in the SUSY sector.  These calculations are useful in the proper determination of the neutralino relic density, specifically in the application of any constraint on the dark matter density similar to $0.1 < \Omega_{\chi} h^2 < 0.3$.  The dark matter density constraint can be used to rule out large portions of parameter space for many models of broken supersymmetry.  Our results encompass the standard calculation, in which both initial state particles are the LSP ($i=j=1$).  Furthermore, our results also allow inclusion of all tree level contributions from coannihilation at least one of the initial state particles is not the LSP ($i$ and/or $j \neq 1$).  Such coannihilation contributions are important in mSUGRA models when the higgsino component of the lightest neutralino is significant and are also important in string-derived models with non-universal gaugino masses.

\section{Acknowledgements}
The authors wish to thank Manuel Drees, Mary K. Gaillard, and Takeshi Nihei for many useful discussions.  This work was supported in part by the Director, Office of Science, Office of High Energy and Nuclear Physics, of the U.S. Department of Energy under Contract DE-AC03-76SF00098, and in part by the National Science Foundation under grant PHY-00-98840.

\end{document}